\documentclass{aa} 
\usepackage{graphicx}
\usepackage{float}
\usepackage{txfonts}
\usepackage{longtable}
\usepackage{booktabs}
\usepackage{makecell}
\usepackage{multirow}
\usepackage{multicol}
\usepackage{xcolor}
\usepackage{natbib}
\usepackage{lscape}
\usepackage{tabularx}
\usepackage{subcaption}

\bibpunct{(}{)}{;}{a}{}{,} 

\newcommand{\MeV}{{\rm MeV~}}
\newcommand{\TeV}{{\rm TeV~}}
\newcommand{\beq}{\begin{equation}}
\newcommand{\eeq}{\end{equation}}

\def \degmark{^\circ}

\def \cmsec{\hbox{cm$^{2}$ s }}

\def \arcmin {\hbox{$^\prime$}}

\def \gray {$\gamma$-ray }
\def \grays {$\gamma$-rays }

\newcommand{\mt}{}
\newcommand{\new}{}
\newcommand{\fv}{ }

\newcommand{\fvr}{ }
\newcommand{\cp}{}
\newcommand{\cpbis}{ } 
\newcommand{\cpr}{ }
\newcommand{\abr}{ }
\newcommand{\abb}{ }

\graphicspath{{converted_graphics/}}
\begin{document}

\title{The Second AGILE Catalog of Gamma-Ray Sources }         

\author{A.~Bulgarelli$^{1}$,  V. Fioretti$^{1}$, N. Parmiggiani$^{1}$, F. ~Verrecchia$^{2,3}$,   C. Pittori$^{2,3}$,  F. ~Lucarelli$^{2,3}$, M. Tavani$^{4,5,6,7}$, A.~Aboudan$^{7,12}$, M. Cardillo$^{4}$, A.~Giuliani$^{8}$,  P.~W.~Cattaneo$^{9}$, A.W.~Chen$^{15}$, G.~Piano$^{11}$, A.~Rappoldi$^{9}$, L. Baroncelli$^{7}$,
A.~Argan$^{4}$,
L.A.~Antonelli$^{3}$,
I. Donnarumma$^{2}$,
F.~Gianotti$^{1}$,
P.~Giommi$^{16}$,
M.~Giusti$^{4,5}$,
F.~Longo$^{13,14}$,
A.~Pellizzoni$^{10}$,
M.~Pilia$^{10}$,
M.~Trifoglio$^{1}$,
A.~Trois$^{10}$,
S.~Vercellone$^{11}$,
A. Zoli$^{1}$
}

\institute{ 
$^{1}$ INAF-OAS Bologna, via Gobetti 93/3, I-40129 Bologna, Italy.\\
$^{2}$  ASI Space Science Data Center (SSDC), Via del Politecnico snc, 00133 Roma, Italy.\\
$^{3}$ INAF-Osservatorio Astronomico di Roma, Via di Frascati 33, I-00078 Monte Porzio Catone, Italy. \\
$^{4}$ INAF-IAPS Roma, via del Fosso del Cavaliere 100, I-00133 Roma, Italy.\\
$^{5}$ Dipartimento di Fisica, Universit\`a Tor Vergata, via della Ricerca Scientifica 1,I-00133 Roma, Italy.\\
$^{6}$ INFN Roma Tor Vergata, via della Ricerca Scientifica 1, I-00133 Roma, Italy.\\
$^{7}$ Consorzio Interuniversitario Fisica Spaziale (CIFS), villa Gualino - v.le Settimio Severo 63, I-10133 Torino, Italy.\\
$^{8}$ INAF-IASF Milano, via E. Bassini 15, I-20133 Milano, Italy.\\
$^{9}$ INFN Pavia, via Bassi 6, I-27100 Pavia, Italy.\\
$^{10}$ INAF-Osservatorio Astronomico di Cagliari, via della Scienza 5, I-09047, Selargius (CA), Italy. \\
$^{11}$ INAF Osservatorio Astronomico di Brera, Milano, Italy.\\
$^{12}$ CISAS, University of Padova, Padova, Italy.\\
$^{13}$ Dipartimento di Fisica, University of Trieste, via Valerio 2, 34127 Trieste\\
$^{14}$ INFN, sezione di Trieste, via Valerio 2, 34127 Trieste\\
$^{15}$ School of Physics, University of the Witwatersrand, 1 Jan Smuts Avenue, Braamfontein, Johannesburg, 2050 South Africa.\\
$^{16}$  ASI, Via del Politecnico snc, 00133 Roma, Italy.\\
}


\offprints{A. Bulgarelli, \email{andrea.bulgarelli@inaf.it} }
\date{accepted}
\authorrunning {A. Bulgarelli et al.}
\titlerunning {Second AGILE Catalog}


\abstract
{} 
{We present the second AGILE-GRID Catalog (2AGL) of \gray sources in the 100 MeV -- 10 GeV  energy range.} 
{With respect to previous AGILE-GRID catalogs, 
the {\mt current} 2AGL Catalog is based on the first 2.3 years of science data from the AGILE mission
(the so called `pointing mode') and incorporates more data and several analysis improvements,
including better calibrations at the event reconstruction level, an updated model for the Galactic diffuse \gray emission, a refined procedure for point-like source detection, and
the inclusion of {\mt a search for} extended \gray sources.}
{The 2AGL Catalog includes 175 high-confidence
sources (above $4\sigma$ significance) with their location regions and spectral properties, and a variability analysis with 4-day light curves for the most significant ones.
Relying on the error region of each source position, including systematic uncertainties,
121 sources are considered as positionally associated with known couterparts at different wavelengths or detected by other \gray instruments. Among the
identified or associated sources, 62 are Active Galactic Nuclei (AGNs) of the blazar class.
Pulsars represent the largest Galactic source class, with 40 associated pulsars, 7 of them with detected 
pulsation; 
8 Supernova Remnants and 4 high-mass X-ray binaries have also been identified.
A substantial number of 2AGL sources are unidentified:
for 54 sources no known  
counterpart is found at different wavelengths. Among these sources, 
we discuss a sub-class of 29 AGILE-GRID-only \gray sources that are not present in 1FGL, 2FGL or 3FGL catalogs; the remaining sources are unidentified in both 2AGL and 3FGL Catalogs. We also present an extension of the analysis of 2AGL sources detected in the 50 -- 100 MeV energy range.}
{}


\maketitle

\section{Introduction}

This paper presents the 2AGL Catalog of high-energy \gray sources 
{\mt detected by the AGILE Gamma-Ray Imager Detector (GRID)
in the energy range 100 MeV -- 10 GeV during the first 2.3 years of operations (2007-2009) in the so called `pointing mode'}.
This {\mt paper follows three previously published papers:} 
the First AGILE-GRID Catalog of \gray sources \citep[1AGL,][]{pittori09}, the Catalog of variable \gray sources during the first \abr{2.3 years} of observations \citep[1AGLR,][]{verrecchia13}, {\mt and a paper dedicated to}
the search of AGILE-GRID TeV source counterparts \citep{rappoldi16}.
{\mt Compared to previous investigations, we have implemented several 
refinements in the analysis of \gray sources }: 

\begin{enumerate}

\item  A new background event filter, called FM3.119, {\cp and new instrument response functions (IRFs), called H0025,
have been used}. The {\cp main} differences relative to the {\cp previous} F4 event filter used for {\cp the 1AGL Catalog} are an improved effective area ($A_\mathrm{eff}$) above 100 MeV and a {\cp better}  
characterisation of the Point Spread Function (PSF) \cite{chen13, sabatini15}. In addition, systematic errors of the IRFs {\cp have been estimated with greater accuracy.} 

\item  We {\cp analyse a larger} data set than in the 1AGL that was based on the first 12 months of observations, from July 13, 2007 to June 30, 2008. The 2AGL Catalog is, indeed, based on the entire `pointing mode' period {\cp corresponding} to 2.3 years from July 13, 2007 to October 15, 2009.

\item This Catalog employs a new diffuse Galactic emission model, in particular for the Galactic central region.

\item We have developed new methods for characterising and localising source 
{\cp candidate} \textit{seeds}, then evaluated for inclusion in the Catalog, using both wavelet techniques and an iterative approach.

\item  In the search for associations of AGILE-GRID sources with counterparts at different wavelengths, we have used new association procedures.

\item  A new version of the AGILE-GRID Science Tools has been used (BUILD25)
{\cpbis 
publicly available from 
the AGILE website at SSDC\footnote{http://agile.ssdc.asi.it}.
}

\item \abr{Energy dispersion \cite{chen13} has been taken into account in the analysis with the new Science Tools.}

\end{enumerate}

{\new The outline of the paper is as follows. In Sect.\ref{sec:prep} we describe the instrument, the {\cp data reduction, and the pointing strategy}. 
In Sect.~\ref{sec:diffusebkg} we describe the AGILE-GRID \gray background models used in the data analysis. 
We then present in Sect.~\ref{sec:construction} the analysis methods used to build the Second AGILE-GRID Catalog of \gray sources, {\cp 2AGL}. Limitations and systematic uncertainties are described in Sect.~\ref{sect:unc}. 
Our results and the list of the 2AGL \gray sources are shown in Sect.~\ref{sec:list},  
{\cpbis where potential counterparts at other wavelengths 
and correspondences with Fermi-LAT Catalog sources are also discussed. 
In Sect.~\ref{sec:notes} 
we comment on some specific 2AGL
sources, divided by classes or sky regions.
}
In Sect.~\ref{sec:list100} we reports an extension of the 2AGL Catalog where the sources detected in the 50 -- 100 MeV energy range are listed. Finally, in Sect.~\ref{sec:concl}, we discuss our results and make some concluding remarks.}

\section{AGILE-GRID instrument, data and observations}
\label{sec:prep}

\subsection{The AGILE-GRID instrument}
AGILE (Astrorivelatore Gamma ad Immagini LEggero) \cite{Tavani,Tavani1}
is a mission of the Italian Space Agency (ASI) devoted to \gray and X-ray astrophysics in the 30 MeV -- 50 GeV, and 18 -- 60 keV energy ranges, respectively.
{\new AGILE was successfully launched on 23 April 2007 in a $\sim 550$ km 
equatorial orbit with low inclination angle, $\sim 2.5\degmark$.
}

AGILE was the only mission entirely dedicated to high energy astrophysics above 30 MeV during the April 2007-June 2008 period.
Later it has been operating together with the Fermi Large Area Telescope (LAT), launched on June 11, 2008 \cite{GLAST,LAT}.
The highly innovative AGILE-GRID instrument is the first of the current generation of high-energy space missions based on solid-state silicon technology. 

\par
The AGILE payload detector consists of
{\new the Silicon Tracker (ST) \citep{2001AIPC..587..754B,2003NIMPA.501..280P, bulgarelli10, cattaneo11}
the Super-AGILE X-ray detector \citep{2007NIMPA.581..728F},
the CsI(Tl) Mini-Calorimeter (MCAL) \citep{labanti09},
and an AntiCoincidence (AC) system \citep{2006NIMPA.556..228P}.
The combination of ST, MCAL and AC forms the Gamma-Ray Imaging Detector (GRID).}
{\cp Accurate timing, positional and attitude information is provided by the 
Precise Positioning System and the two Star Sensors units.}
The ST is the core of the AGILE-GRID and
plays  \abr{two roles at the same time: it converts the \grays in heavy-Z material layers ($245 \: \mathrm{mm}$ of Tungsten, $0.07$ radiation length), where the photon interacts producing an $\mathrm{e^+e^-}$ pair in the detector, and records the electron/positron tracks by a sophisticated combination of Silicon microstrip detectors and associated readout, providing  3D hits.
The ST consists of a total of 12 trays, the first 10 with the Tungsten converter foil followed by two layers of 16 single-sided, $410 \: \mathrm{\mu m}$ thick, $9.5 \times 9.5 \: \mathrm{cm^2}$ silicon detectors} with strips orthogonal to each other, the last two trays consisting only of the silicon detectors. 
\abr{MCAL is composed of 30 CsI(Tl) scintillator bars each one $15 \times 23 \times 375 \: \mathrm{mm^3}$ in size, arranged in two orthogonal layers, for a total thickness of 1.5 radiation lengths. In each bar the readout of the scintillation light is accomplished by two custom PIN Photodiodes (PD) coupled one at each small side of the bar.}
\abr{The AC system is aimed at a very efficient charged particle background rejection. It completely surrounds all AGILE detectors (Super-AGILE, ST and MCAL). Each lateral face is segmented in three plastic scintillator layers (0.6 cm thick) connected to photomultipliers placed at the bottom of the panels. A single plastic scintillator layer (0.5 cm thick) constitutes the top-AC whose signal is read by four light photomultipliers placed at the four corners of the structure frame.}
The AGILE-GRID event processing is operated by on-board trigger logic algorithms
\citep{PDHU} and by on-ground event filtering (see Sect. \ref{sec:response}).  

\subsection{AGILE-GRID Response Characteristics}
\label{sec:response}


\textbf{Energy estimation and direction reconstruction.} The track reconstruction for energy estimation and event direction reconstruction is carried out by an AGILE-GRID specific implementation of the Kalman Filter technique \citep{giulianik1} and provides the incident direction and the energy of the events in the AGILE-GRID reference system.

\textbf{On-ground background event filter.} The FM3.119 is the currently used on-ground background event filter for the scientific analysis of the AGILE-GRID data. The filter assigns a classification flag to each event depending on whether it is recognised as a \gray event, a charged particle, a "single-track" event, or an event of uncertain classification (limbo).
The filter is based on a Boosted Decision Tree (BDT) technique; this technique is used with success in High Energy Physics (HEP) experiments \citep{yang05} in order to select events of interest, the so called signal events, out of numerous background events. The BDT technique maximises the signal-to-background ratio, efficiently suppressing the background events and, in the meanwhile, keeping a high signal detection efficiency. The selection is done on a majority vote on the result of several decision trees, which are all derived from the same training sample by supplying different event weights during the training. For the development of the FM3.119 filter these techniques have been tuned with one sample of Monte Carlo events, the training sample, and then tested with an independent Monte Carlo sample, the testing sample. From  these simulations 182 descriptor parameters of the interacting event inside the AGILE-GRID are extracted and used for training, with the aim of selecting a subset of these descriptor parameters as discriminant input variables for optimising the event separation. A final set of 57 discriminant variables has been selected, with an additional post-fitting set of cuts to further improve the signal-to-noise ratio, comprising also the previously developed more stringent F4 on-ground background event filter, optimised for a good pattern recognition of a subclass of \gray events. 



\textbf{Instrument Response Functions}. The effective area ($A_\mathrm{eff}$), the Point Spread Function (PSF), and the Energy Dispersion Probability (EDP), collectively referred to as the Instrument Response Functions (IRFs), depend on the direction of the incoming \gray in instrument coordinates. New efforts in the development of the background rejection filter FM3.119 led to reprocessing all AGILE-GRID data with the new IRFs H0025. The $A_\mathrm{eff}$ above 100 MeV is improved, with a precise characterization of the PSF with flight data \cite{chen13, sabatini15}. The IRFs I0023 analysed in \citep{chen13} are the same as H0025, except for a different boundary of two energy channels: we have 100 -- 400 MeV, 400 -- 1000 MeV in I0023, and  100 -- 300 MeV, 300 -- 1000 MeV in H0025. Both AGILE-GRID PSF and $A_\mathrm{eff}$ are characterised 
by a very good off-axis performance and are well calibrated up to almost $60\degmark$,
showing a very smooth variations with the angle relative to the instrument axis \citep{chen13}.
On-ground calibrations have been used also to characterise the performances of the FM3.119 filter and to validate the new IRFs \citep{cattaneo18}. In addition, systematic errors of the IRFs are better characterised (see Sect.~\ref{sect:systirf}).

\abr{
The scientific performances of the AGILE-GRID can be summarised as follows: $A_\mathrm{eff} \sim 400$ cm$^2$, FoV $\sim$ 2.5 sr, energy range 30 MeV -- 50 GeV, and  a PSF at $30\degmark$ off-axis for $E>100$ MeV of $2.1\degmark$, for $E > 400$ MeV of $1.1\degmark$, and for $E > 1$ GeV of $ 0.8\degmark$.
}

\subsection{Data reduction}

All AGILE-GRID data are routinely processed using the scientific
data reduction software tasks developed by the AGILE team and integrated into an automatic
pipeline system developed at the ASI Space Science Data Center (ASI/SSDC).
The first step of the data reduction pipeline converts on a contact-by-contact basis
{\new the satellite 
data time into Terrestrial Time (TT), and performs some
preliminary calculations and {\cp unit} conversions.}
A second step consists in the \gray event reconstruction with the
AGILE-GRID implementation of the Kalman Filter technique. 
The background event filter FM3.119 is then applied and a classification flag is assigned to each event.  
An AGILE auxiliary file (LOG) is then created, containing all the spacecraft information relevant
to the computation of the effective exposure and GTI (Good Time Interval).
Finally, the event direction in sky coordinates is reconstructed
and reported in the AGILE event files (EVT), excluding events
flagged as charged background particles.
This step produces the 
Level-2 (LV2) archive of LOG and EVT files, that have been used for the construction of this Catalog. The AGILE-GRID data obtained both in pointing and in spinning mode are publicly available from the ASI/SSDC\footnote{https://www.asdc.asi.it/mmia/index.php?mission=agilemmia}.

\subsection{Observations}
\label{sect:obs}

\begin{figure*}
	\includegraphics[width=\linewidth]{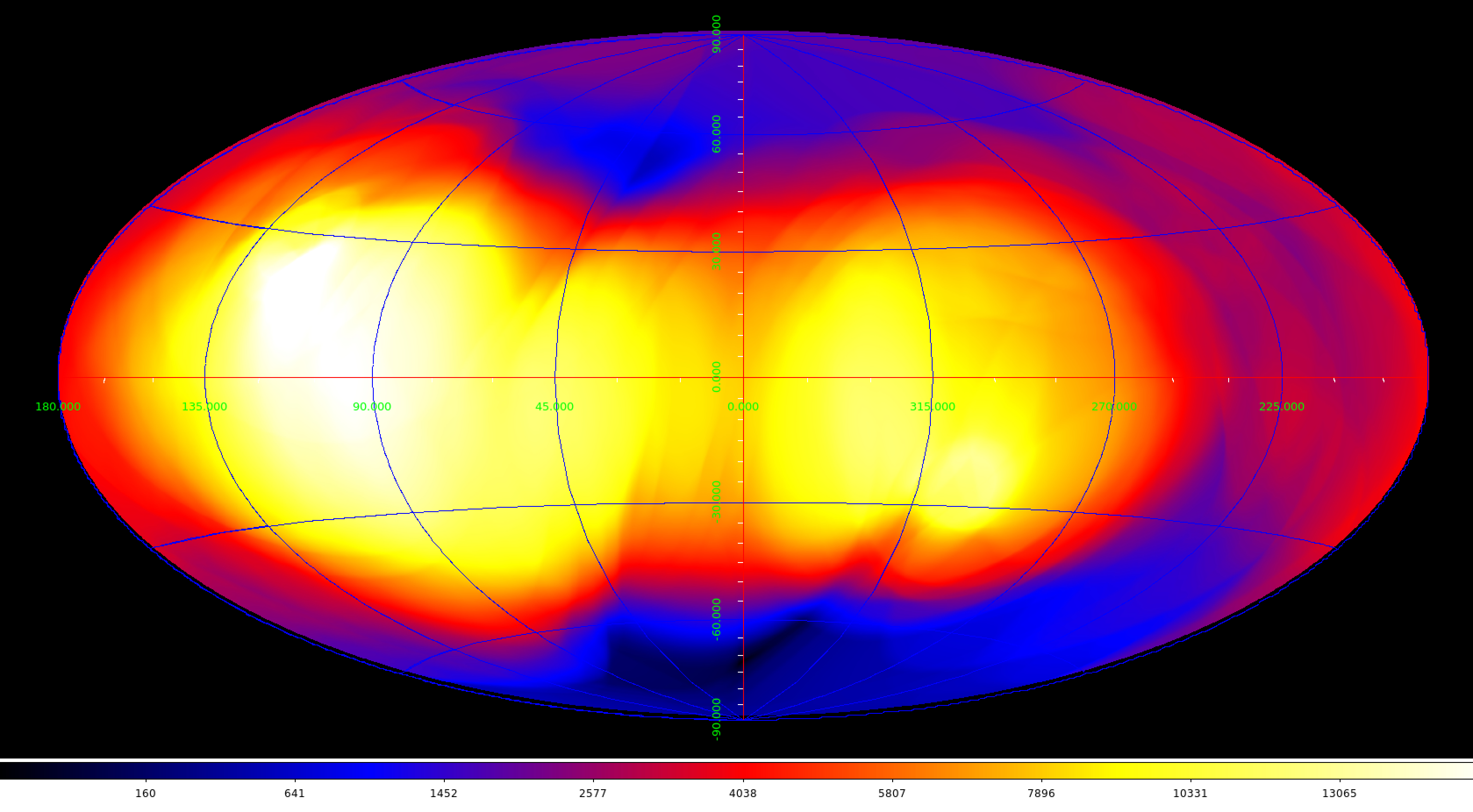}
    \caption{The squared root scaled exposure sky map in the 100 MeV -- 10 GeV energy range in Galactic coordinates and Hammer-Aitoff projection for the 2.3-year period analysed for the 2AGL Catalog (expressed in units of $\mathrm{cm^2 \ s \ sr}$). {\cpbis Bin size = $0.1\degmark$.}
    }
\label{fig:expo}
\end{figure*}

\begin{figure*}
    \includegraphics[width=\linewidth]{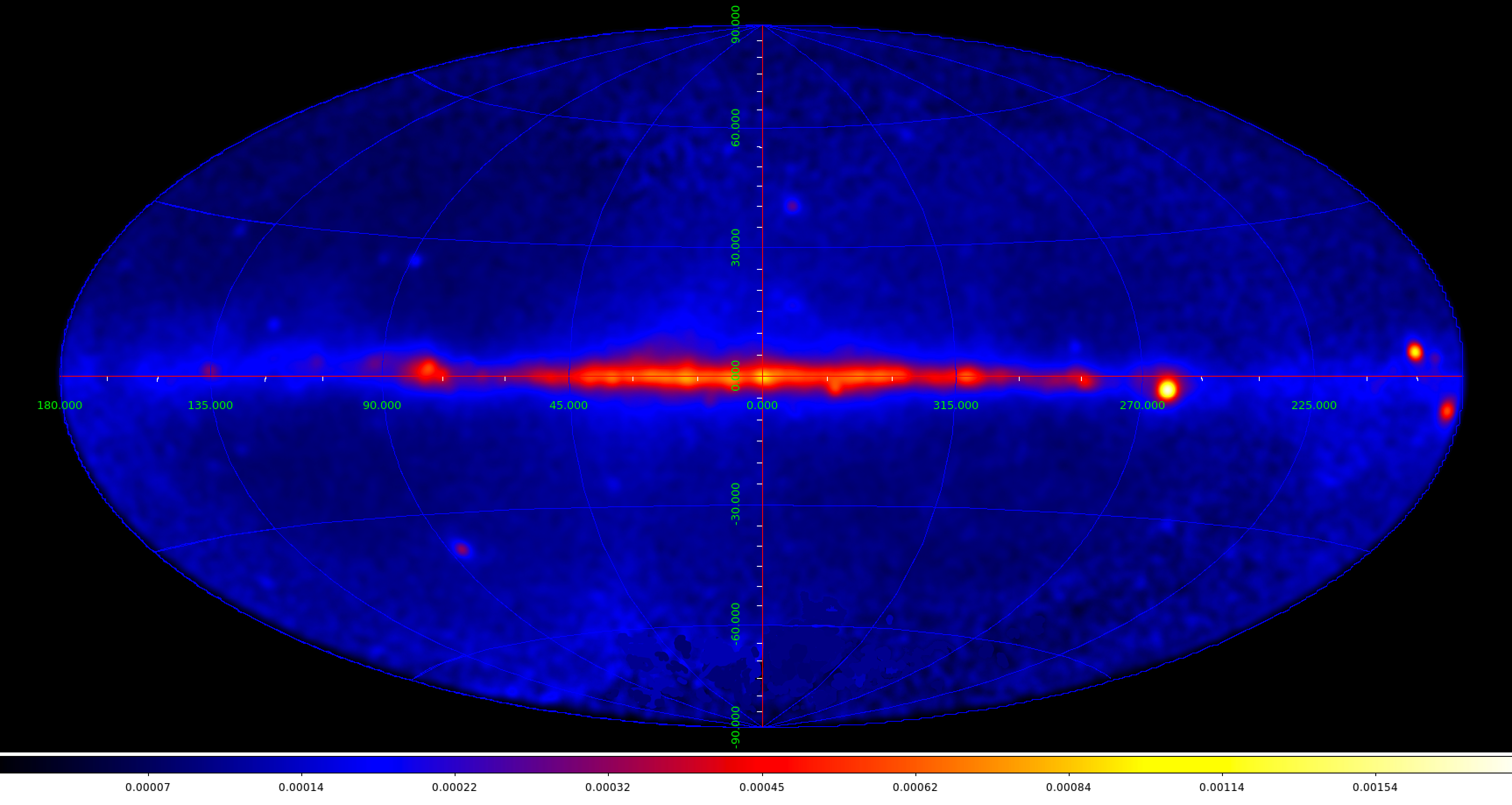}
    \caption{The intensity map in the 100 MeV -- 10 GeV energy band in Galactic coordinates and Hammer-Aitoff projection for the 2.3-year period analysed for the 2AGL Catalog (expressed in units of $\mathrm{ph \ cm^{-2} s^{-1} sr^{-1}}$). {\cpbis Bin size = $0.1\degmark$.}
    }
\label{fig:int}
\end{figure*}

The 2AGL Catalog sensitivity is not uniform, reflecting the in-homogeneous AGILE-GRID sky coverage {\cp during the `pointing period', 
with a mean exposure} focused mainly towards the Galactic plane: this means that the Catalog covers the entire sky with {\cp this} observational bias. In addition, the sensitivity is not intrinsically uniform over the sky due to the large range of brightness of the foreground diffuse Galactic \gray emission. The total \gray exposure and intensity maps 
obtained over the selected period with the FM3.119 filter, in Hammer-Aitoff projection and Galactic coordinates, are shown in Fig.~\ref{fig:expo} and Fig.~\ref{fig:int}, respectively. \abr{Exposure values span from 1 to 20235 $\mathrm{cm^2 \ s \ sr}$; $10\%$ of the pixels have a value of less than 1200 $\mathrm{cm^2 \ s \ sr}$},
{\cpbis   
which corresponds to about 16 days of effective exposure. This is the minimum exposure value corresponding to a 2AGL source detection.}

The AGILE Commissioning {\cp 
ended on July 9, 2007, and the following science verification phase (SVP)} lasted about four months, up to November 30, 2007. On December 1, 2007 the baseline nominal observations and pointing plan of AO Cycle-1 (AO-1) started with the Guest Observer program, with the AGILE spacecraft operating in 
{\cp 
pointing mode until October 15, 2009, and completing 101 pointings called observation blocks (OBs), see Table~\ref{tab:pointings}.  

The AGILE pointings are subject to illumination constraints requiring
that the fixed solar panels always be oriented within $3\degmark$ 
from the Sun direction. OBs usually consisted of predefined long exposures, drifting about $1\degmark$ per day
with respect to the initial boresight direction to obey solar panels
constraints.}
{\cpbis
The strategy that drove the pointing history during the first two years of observations 
(Cycle-1 and Cycle-2) reflected the need to achieve a good
balance between Galactic and extra-galactic targets as well as optimal
observability from both space- and ground-based facilities.

The AGILE Pointing Plan has been prepared taking into account 
several scientific and operational requirements such as:
\begin{itemize}
\itemsep0em
\item maximisation of the overall sky exposure factor 
 by limiting the observation of the sky regions more affected 
by Earth occultation;
\item substantial exposure of the Galactic plane and in 
particular of the Galactic Center and
of the Cygnus regions during Cycle-1 in order to achieve
long-time scale monitoring of Galactic \gray and hard X-ray sources;
\item maximisation of the scientific output of the mission during Cycle-2
in co-presence with the Fermi satellite, looking for
confirmation of transient activity from several candidates 
detected during Cycle-1.
\end{itemize}
The AGILE Pointing Plan was aimed in particular at reaching specific 
scientific goals, including:
\begin{itemize}
\itemsep0em
\item large photon counting statistic for \gray pulsar candidates;
\item improved positioning of the majority of unidentified \gray sources concentrated in the galactic plane;
\item micro-quasar studies with simultaneous hard X-ray and \gray data;
\item determination of the origin of \gray emission associated with a selected list of supernova remnants;
\item an improvement of the \gray Galactic diffuse emission model and of the Galactic cosmic-ray propagation and interaction in specific regions.
\end{itemize}
Because of the transient nature of the majority of 
extragalactic \gray sources and of many new \gray 
candidates in our Galaxy, and taking into account the large field of view of the AGILE-GRID,
the general strategy to reconcile extragalactic 
and Galactic investigations within a single observing plan, 
was to carry out 4-6 Target of Opportunity
(ToO) repointings per year due to source flaring activity. 
Nine repointings where actually carried out
due to Galactic or extra-galactic flaring activity, for a total
of 12 ToO Observation Blocks out of 101 
(3 of which have been extension of previous ones.)
These 9 ToO repointings had a minimal impact (about $9\%$) on the long-duration 
baseline coverage of the Galactic plane.
{\cpbis 
There is however an observational bias regarding a few well-known
high-latitude blazars such as 3C279 and PKS 0537-441, 
which in the 2AGL have an high average flux value because they
have been mainly observed in ToO pointings during flares.
}

A \gray flare monitoring program is active on daily basis since the beginning of the mission, with a dedicated alert system that is implemented within the AGILE Ground Segment and a Flare Advocate {\cpbis
working group. Details are reported in \citep{bulgarelli14, pittori13}.
}

Since November 2009}, due to a failure of the spacecraft reaction wheel, the attitude control system was reconfigured and the scientific mode of operation was changed. Currently the instrument operates in `spinning mode', i.e. the instrument 
scans the sky with an angular velocity of about $0.8\degmark \mathrm{s}^{-1}$, resulting in an exposure of about $7 \times 10^{6}$ \cmsec for about $70\%$ of the sky in one day.

\onecolumn

\fontsize{8}{8}\selectfont
{\setlength{\tabcolsep}{4pt}

\begin{longtable}{lrcccccc}
 
  \midrule[0.6pt]
  \toprule[1pt]

\label{tab:pointings} 
& & & & & & &  \\
\multicolumn{1}{l}{Region Name} &
\multicolumn{1}{c}{OB Number} &
\multicolumn{1}{c}{Starting RA, Dec} &
\multicolumn{1}{c}{Starting LII, BII} &
\multicolumn{1}{c}{Observation Start} &
\multicolumn{1}{c}{Observation End} &
\multicolumn{1}{c}{Observation Start} &
\multicolumn{1}{c}{Observation End} \\[3pt]
\multicolumn{1}{l}{} &
\multicolumn{1}{c}{} &
\multicolumn{1}{c}{J2000 (deg)} &
\multicolumn{1}{c}{(deg)} &
\multicolumn{1}{c}{(UTC)} &
\multicolumn{1}{c}{(UTC)} &
\multicolumn{1}{c}{(MJD)} &
\multicolumn{1}{c}{(MJD)} \\[3pt]
\toprule[0.6pt]


\endhead
\toprule[1pt]
\caption{The Observation Blocks (OB) of AGILE observations in "pointing mode".}
\endfoot

3C279 Region   &   900       &   195.596  ,  -6.649    &     307.8118 ,  56.1183  & 	 2007-07-09 12:00   & 2007-07-13 12:00 & 54290.500 & 54294.500 \\ [4pt]
VELA Region    &   1000      &   157.979  ,  -60.214   &     286.4188 ,  -1.8951  & 	 2007-07-13 12:00   & 2007-07-24 12:00 & 54294.500 & 54305.500 \\ [4pt]
ToO 3C 454.3   &   1100      &   17.829   ,  36.694    &     127.3645 , -26.0059  & 	 2007-07-24 12:00   & 2007-07-30 12:00 & 54305.500 & 54311.500 \\ [4pt]
ToO 3C 454.3   &   1150      &   17.829   ,  36.694    &     127.3645 , -26.0059  & 	 2007-07-24 12:00   & 2007-07-30 12:00 & 54305.500 & 54311.500 \\ [4pt]
VELA Region    &   1200      &   150.836  , -70.19    &     289.5293 , -11.8265  & 	 2007-07-30 12:00   & 2007-08-01 12:00 & 54311.500 & 54313.500 \\ [4pt]
SA Crab -45    &   1300      &   37.097   , 12.712    &     156.5885 , -43.7329  & 	 2007-08-01 12:00   & 2007-08-02 12:00 & 54313.500 & 54314.500 \\ [4pt]
VELA Region    &   1400      &   176.006  , -66.063   &     296.1593 ,  -4.0824  & 	 2007-08-02 12:00   & 2007-08-12 12:00 & 54314.500 & 54324.500 \\ [4pt]
SA Crab -35    &   1500      &   47.41    , 16.075    &     164.8343 , -35.3162  & 	 2007-08-12 12:00   & 2007-08-13 12:00 & 54324.500 & 54325.500 \\ [4pt]
VELA Region    &   1600      &   195.551  , -66.564   &     304.0044 ,  -3.7154  & 	 2007-08-13 12:00   & 2007-08-22 12:00 & 54325.500 & 54334.500 \\ [4pt]
SA Crab -25    &   1700      &   57.139   , 18.566    &     171.0790 , -27.3115  & 	 2007-08-22 12:00   & 2007-08-23 12:00 & 54334.500 & 54335.500 \\ [4pt]
VELA Region    &   1800      &   216.979  , -64.437   &     313.1071 ,  -3.4890  & 	 2007-08-23 12:00   & 2007-08-27 12:00 & 54335.500 & 54339.500 \\ [4pt]
Galactic Plane   &   1900    &   236.570  , -41.874   &     334.4369 ,  10.0581  & 	 2007-08-27 12:00   & 2007-09-01 12:00 & 54339.500 & 54344.500 \\ [4pt]
SA Crab (15,15)  &   2000    &   69.483  , 5.592     &     190.8962 , -26.2858  & 	 2007-09-01 12:00   & 2007-09-02 12:00 & 54344.500 & 54345.500 \\ [4pt]
SA Crab (0,15)   &   2100    &   68.205   , 20.566    &     177.1349 , -18.2781  & 	 2007-09-02 12:00   & 2007-09-03 12:00 & 54345.500 & 54346.500 \\ [4pt]
SA Crab (-15,15) &   2200    &   66.651   , 35.559    &     164.6334 ,  -9.3529  & 	 2007-09-03 12:00   & 2007-09-04 12:00 & 54346.500 & 54347.500 \\ [4pt]
Field 8         &   2300     &   51.408   , 71.022    &     134.8816 ,  11.8210  & 	 2007-09-04 12:00   & 2007-09-12 12:00 & 54347.500 & 54355.500 \\ [4pt]
SA Crab (0,5)   &   2400     &   78.535   , 21.730    &     182.1630 ,  -9.8874  & 	 2007-09-12 12:00   & 2007-09-13 12:00 & 54355.500 & 54356.500 \\ [4pt]
Field 8         &   2500     &   74.882   , 58.334    &     150.9906 ,   9.7255  & 	 2007-09-13 12:00   & 2007-09-15 12:00 & 54356.500 & 54358.500 \\ [4pt]
SA Crab (45,0)   &   2600    &   84.212   , -23.014   &     226.7035 , -26.1161  & 	 2007-09-15 12:00   & 2007-09-16 12:00 & 54358.500 & 54359.500 \\ [4pt]
SA Crab (5,0)   &   2700     &   82.987   , 16.983    &     188.5217 ,  -8.9833  & 	 2007-09-16 12:00   & 2007-09-17 12:00 & 54359.500 & 54360.500 \\ [4pt]
SA Crab (0,0)   &   2800     &   83.774   , 22.026    &     184.6179 ,  -5.6675  & 	 2007-09-17 12:00   & 2007-09-18 12:00 & 54360.500 & 54361.500 \\ [4pt]
SA Crab (-5,0)    &   2900   &   84.62    , 27.048    &     180.7737 ,  -2.3343  & 	 2007-09-18 12:00   & 2007-09-19 12:00 & 54361.500 & 54362.500 \\ [4pt]
SA Crab (-15,0)   &   3000   &   85.347   , 37.089    &     172.5873 ,   3.5179  & 	 2007-09-19 12:00   & 2007-09-20 12:00 & 54362.500 & 54363.500 \\ [4pt]
SA Crab (-25,0)   &   3100   &   86.174   , 47.118    &     164.2603 ,   9.2213  & 	 2007-09-20 12:00   & 2007-09-21 12:00 & 54363.500 & 54364.500 \\ [4pt]
SA Crab (-35,0)   &   3200   &   87.140   , 57.126    &     155.6110 ,  14.6016  & 	 2007-09-21 12:00   & 2007-09-22 12:00 & 54364.500 & 54365.500 \\ [4pt]
SA Crab (-45,0)   &   3300   &   88.348   , 67.136    &     146.4473 ,  19.4825  & 	 2007-09-22 12:00   & 2007-09-23 12:00 & 54365.500 & 54366.500 \\ [4pt]
SA Crab (0,-5)    &   3400   &   90.097   , 22.143    &     187.5419 ,  -0.5862  & 	 2007-09-23 12:00   & 2007-09-24 12:00 & 54366.500 & 54367.500 \\ [4pt]
SA Crab (15,0)    &   3500   &   91.034   , 7.141     &     201.1056 ,  -7.1395  & 	 2007-09-24 12:00   & 2007-09-25 12:00 & 54367.500 & 54368.500 \\ [4pt]
SA Crab (25,0)    &   3600   &   91.838   , -2.882    &     210.4602 , -11.1195  & 	 2007-09-25 12:00   & 2007-09-26 12:00 & 54368.500 & 54369.500 \\ [4pt]
SA Crab (35,0)    &   3700   &   92.502   , -12.926   &     220.0176 , -14.9489  & 	 2007-09-26 12:00   & 2007-09-27 12:00 & 54369.500 & 54370.500 \\ [4pt]
Crab Nebula       &   3800   &   94.323   , 22.050    &     189.5211 ,   2.7938  & 	 2007-09-27 12:00   & 2007-10-01 12:00 & 54370.500 & 54374.500 \\ [4pt]
SA Crab (0,-15)   &   3900   &   98.552   , 21.875    &     191.4932 ,   6.1922  & 	 2007-10-01 12:00   & 2007-10-02 12:00 & 54374.500 & 54375.500 \\ [4pt]
SA Crab (-15,-15) &   4000   &   100.839  , 36.784    &     178.6417 ,  14.3544  & 	 2007-10-02 12:00   & 2007-10-03 12:00 & 54375.500 & 54376.500 \\ [4pt]
SA Crab (15,-15)  &   4100   &   99.566   , 6.788     &     205.3927 ,   0.1791  & 	 2007-10-03 12:00   & 2007-10-04 12:00 & 54376.500 & 54377.500 \\ [4pt]
Crab Field        &   4200   &   101.724  , 21.699    &     192.9681 ,   8.7550  & 	 2007-10-04 12:00   & 2007-10-12 12:00 & 54377.500 & 54385.500 \\ [4pt]
SA Crab (0,-25)   &   4300   &   110.131  , 20.718    &     197.2281 ,  15.4667  & 	 2007-10-12 12:00   & 2007-10-13 12:00 & 54385.500 & 54386.500 \\ [4pt]
Gal. Center       &   4400   &   290.920  , -18.896   &      19.2683 , -15.4110  & 	 2007-10-13 12:00   & 2007-10-22 12:00 & 54386.500 & 54395.500 \\ [4pt]
SA Crab (0,-35)   &   4500   &   120.494  , 18.879    &     203.0392 ,  23.7444  & 	 2007-10-22 12:00   & 2007-10-23 12:00 & 54395.500 & 54396.500 \\ [4pt]
Gal. Center Reg.  &   4600   &   301.173  , -17.107   &      25.0972 , -23.6663  & 	 2007-10-23 12:00   & 2007-10-24 08:00 & 54396.500 & 54397.333 \\ [4pt]
ToO 0716+714      &   4610   &   148.939  , 67.888    &     143.3642 ,  41.5875  & 	 2007-10-24 08:00   & 2007-10-29 12:00 & 54397.333 & 54402.500 \\ [4pt]
ToO Extended   &   4630   &   157.461  , 66.942    &     141.5537 ,  44.7248  & 	 2007-10-29 12:00   & 2007-11-01 12:00 & 54402.500 & 54405.500 \\ [4pt]
SA Crab (0,-45)   &   4700   &   130.614  , 16.339    &     209.7914 ,  31.7351  & 	 2007-11-01 12:00   & 2007-11-02 12:00 & 54405.500 & 54406.500 \\ [4pt]
Cygnus Region     &   4800   &   296.880  , 34.501    &      69.5937 ,   4.6227  & 	 2007-11-02 12:00   & 2007-12-01 12:00 & 54406.500 & 54435.500 \\ [4pt]
Cygnus Field 1    &   4900   &   304.432  , 53.552    &      88.8156 ,   9.9272  & 	 2007-12-01 12:00   & 2007-12-05 09:00 & 54435.500 & 54439.375 \\ [4pt]
Cygnus Repointing &   4910   &   322.496  , 38.244    &      85.1187 ,  -9.4171  & 	 2007-12-05 09:00   & 2007-12-16 12:00 & 54439.375 & 54450.500 \\ [4pt]
Cygnus Repointing &   4920   &   322.496  , 38.244    &      85.1187 ,  -9.4171  & 	 2007-12-05 09:00   & 2007-12-16 12:00 & 54439.375 & 54450.500 \\ [4pt]
Virgo  Field      &   5010   &   173.433  , -0.437    &     265.6464 ,  56.7005  & 	 2007-12-16 12:00   & 2008-01-08 12:00 & 54450.500 & 54473.500 \\ [4pt]
Vela  Field       &   5100   &   147.060  , -62.517   &     283.4703 ,  -6.7881  & 	 2008-01-08 12:00   & 2008-02-01 12:00 & 54473.500 & 54497.500 \\ [4pt]
South Gal Pole    &   5200   &    58.347  , -37.795   &     240.3889 , -50.5780  & 	 2008-02-01 12:00   & 2008-02-09 09:00 & 54497.500 & 54505.375 \\ [4pt]
ToO MKN 421       &   5210  &   250.974  ,  50.293   &      77.3096 ,  40.6278  & 	 2008-02-09 09:00   & 2008-02-12 12:00 & 54505.375 & 54508.500 \\ [4pt]
South Gal Pole Repointing &   5220 & 65.660 , -35.714   &     237.5007 , -44.6737  & 	 2008-02-12 12:00   & 2008-02-14 12:00 & 54508.500 & 54510.500 \\ [4pt]
Musca Field       &   5300   &   191.934  , -71.893   &     302.6408 ,  -9.0241  & 	 2008-02-14 12:00   & 2008-03-01 12:00 & 54510.500 & 54526.500 \\ [4pt]
Gal. Center 1     &   5400   &   243.596  , -50.979   &     332.1063 ,   0.0207  & 	 2008-03-01 12:00   & 2008-03-16 12:00 & 54526.500 & 54541.500 \\ [4pt]
Gal. Center 2     &   5450   &   265.781  , -28.626   &     359.9782 ,   0.6280  & 	 2008-03-16 12:00   & 2008-03-30 12:00 & 54541.500 & 54555.500 \\ [4pt]
Anti-Center 1     &   5500   &   100.944  ,  21.711   &     192.6369 ,   8.1084  & 	 2008-03-30 12:00   & 2008-04-05 12:00 & 54555.500 & 54561.500 \\ [4pt]
SA Crab (8,24)    &   5510   &   108.283  ,  28.625   &     188.9607 ,  16.9953  & 	 2008-04-05 12:00   & 2008-04-07 12:00 & 54561.500 & 54563.500 \\ [4pt]
SA Crab (15,26)   &   5520   &   111.762  ,  35.688   &     183.0072 ,  22.2023  & 	 2008-04-07 12:00   & 2008-04-08 12:00 & 54563.500 & 54564.500 \\ [4pt]
Anti-Center 2     &   5530   &   110.404  ,  20.758   &     197.2962 ,  15.7167  & 	 2008-04-08 12:00   & 2008-04-10 12:00 & 54564.500 & 54566.500 \\ [4pt]
Vulpecula Field   &   5600   &   286.259  ,  20.819   &      53.0394 ,   6.4733  & 	 2008-04-10 12:00   & 2008-04-30 12:00 & 54566.500 & 54586.500 \\ [4pt]
North Gal Pole    &   5700   &   250.075  ,  72.497   &     104.8522 ,  35.4379  & 	 2008-04-30 12:00   & 2008-05-10 12:00 & 54586.500 & 54596.500 \\ [4pt]
Cygnus Field 2    &   5800   &   304.286  ,  35.974   &      74.0497 ,   0.2720  & 	 2008-05-10 12:00   & 2008-06-09 18:00 & 54596.500 & 54626.750 \\ [4pt]
ToO WComae ON+231 &   5810   &   182.285  ,  29.614   &     195.5016 ,  80.3738  & 	 2008-06-09 18:00   & 2008-06-15 12:00 & 54626.750 & 54632.500 \\ [4pt]
Cygnus Repointing &   5820   &   323.248  ,  50.079   &      93.6645 ,  -1.1664  & 	 2008-06-15 12:00   & 2008-06-30 12:00 & 54632.500 & 54647.500 \\ [4pt]
 Antlia Field      &   5900   &   161.83 ,  -47.73   &   282.31 ,  10.11  &  2008-06-30 12:00  & 2008-07-25 18:00 & 54647.500 & 54672.750 \\ [4pt]
 TOO 3C 454.3      &   5910   &    19.37 ,  38.09    &   128.56 , -24.49  &  2008-07-25 18:00  & 2008-07-31 12:00 & 54672.750 & 54678.500 \\ [4pt]
 Extension TOO 3C454.3    &   5920   &    25.09 ,  40.12    &   330.46 ,  28.98  &  2008-07-31 12:00  & 2008-08-15 12:00 & 54678.500 & 54693.500 \\ [4pt]
 Musca Field 2     &   6010   &   175.31 ,  -74.13   &   298.10 , -11.92  &  2008-08-15 12:00  & 2008-08-31 12:00 & 54693.500 & 54709.500 \\ [4pt]
 ToO SGR 0501+4516 &   6110   &    61.87 ,   44.06   &   333.90 ,  27.26  &  2008-08-31 12:00  & 2008-09-10 12:00 & 54709.500 & 54719.500 \\ [4pt]
 Gal. Center 3     &   6200   &   256.55 ,  -28.53   &   355.51 ,   7.40  &  2008-09-10 12:00  & 2008-10-10 12:00 & 54719.500 & 54749.500 \\ [4pt]
 ToO PKS 0537-441  &   6210   &    98.80 , -46.77    &   255.44 , -22.05  &  2008-10-10 12:00  & 2008-10-17 12:00 & 54749.500 & 54756.500 \\ [4pt]
 Aquila Field      &   6310   &   290.97 ,  10.10    &    45.62 ,  -2.51  &  2008-10-17 12:00  & 2008-10-31 12:00 & 54756.500 & 54770.500 \\ [4pt]
 Cygnus Field 3    &   6400   &    295.52 ,  35.64   &    70.03 ,   6.15  &  2008-10-31 12:00  & 2008-11-30 12:00 & 54770.500 & 54800.500 \\ [4pt]
 Cygnus Field 4    &   6500   &    320.40 ,  35.50   &    81.95 , -10.17  &  2008-11-30 12:00  & 2008-12-20 12:00 & 54800.500 & 54820.500 \\ [4pt]
 Cygnus Field 5    &   6600   &    334.10 ,  44.05   &    95.70 , -10.47  &  2008-12-20 12:00  & 2009-01-12 18:00 & 54820.500 & 54843.750 \\ [4pt]
 ToO Carina Field  &   6610   &    161.67 , -59.86   &   287.86 ,  -0.69  &  2009-01-12 18:00  & 2009-01-19 18:00 & 54843.750 & 54850.750 \\ [4pt]
 Cygnus Field 6    &   6710   &    325.75 ,  68.11   &   106.75 ,  11.37  &  2009-01-19 18:00  & 2009-02-28 12:00 & 54850.750 & 54890.500 \\ [4pt]
 Gal.Center 4      &   6800   &    247.20 , -29.03   &   349.85 ,  13.43  &  2009-02-28 12:00  & 2009-03-25 12:00 & 54890.500 & 54915.500 \\ [4pt]
 Gal.Center Prolonged  &   6810  &    275.73 , -30.50   &     2.59 ,  -7.83  &  2009-03-25 12:00  & 2009-03-31 12:00 & 54915.500 & 54921.500 \\ [4pt]
 Crab Field        &   6910   &    102.70 ,  31.71   &   184.07 ,  13.75  &  2009-03-31 12:00  & 2009-04-07 12:00 & 54921.500 & 54928.500 \\ [4pt]
 Aquila Field 1    &   7010   &    288.88 , -19.31   &    18.06 , -13.82  &  2009-04-07 12:00  & 2009-04-15 12:00 & 54928.500 & 54936.500 \\ [4pt]
 Aquila Field 2    &   7100   &    290.88 ,  16.16   &    50.92 ,   0.44  &  2009-04-15 12:00  & 2009-04-30 12:00 & 54936.500 & 54951.500 \\ [4pt]
 Cygnus Field 7    &   7200   &    299.11 ,  29.78   &    66.49 ,   0.59  &  2009-04-30 12:00  & 2009-05-15 12:00 & 54951.500 & 54966.500 \\ [4pt]
 Vela Field 2      &   7300   &    127.35 , -37.14   &   256.41 ,   1.08  &  2009-05-15 12:00  & 2009-05-25 18:00 & 54966.500 & 54976.750 \\ [4pt]
 3rd ToO 3C454.3   &   7310   &    328.44 ,  10.91   &    68.32 , -32.54  &  2009-05-25 18:00  & 2009-05-29 12:00 & 54976.750 & 54980.500 \\ [4pt]
 Restart Vela Field 2 &   7320  &    136.50 , -40.71   &   263.67 ,   4.38  &  2009-05-29 12:00  & 2009-06-04 12:00 & 54980.500 & 54986.500 \\ [4pt]
 Virgo Field 2     &   7410   &    167.14 ,  10.71   &   242.13 ,  60.76  &  2009-06-04 12:00  & 2009-06-15 12:00 & 54986.500 & 54997.500 \\ [4pt]
 Cygnus Field 8    &   7500   &    330.22 ,  43.11   &    92.83 ,  -9.58  &  2009-06-15 12:00  & 2009-06-25 12:00 & 54997.500 & 55007.500 \\ [4pt]
 Cygnus Field 9    &   7600   &    344.77 ,  37.90   &    99.66 , -19.84  &  2009-06-25 12:00  & 2009-07-15 12:00 & 55007.500 & 55027.500 \\ [4pt]
 Cygnus Field 10   &   7700   &    330.35 ,  64.26   &   105.73 ,   7.23  &  2009-07-15 12:00  & 2009-08-12 12:00 & 55027.500 & 55055.500 \\ [4pt]
 Vela Field 3      &   7800   &    202.30 , -62.10   &   307.33 ,   0.45  &  2009-08-12 12:00  & 2009-08-31 12:00 & 55055.500 & 55074.500 \\ [4pt]
 Norma Field       &   7900   &    243.77 , -35.45   &   343.05 ,  11.11  &  2009-08-31 12:00  & 2009-09-10 12:00 & 55074.500 & 55084.500 \\ [4pt]
 SA Crab (15,6)    &   8000   &     78.33 ,  6.66    &   195.06 , -18.31  &  2009-09-10 12:00  & 2009-09-13 12:00 & 55084.500 & 55087.500 \\ [4pt]
 SA Crab (25,3)    &   8100   &     81.78 , -3.12    &   205.88 , -20.15  &  2009-09-13 12:00  & 2009-09-16 12:00 & 55087.500 & 55090.500 \\ [4pt]
 Galactic Center 5 &   8200   &    263.19 , -23.49   &   232.78 , -28.89  &  2009-09-16 12:00  & 2009-09-30 12:00 & 55090.500 & 55104.500 \\ [4pt]
 Aquila Field 3    &   8300   &    278.13 , -23.22   &    10.11 ,  -6.45  &  2009-09-30 12:00  & 2009-10-15 12:00 & 55104.500 & 55119.500 \\ [4pt]

\end{longtable}

}

\normalsize

\twocolumn



\section{Background Modeling}
\label{sec:diffusebkg}

\subsection{Galactic diffuse \gray background}
\label{sec:bkggal}

The diffuse \gray background is the primary component of the background. It is assumed to be produced by the interaction of Cosmic Rays (CR) with the Galactic interstellar medium, the Cosmic Microwave Background (CMB) and the InterStellar Radiation Field (ISRF) through three physical processes:
hadron-hadron collision, Bremsstrahlung and inverse Compton emission. 
The model for the diffuse \gray background has been updated for the 2AGL Catalog, with an update of the Galactic Centre region diffuse \gray emission and convolved with the new IRFs H0025.


The AGILE diffuse emission model \cite{giuliani04} substantially improves the previous EGRET model by using Neutral Hydrogen (HI) and CO
updated maps in order to model the matter distribution in the Galaxy.
It is based on a 3-D grid with $0.1\degmark \times 0.1\degmark$  binning in Galactic longitude and latitude, and a 0.2 kpc step in distance along the line of sight. 
Concerning the distribution of neutral hydrogen, we use the Leiden-Argentine-Bonn (LAB) survey of Galactic HI \cite{kalberla05}. The LAB survey improves 
the previous results especially in terms of sensitivity (by an order of magnitude),
velocity range and resolution. 
In order to properly project the velocity-resolved radio data, we use the Galactic rotation curves parameterised by \citep{clemens85}.
The  detailed and relatively
high-resolution distribution of molecular hydrogen is obtained from the CO observations described in \cite{dame01}. The CO is assumed to be a tracer of molecular hydrogen, through a known ratio between hydrogen density and CO radio emissivity.

\begin{figure*}
    \includegraphics[width=\linewidth]{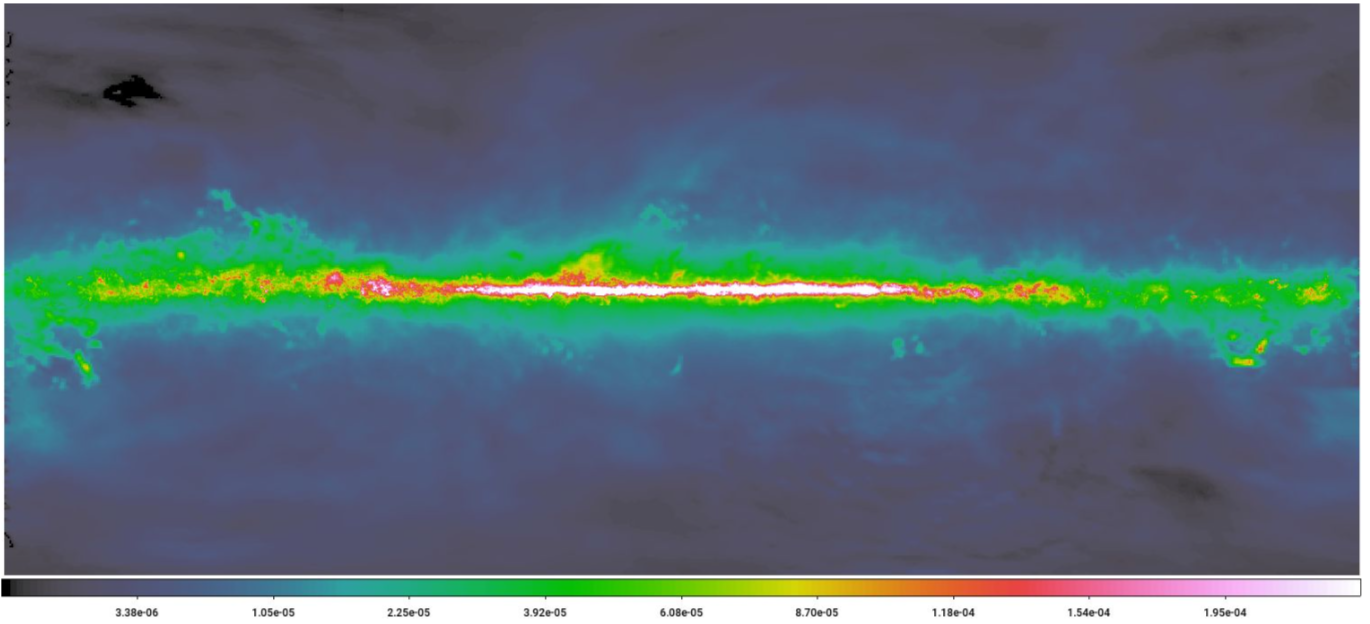}
     \caption{AGILE diffuse \gray background emission model \cite{giuliani04} in the 300 MeV -- 1 GeV energy range, in $\mathrm{ph \: cm^{-2} s^{-1} sr^{-1}}$, in Plate Carree projection with a  bin size of {\cpbis $0.1\degmark$.}
     } 
     \label{fig:diff}
\end{figure*}

Cosmic rays can emit \grays through the inverse Compton mechanism due to their interaction with photons of the CMB and the ISRF. In order to account for the latter component we use the analytical model proposed by \cite{chi91}. It describes the ISRF as the result of three main contributions: far
infrared (due to dust emission), near infrared, and optical/UV (due to stellar emission). The CR distribution (both protons and electrons) in the Galaxy is obtained using the GALPROP CR model \citep{strong00}. 
As {\cpbis an} example, Fig.~\ref{fig:diff} reports the AGILE diffuse \gray background emission model convolved with PSF and energy dispersion in the 300 MeV -- 1 GeV energy range.

\subsection{Isotropic background}
\label{sec:bkgiso}
{\cpbis The (quasi) isotropic background includes both a 
contribution from the cosmic extragalactic diffuse emission 
as well as a component of noise due to residual cosmic-ray
induced backgrounds at the detector level.
This residual particle background is dominant 
in the AGILE data used for this analysis 
(based on the standard filter FM3.119)\footnote{The estimate of the 
pure cosmic extragalactic diffuse emission 
is out of the scope of this paper, and it would require much more
stringent requirements on the purity of the
gamma-ray event selection \cite{ackermann15}.}. 
We evaluate the isotropic background with the Maximum Likelihood Estimator
(MLE) method in each region within the radius of analysis 
of each gamma-ray candidate source (see Sect. 4.1). 
This background is represented by a parameter for each energy bin in the MLE
(that could be left free or fixed after its evaluation), 
taking typical values between $\mathrm{4-8 \times 10^{-5}\, cts \ cm^{-2} s^{-1} sr^{-1}}$ in the energy range 100 MeV -- 10 GeV.}


\subsection{Residual Earth limb}

At the $\sim 550$ km altitude of the (equatorial nearly-circular) orbit of AGILE, the limb of the Earth is an intense source of \grays from CR collisions with the upper atmosphere, and during the observations the AGILE-GRID Field of View generally subtended part of the Earth limb. Even if a residual component of limb emission remains in the data, thanks to an effective on-board background rejection filtering \citep{PDHU}, {\cpbis to further reduce 
gamma-ray Earth-albedo contamination we limit the data selection and exposure calculations  excluding photons coming within 80$\degmark$ from 
the reconstructed satellite-Earth vector (\textit{albrad}=80), as in the 1AGL and 1AGLR catalogs.
%
The AGILE TeVCat \citep{rappoldi16} uses a more conservative angle cut with a value of \textit{albrad}=85$\degmark$.
}

\section{Construction of the Catalog}
\label{sec:construction}

The 2AGL Catalog is a catalog of point-like and extended sources. In this section we report the procedure to construct the 2AGL Catalog, emphasising the differences with respect to the 1AGL Catalog. Most of the procedure reported here is relevant for point-like sources. The analysis of extended sources is described in {\cp Sect.~\ref{sect:extended}.}

The basic analysis steps are source detection and localization, significance estimation, and spectral shape {\cp determination}. Each step has been performed with the AGILE-GRID Science Tools (version BUILD25).

\subsection{General analysis method}
\label{sec:mle2}

All analyses are performed using a binned MLE (Maximum Likelihood Estimator) method. The analysis depends on the \abr{isotropic and} Galactic diffuse emission, the \gray photon statistics, the IRFs as a function of energy and off-axis angle, 
and on the background filtering.

A likelihood ratio test is used to compare two ensembles of models, each model is a linear combination of parameters for point-like and extended sources, \abr{isotropic and} Galactic diffuse \gray background components of the \gray emission, and adding also point-like sources with fixed \gray emission; the two ensembles of models are one a parameter subset of the other. For each ensemble of models the set of free parameters is estimated by fitting the model with the data and computing the maximum likelihoods, where $L_0$ is the likelihood for the null hypothesis and $L_1$ is the likelihood for the alternative hypothesis. The likelihood ratio test is ${L_0}/{L_1}$, and the test statistic $TS$ is defined as $TS = - 2 \ln ({L_0}/{L_1})$ \citep{1996ApJ...461..396M, bulgarelli12a}. The test statistics $TS$ is used for quantifying how significantly a source emerges from the background \citep{wilks38}. 

To describe a single point-like source, between four and six parameters are used (two for the position, one for the predicted counts and the remaining for the shape parameters of the spectral model, see Sect.~\ref{sect:spectralshape}): the results are the predicted source counts, the values of the spectral shape parameters, and the position of the source in Galactic coordinates.     
The parameters 
reported in this paper are estimated by a likelihood analysis of the $10\degmark$ field surrounding the sources and considering nearby sources.

Among the parameters \abr{evaluated by MLE}, the background is described by the coefficients of the 
{\cpbis
Galactic}
diffuse (see Sect.~\ref{sec:bkggal}) and isotropic (see Sect.~\ref{sec:bkgiso}) background. We have two parameters for each energy bin to describe the Galactic (diffuse) and isotropic \gray emission: (i) $g_\mathrm{gal}$, the coefficient of the Galactic diffuse emission model, and (ii) $g_\mathrm{iso} \times 10^{-5} \mathrm{cts \ cm^{-2} s^{-1} sr^{-1}}$, the isotropic diffuse intensity. A value of $g_\mathrm{gal} < 1$ is expected if the Galactic diffuse emission model is correct. 
{\cpbis
An average value of $\Bar{g}_\mathrm{gal}=(0.46 \pm 0.09)$ is obtained
in this analysis.}


\abr{The parameters kept free are estimated by the MLE}. It is possible to keep each parameter either free or fixed; a free parameter is allowed to vary to find the maximum likelihood. The point-like source parameters are varied  with the following possible combinations: (1) variation of only the flux, (2) variation of position and flux, (3) variation of spectral shape and flux, or (4) variation of all parameters. For the 2AGL Catalog the position, the flux and the spectral model parameters are estimated in the same procedure with a global fitting that takes care, at the same time, e.g. that a shifted position would affect the spectral models or the positions of nearby sources. For some cases described hereafter the number of free parameters is reduced (see Sect.~\ref{sect:loc} and \ref{sect:spectralshape}).

To describe an extended source we produce a template of the shape of the extended emission at the expected position and we fit this shape with data, with a fixed spectral index $\alpha$ (we use $\alpha=2.1$) of a Power Law (the only available spectral shape for extended source provided by the AGILE Science Tools), allowing the predicted counts to vary. The photons are binned into FITS count maps. The \gray exposure maps, and Galactic diffuse emission maps are then used to calculate the parameters of the models. 
Particular care is required to carry out the analysis
in regions of the Galactic plane that are characterised by a relatively high and structured 
flux of the diffuse Galactic emission, as well as in regions near bright
\gray sources leading to possible source confusion.

\subsubsection{Localisation}
\label{sect:loc}
The position of each source is determined by maximising the likelihood with respect to its position, keeping the other parameters of the point-like source free. 
For each source we evaluate the $95\%$ elliptical and circular confidence regions. 

The AGILE Science Tools perform the positional and spectral shape optimisation at the same time, but sometimes the contour is not evaluated during the spectral shape evaluation, due to the high number of free parameters; as consequence, it is not possible to obtain confidence regions optimised with the spectral shape, even if the best position of the source is correctly evaluated by MLE. To overcome this problem we evaluate the best position and elliptical confidence region reducing the number of free parameters: all these sources are marked with flag 5 (see Table~\ref{tab:cat2flags}).



\subsubsection{Spectral Models}
\label{sect:spectralshape}

We perform a full energy band spectral fit of the data to incorporate the constraint that the spectral shape should smoothly vary with energy. The 1AGL and 1AGLR Catalogs considered only Power Law (PL) spectra; this was a simpler approach but not a good spectral representation for bright sources. With the exposure increasing, the discrepancies between PL and curved spectra could affect the global fit of the source, altering the spectra of nearby sources. Increasing the number of free parameters means that finding the true best fit is more difficult and, therefore, only spectra with one or two additional parameters are considered. The spectral representations used in the 2AGL Catalog are Power Law, exponential cut-off Power Law, super-exponential cut-off Power Law, and Log Parabola.

The Power Law spectral model (PL) is used for all sources not significantly curved and with low exposure:
\begin{equation}
\frac{dN}{dE} = N_0 E^{-\alpha},
\end{equation}
{\cpbis
where $N_0$ is the prefactor, and alpha is the index explicitly evaluated by the MLE method. Our MLE spectral fitting does not explicitly output the prefactor  value, which is internally calculated by the numerical procedure.}

The majority of the 2AGL sources are described by a Power Law. With the exception of the brightest sources, the AGILE-GRID analysis may not be spectrally resolved due to low statistics. In this case the Power Law spectral model is assumed and in general, a fixed spectral index $\alpha=2.1$ is adopted for the initial step of the MLE analysis. 
The exponential cut-off Power Law spectral model (PC) is 

\begin{equation}
\frac{dN}{dE} = N_0 E^{-\alpha} \exp \Bigg(-\frac{E}{E_\mathrm{c}}\Bigg),
\end{equation}

where $N_0$ is the prefactor,  $\alpha$ is the index, and $E_\mathrm{c}$ is the cut-off energy. $E_\mathrm{c}$ and $\alpha$ are explicitly provided by the MLE method. The super exponential cut-off Power Law spectral model (PS) is 


\begin{equation}
\frac{dN}{dE} = N_0 E^{-\alpha} \exp \Bigg(-\Bigg(\frac{E}{E_c}\Bigg)^\beta \Bigg),
\end{equation}

where $N_0$ is the prefactor, 
$\alpha$ is the first index, $\beta$ the second index, and $E_c$ is the cut-off energy. The parameters $\alpha$, $E_\mathrm{c}$ and $\beta$ are explicitly provided by the MLE method. 

The Log Parabola spectral model (LP) is 


\begin{equation}
\frac{dN}{dE} = N_0 E^{-\alpha - \eta \;\ln(E/E_0)},
\end{equation}

where $N_0$ is the prefactor, 
$\alpha$ is the first index, $\eta$ the curvature. The parameters $\alpha$, $E_\mathrm{c}$ and $\eta$ are explicitly provided by the MLE method. 

In order to select the best spectral shape for every source, we perform a full spectral fit of the data with the spectral representations listed in this section. The MLE estimator does not converge with all spectral shapes: this can be due to poor statistics or to the presence of too many parameters in the spectral model. Another common problem is that, even if there is a fit convergence, the estimated parameters are too close to their limits or their errors are greater than the values of the parameters themselves: in these cases the fit is discarded. Our selection of curved spectra follows the acceptance criteria described in \citep{nolan2012}. Briefly, a source is considered significantly curved if $TS_\mathrm{curved} > 16$, where $TS_\mathrm{curved} = 2\times(\log L(\textit{curved spectrum}) - \log L(\textit{power law})$, where L is the likelihood function obtained changing only the spectral representation of that source and refitting all free parameters.


\subsubsection{Upper limit calculation}

Upper limits are calculated using the same technique used for the asymmetrical errors for detected sources. We find the point-like source flux which maximises the likelihood. 


\abr{The calculated upper limit is a conservative value guaranteed to be at or above the upper limit 
{\cpbis
of the confidence interval.
}
It is calculated using the following simple formula: $UL = \Delta F_+ + |F|$ where $\Delta F_+$ is the positive error in the flux and $|F|$ is the absolute value of the flux. In the 2AGL Catalog we report the $2\sigma$ upper limits. }

For very faint sources (in a single energy band or for the full energy band during the variability analysis) when $TS < 1$ we have used a Bayesian method \citep{helene1983}. The upper limit is found by integrating the likelihood from 0 up to the flux that encompasses $95\%$ of the posterior probability: in this way the upper limits calculated with both methods are similar for sources with $TS = 1$.

\subsection{Binned sky maps preparation}

In order to merge the data from different observing periods over the whole sky, we have produced sets of sky maps 
in the ARC projection \citep{2002A&A.395..1077C} in Galactic coordinates. Cataloged sources are detected by merging all the available data over the entire time period.

Different sets of counts and exposure maps were produced with the AGILE-GRID standard software package. We report choice of parameters for maps generation of all sets, reporting in parenthesis the parameters values to be used in the software package distributed to AGILE Guest Observers.

To reduce the particle background contamination, only events tagged as con\-fir\-med
\gray events were selected ($filtercode=5$). The South Atlantic Anomaly data were excluded ($phasecode=6$) and all \gray events whose reconstructed directions with respect to
 the satellite-Earth vector is smaller than $80\degmark$ ($albrad=80$) were
also rejected, in order to eliminate the Earth albedo contamination. 

The {\cp considered energy range for the 2AGL source analysis} is 100 MeV -- 10 GeV. To reduce the uncertainty in the reconstruction of events, we have selected only photons with a reconstructed direction within $50\degmark$ from the boresight ($fovradmax=50$). 

\subsection{Determination of seeds }
\label{sec:seeds}

The detection and localization procedure is basically iterative, starting from a list of seeds. The seeds are the initial sky positions of the candidate point-like sources.
The process starts without any set of input sources, to avoid any kind of biases from different data sets. 

A tiling of the sky is created ("pixelization"). For each region of the sky, the initial set of candidate sources has been determined using blind search techniques as a wavelet-based method ("wavelet algorithm") and generating significance maps ($TS$  maps) iteratively.
Each region has been optimised independently. At the end of this step we get an independent list of seeds for each region of the sky. 

\textbf{Pixelization.} To create a tile of the sky with a sufficient resolution, we have used 3072 circular regions (hereafter called {\cp``}rings") centred on points defined by HEALPix (Hierarchical Equal Area isoLatitude Pixelization) \citep{2005ApJ...622..759G} tessellation with $N_\mathrm{side} = 16$. We have produced binned maps of $0.5\degmark$ (used only for test) and $0.1\degmark$ bin size with a side of $30\degmark$ for each tile 
{\cpbis in Galactic coordinates,}
whose centres lay at constant latitude, with a unique energy bin of 100 MeV -- 10 GeV.
The tiles are discrete, overlapping and not independent. HEALPix algorithm produces a subdivision of a spherical surface in which each pixel covers the same surface area as every other pixel.
Note, however, that here we do not use the HEALPix projection but only a property of its grid; the pixel centres occur on a discrete number of rings of constant latitude in order to represent all-sky binned \gray data.

\textbf{Wavelet algorithm.} A Continuous Wavelet Transform (CWT) is used to determine the first list of seeds. CWT analyses a signal at different scales and is computed convolving the signal under investigation with the dilated and translated version of a wavelet function. When the support of the wavelet is small the CWT reacts mainly to high frequencies while, as the dilation increases, the wavelet support increases and the CWT is able to detect the lower frequency components of the signal \citep{louis97}.  In this work we used the negative of the Laplacian of the Gaussian, called Marr or Mexican Hat wavelet:

\begin{equation}
\psi(x)=(2-\|x\|^2)\exp\left(2-\frac{\|x\|^2}{2}\right)
\end{equation}

This wavelet has a positive kernel surrounded by a negative annulus. Its positive kernel has a Gaussian-like shape very similar to the AGILE-GRID Point Spread Function (PSF) hence it is effective in detecting point-like sources. It has a limited extent both in spatial and Fourier domains that guarantees good localization performance and limited aliasing effects \citep{freeman02}.

We use binned maps of $0.1\degmark \times 0.1\degmark$. The CWT of these maps consists of a three-dimensional grid of pixels with the third dimension corresponding to the scales at which the transform is computed. We use a dyadic scale starting from one pixel up to the map size. The CWT at scales between 1 to 5 pixels provides evidence of point-like sources, some extended sources or clusters of sources are evident at scales between 5 and 10 pixels, while at higher scales the background is clearly identified.

\begin{figure*}
    \includegraphics[width=\linewidth]{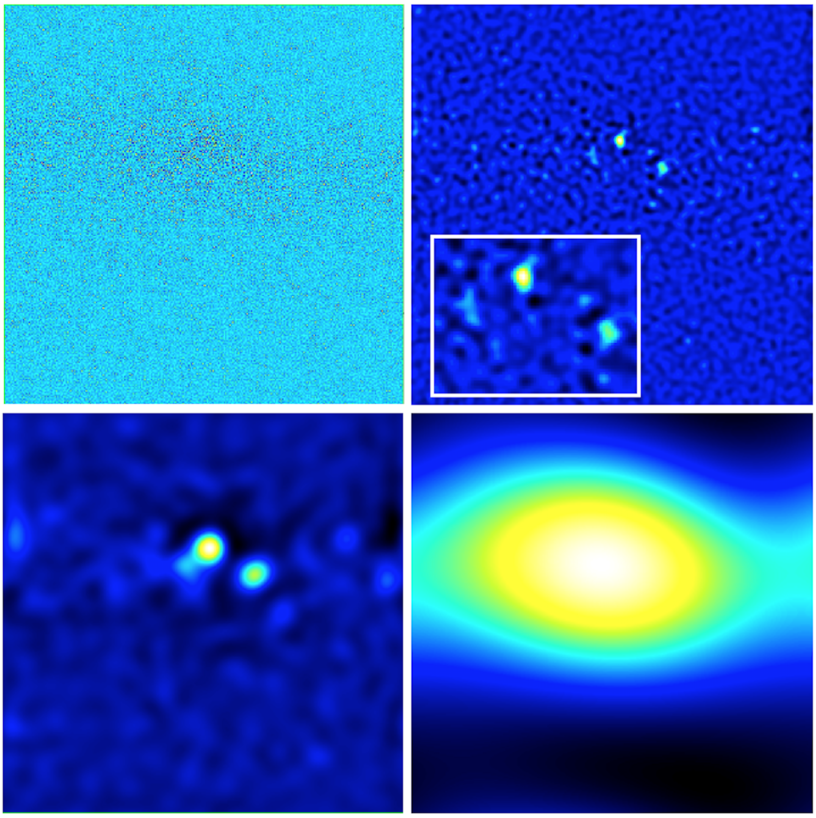}
    \caption{Scales 10, 20, 25, and 35 of the wavelet algorithm applied on the Cygnus region, from top to bottom, from left to right.  The  maps of scale 20 (top, right) contains a zoom of the Cygnus region.}\label{fig:CWT}
\end{figure*}

The detection of sources is related to the probability of correctly classifying each pixel of the CWT as belonging to a source or to the cosmic background. The source detection threshold of each pixel is derived simulating several background sky-maps and computing the CWT. In general, if a group of pixel is selected at any given scale then a similar group will exist both at a finer and at a coarser scale. Each connected region of CWT pixels in a given range of scales can be considered as a putative source characterized by its centroid (spatial position) and scale extension. An example of 4 scales applied to the Cygnus region is shown in Fig.~\ref{fig:CWT}.

\textbf{Iterative procedure on significance $TS$ maps.} The second list of seeds is determined to compute the significance $TS$ map for each ring with an iterative procedure.

\begin{figure*}
    \includegraphics[width=1\linewidth]{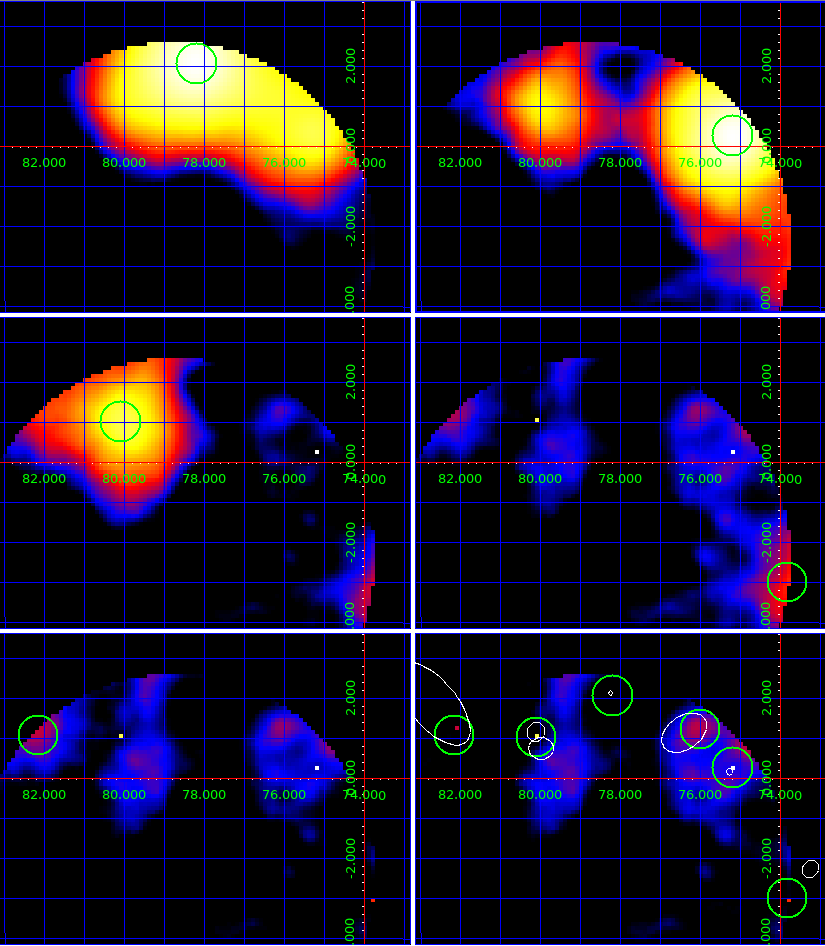}
    \caption{\label{fig:tsmap}The first six steps of the iterative procedure on the significance $TS$ maps {\cpbis described in Sect.~\ref{sec:seeds} applied to} the Cygnus region, 
    {\cpbis panels} from top to bottom, from left to right. Green circles are centred on the pixels with the maximum $TS$ found in each step. The last map (bottom, right) contains all the green circles of the first six steps and the final $95\%$ confidence region of the 2AGL sources. \abr{Note that even if border seeds are removed (see seeds found in {\cpbis panels} 4 and 5), they are still present in the final version of the Catalog because they {\cpbis may} be located in the overlapping nearby rings with a position {\cpbis closer} to the center, and 
    sometimes the final position could {\cpbis slightly} change (see seed found in {\cpbis panel} 4).}}
\end{figure*}

The first step starts without sources and, for each bin of the map, we perform a MLE, adding a point-like source in the ensemble of models at the centre of the bin under evaluation, with the flux parameter kept free assuming a Power-Law spectral index $\alpha=2.1$. We have used binned maps of $0.1\degmark \times 0.1\degmark$ and a radius of search of $5\degmark$. 
\abr{For each interaction we may obtain a set of neighbouring bins with high significance}.
We select a new candidate seed selecting the bin with the maximum $TS$ value and only if $TS > 9$: the position of the seed is in the centre of the bin. This seed is added to the ensemble of models for the next iteration with the list of seeds identified in the previous steps. The seeds are kept with flux and position parameters fixed.
The iterative procedure stops if no detection is found with $TS>9$ or if the maximum allowed number of iterations is reached. At the end all seeds are merged, combining in one single seed the overlapping ones. 
{\cpbis 
Seeds close to the boundary of the circle of search in are removed, but the seeds can still survive because they are found in the neighbouring and overlapping rings with a position closest to the centre of the ring.
}
An example of this iterative procedure on Cygnus region is shown in Fig.~\ref{fig:tsmap}.

\textbf{Final list of seeds} The list of seeds obtained from the iterative procedure on the significance $TS$ is added to the list of seeds obtained with the wavelet technique. The detections present only in the wavelet list are checked with a manual analysis and added to the final list if the detection has $\sqrt{TS} \geq 3$.
The final list of seeds results in 912 candidate point-like sources. 

\subsection{Iterative analysis of seeds}

The main purpose of this step is to reduce the number of seeds obtaining, at the end, a list of candidate sources for a refined analysis.

\textbf{Pixelisation.} We create a tiling of the sky for this step of analysis using 192 rings centred on points defined by HEALPix tessellation with $N_\mathrm{side} = 4$. We produce binned maps of $0.1\degmark$ bin size with a side of $50\degmark$ for each tile oriented with the north Galactic pole facing upward, whose centres are at a constant latitude with a single energy bin 100 MeV -- 10 GeV. We use a lower number of rings for this step in order to reduce the border effects of the rings.

\textbf{Iterative automated analysis of seeds} This stage starts from the list of seeds, ordered according to the estimated flux. 
For each iteration, an automated procedure selects one seed and the best ring for analysis, adding to the ensemble of models the seeds obtained with the previous iterations of this procedure and within $25\degmark$ from the centre of the selected ring. The seeds within $5\degmark$ from the selected seed under analysis are kept with flux free, and the seed under analysis is kept with position and flux free. The MLE method interactively optimises the position and flux of all the seeds of the region at the same time, with a radius of analysis of $10\degmark$. The localisation procedure of point-like sources provides the position, the $95\%$ elliptical confidence region, and the best evaluation of the significance, using for all sources a Power Law spectral shape with $\alpha=2.1$. 

At the end of each iteration the flux and position of a single seed are optimised and the list of seeds is updated  with the new flux and position if the detection has $TS > 9$, otherwise the seed is removed from the list. 
Since neighbouring regions are coupled, sharing data and sources, we repeat this step until the likelihoods are jointly optimised.
With this procedure detections above $TS > 9$ significance are considered during the analysis. 

\subsection{Manual analysis}

For the most complex {\cp and crowded} regions of the sky, additional manual analysis is  performed to add new candidate sources, or to verify the results obtained with the previous step. The list of candidate 2AGL sources is then updated with \abr{new 16 candidate point-like sources} obtained with this step. The final list of candidate 2AGL sources results in 318 candidate point-like sources with $TS > 9$. 




\subsection{Refined analysis}
\label{sect:refanal}

The main purpose of this step is to confirm candidate sources identified in the last step, obtaining at the end the final list of 2AGL sources. \abr{The 2AGL Catalog includes sources above $TS > 16$ significance (corresponding to $4\sigma$ with one free parameter).}

\textbf{Pixelisation.} The sky is tiled using a refined HEALPix tessellation used for the seeds with $N_\mathrm{side}=16$ corresponding to 3072 pixel with a mean spacing of $3.6645\degmark$. We produce binned maps of $0.1\degmark \times 0.1\degmark$ bin size with a radius of $15\degmark$ and with the following energy bins: 100 -- 300, 300 -- 1000, 1000 -- 3000, 3000 -- 10000 MeV. Additional bins of 30 -- 50 and 50 -- 100 MeV are produced by the same procedure.

\textbf{Best position, Spectral determination and Significance Thresholding.} The list of candidate sources obtained with the previous steps is analysed as follows:
\begin{enumerate}
\item this stage starts from the list of candidate sources, ordered according to the estimated flux;
\item extraction of the first candidate source, keeping free all its parameters (flux, position, spectral shape parameters), keeping free the flux and fixed the position of the sources that are within $3\degmark$ from the source under analysis, and fixing flux and position of the remaining sources within the sky map. The spectral shape is evaluated trying all available spectral models and selecting the best fit based on the $TS_\mathrm{curved}$. In some cases the Power Law index could results close to the boundaries: in this case, we fix the index $\alpha=2.1$. An analysis flag (see Sect.~\ref{sect:flags}) is specified for this problem;
\item update of the list of candidate sources with new position, flux and spectral model. \abr{A final run with all parameters fixed except the flux is performed and a significance threshold $TS \geq 16$ 
{\cpbis
(corresponding to $4\sigma$ with one free parameter)}
is then applied for the final selection of candidate sources.} Restart with the next source of the list.
\end{enumerate}

\begin{figure}
    \includegraphics[width=\linewidth]{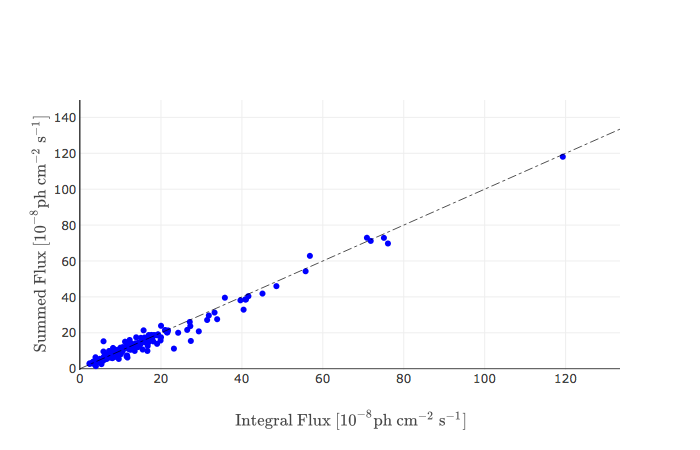}
    \caption{Comparison of the energy flux in the 100 MeV -- 10 GeV energy range estimated from the sum of bands (ordinate) and the fit to the full band (abscissa) for all 2AGL sources, in $\mathrm{10^{-8}\ ph\ cm^{-2} s^{-1}}$. No obvious bias can be observed. }\label{fig:figintegralvssummedflux}
\end{figure}

The source photon fluxes are reported in four energy bands: 100 -- 300 MeV, 300 -- 1000 MeV, 1 -- 3 GeV, and 3 -- 10 GeV. We perform a global fit over the full range, as described in Sect.~\ref{sect:spectralshape}. The fluxes in each band are obtained by freezing the spectral shape parameters to those obtained in the fit over the full range and adjusting the normalization in each spectral band. If in any band $\sqrt(TS) < 3$, the upper limit is selected.
Fig. \ref{fig:figintegralvssummedflux} reports a comparison between the energy flux estimated in the 100 MeV -- 10 GeV energy range from the sum of bands and the one estimated from the fit to the full range for all 2AGL sources. No obvious bias can be observed.

\subsection{Residual $TS$ significance maps.} 
The $TS$ significance maps is used to compute the residual significance maps at the end of the analysis procedure to search for missing point-like sources. Each residual $TS$ found in this way has been evaluated with an additional step of the refined analysis. Five new sources have been found and added, and neighbouring sources within a radius of $5\degmark$ reevaluated. 

\subsection{Variability}
\label{variab}
A temporal variability analysis is performed on the 2AGL Catalog sources. Temporal variability is common to different classes of \gray sources and it is important to determine a variability index. For each source we split the AGILE-GRID data in 4-days time intervals: this is a compromise between the duration of an observation (see Table~\ref{tab:pointings}) and a useful exposure time. For each time interval we produce sky maps of $0.5\degmark$ bin size and of $50\degmark$ diameter, centred in the position of the source under analysis.


We produce a light curve for each source in the Catalog. Due to the already mentioned observation constraints of the pointing mode, the number of bins for each light curve could be different.
To define a variability index, first we analyse each time bin keeping free only the flux of the source under analysis, adding the neighbouring 2AGL sources within the sky map. To avoid large error bars, the position and the spectral parameters of the source are frozen, assuming spectral variability to be negligible; in addition we evaluate the diffuse \gray background emission and the isotropic background over the entire Observing Block. 


Let $TS^i_1$ the value of the $TS$ obtained optimising the flux in each period of time $i$ and $TS_0$ the value of $TS$ estimated evaluating all the time bins at the same time but considering a constant flux, $TS_\mathrm{var}$ is described by the following relation:

\begin{equation}
TS_\mathrm{var} =  \sum_{i=1}^N TS^i_1 - TS_0. 
\end{equation}

If the null hypothesis is correct (the source has a constant flux) $TS_\mathrm{var}$ is distributed as $\chi^2$ with N-1 degrees of freedom, and a value of $TS_\mathrm{var} > h(N-1)$ is used to identify variable sources at $99\%$ confidence level, where $h(N)$ gives the threshold corresponding to a $99\%$ confidence level in a $\chi^2$ distribution with $N$ degrees of freedom.

It is possible to introduce a corrective factor (similar to \citep{nolan2012}) to take into account the systematic error:

\begin{equation}
TS_i^\mathrm{corr}  = F_{\sigma_i}^2 / ( F_{\sigma_i}^2 + f^2 * F_0^2 ).
\end{equation}

\abr{$F_{\sigma_i}$ is the error on the flux  in each period of time $i$, $F_0$ is the flux estimated evaluating all the time bins at the same time but considering a constant flux. } We consider $f = 0.02$ in our analysis ($2\%$ of systematic error). \abr{This value was found sufficient such that no AGILE-GRID pulsars are above threshold, excluding the Crab which have a highly variable nebular component at AGILE-GRID energies \citep{tavani11, abdo11}. } The corrected $TS_\mathrm{var}$ is

\begin{equation}
TS^{*}_\mathrm{var} =  \sum_{i=1}^N ( TS_i^\mathrm{corr} * TS^i_1) - TS_0.
\end{equation}

The variability index VI assumes the value 1 if $TS^{*}_\mathrm{var} > h(N-1)$.
Upper limits calculated through the MLE method are handled by the procedure described above. In this procedure, all the upper limits are those obtained with the MLE method, no Bayesian procedure was used.

To be conservative, we evaluate the VI if $N>12$.
Some variable light curves (based on VI) are reported in Fig.~\ref{fig:lc_PKS1510M089} to Fig.~\ref{fig:lc_3C4543}. 
{\cp
}

\subsection{Extended sources}
{\cp}
\label{sect:extended}

The procedure described in the previous sections is related to point-like sources. We have modelled a list of sources as spatially extended sources, using a 2D Gaussian model. Nearby point-like sources have been fixed in position and spectral shape, keeping only the flux free for sources within $3\degmark$ from the extended source under analysis and removing only sources inside the extended template that have no association with known point-like sources. 

The list of analysed sources is reported in Table \ref{tab:catextended}. The first column reports the region name, the second and third report the Galactic coordinates of the centre of the region, $l_\mathrm{e}$ and $b_\mathrm{e}$, the fourth and the fifth report the two radii, $r_1$ and $r_2$, used for the analysis, that indicate the dispersion for 2D Gaussian sources. \abr{
$r_1$ is chosen considering the observed extension at GeV and TeV energies (or considering a mean value if the shape of the extended region is not circular), $r_2$ is usually the double of $r_1$ with some exceptions (i.e. difference in radius between GeV and TeV energies or very eccentric shape). In particular, 3FGL \citep{acero15} and TeVCat Catalogs have  been used for the selection of the two values.  The online TeVCat catalog\footnote{http://tevcat.uchicago.edu/ (Wakely, S., and Horan, D.)}  is continuously updated with new sources detected by TeV experiments. At the time of writing (January 2019), the TeVCat catalog contains a total of 219 TeV sources.}

\begin{table*}[!ht]
\begin{center}
\caption{Definitions of the analysis flags} 
\label{tab:cat2flags} 
\begin{tabularx}{\textwidth}{| c | X |}
\hline
\multicolumn{1}{|c}{Flag} & 
\multicolumn{1}{c|}{Description} \\[3 pt]
\hline
 & \\
1 & $\alpha$ fixed to 2.1 \\
2 & 2 upper limits over 4 in the energy bands 100--300, 300--1000, 1000--3000, 3000--10000 MeV \\
3 & 3 upper limits over 4 in the energy bands 100--300, 300--1000, 1000--3000, 3000--10000 MeV \\
4 & 4 upper limits over 4 in the energy bands 100--300, 300--1000, 1000--3000, 3000--10000 MeV \\
5 & Optimisation of position and spectral shape in two different steps \\[3 pt]
\hline
\end{tabularx}
\end{center}

\end{table*}

\subsection{Exposure uniformity within the region of the MLE analysis}
Due to the non-homogeneous sky coverage of the AGILE observations during the first 2.3 years, it might occur that some candidate sources lay near the borders of certain pointings.
{\cpbis
In order to have an unbiased estimate of the coefficients of the Galactic 
diffuse emission and isotropic background that could lead to 
to an incorrect evaluation of the flux and position of the source, 
exposure uniformity within the region of the analysis is required.

We apply a specific check to verify the uniformity of the exposure within
the $10$-degree radius of the AGILE MLE analysis
centred at each source candidate position, over the considered timescale. }
The fraction of pixels of the exposure map within the region of analysis having a value
below a pre-defined threshold is calculated, and if it is more
than 10\% the region is considered unreliable and the candidate
is discarded. The exposure threshold value is evaluated 
by calculating the mean exposure of the observation over 
the full FoV area and comparing it with the
values of some reference good exposures.

\section{Limitations and Systematic Uncertainties}
\label{sect:unc}

\subsection{Instrument response functions systematic uncertainties}
\label{sect:systirf}

%

\subsubsection{Systematic uncertainties on flux}

\abr{In order to estimate systematic effects due to changes in instrument characteristics, inaccuracies in the instrument response functions, and uncertainties in the galactic diffuse model, we compared the behaviour of the residual for 100 MeV - 10 GeV near the peak of the Vela pulsar as a function of Observation Blocks (see Table \ref{tab:pointings}). For each Observation Block we constructed a model consisting of three components: the point spread function at the position of Vela with the flux and super exponential cut-off Power Law spectral model  listed in the 2AGL Catalog, an isotropic component of $5.9 \times 10^{-5} \:  \mathrm{ph \: cm^{-2} s^{-1}}$, and a galactic diffuse component with coefficient 0.5, evaluated during the 2AGL Catalog analysis. We then calculated the $residual = (model - counts) / exposure$ and the residual error $residual_{error} = \sqrt{2 \times model} / exposure$. We then binned the residual in annuli of $0.5\degmark$ and observed the behavior of the residual in the innermost ring as a function of Observation Block. 
For most of the OBs the residual is consistent with a value slightly less than $1 \times \mathrm{10^{-8} \: ph \: cm^{-2} s^{-1}}$ but there is one OB, the OB 4100, where the residual is in the opposite direction and slightly higher. In any case the errors are very small compared to the statistical errors in the flux even for Vela, the brightest point source in the Catalog.}

\subsubsection{Systematic uncertainties on spectral index}
\label{sect:sysindex}
\abr{In addition to the effect on the source flux the systematic uncertainties on the IRFs affect also the source spectral index.
There are two relevant sources of systematic uncertainties: the 
shape of $A_\mathrm{eff}(E_\gamma)$ and the EDP.
The former is shown in Fig.1 of \cite{chen13} and is obtained by simulation. Estimating the systematic errors from the simulation alone is unreliable, while in \cite{cattaneo18} the unreliability of the experimental estimation by the calibration under beam is discussed. An alternative approach is generating Monte Carlo spectra with a given shape of $A_\mathrm{eff}(E_\gamma)$ and index equal to -2.10, and fit the spectra with another shape. The most conservative choice is to use the $A_\mathrm{eff}(E_\gamma)$ from \cite{chen13} and fit assuming a flat $A_\mathrm{eff}(E_\gamma)$ and vice versa; that is much larger than any possible error and we can estimate to cover a variation of the index of no less than $\pm2\sigma^{A_\mathrm{ eff}}_\alpha$. Out of 100 Monte Carlo experiments, the average index from the fit is -2.01 in the first case and -2.25 in the second. Therefore the systematic error can be estimated as $\sigma^{A_\mathrm{eff}}_\alpha = 0.24/4 = 0.06$.\\
The latter systematic uncertainty from EDP can be estimated analogously generating Monte Carlo spectra with EDP and fitting the index without and vice versa.
That is a very conservative estimation because the calibration under beam \cite{cattaneo18} provides a measure of the EDP consistent with expectations and therefore any possible systematic error can be only a fraction of the EDP itself. Following the previous approach the systematic error due to EDP can be estimated as $\sigma^\mathrm{EDP}_\alpha = 0.07$.\\
Finally, the Monte Carlo spectra can be generated with $A_\mathrm{eff}(E_\gamma)$ and EDP and fitted without either and vice versa. The overall systematic errors is $\sigma_\alpha = 0.10$ consistent with the quadratic sum of the separate contributions.
}







\subsection{Limitations on extended source analysis}

There are some limitations in the analysis of extended regions that reduce the number of extended sources of the 2AGL Catalog: (i) only a 2D Gaussian model is used as spatially extended sources model: this is not true for all extended shapes; (ii) due to limitations of the analysis tools, the analysis is performed with only a unique sky map integrated in the 100 MeV -- 10 GeV energy range: this strongly limits the identification of extended sources where the emission peaks at higher energies; (iii) only the Power Law spectral model with a fixed spectral index $\alpha=2.1$ is used.

\subsection{Limitations of the variability index}
\abb{The variability index VI is described in Sect. \ref{variab}. The main limitation of this index is in the reduced number of temporal bins for each light curve, with a number of upper limits that is not negligible in many cases (even if the procedure handles this). The last limitation is that the variability index is not provided for light curves with less than 12 temporal bins. For these reasons, this index is not included in the identification criteria of counterparts.}

\subsection{Analysis flags}
\label{sect:flags}
Some peculiar conditions that require caution in order to assess the confidence of a source are described in Table~\ref{tab:cat2flags}.

Flag=1 indicates when, keeping the spectral index of the Power Law free, the value moves close to the boundaries of the search space of the spectral parameters. In this case a Power Law with fixed $\alpha=2.1$ is assumed. \abr{This happens  during the step 2 of the refined analysis (Sect. \ref{sect:refanal}).  Only 4 sources have this flag.}

Flags from 2 to 4 indicate that there are 2, 3 or 4 upper limits over 4 energy bins (100-300, 300-1000, 1000-3000, 3000-10000 MeV). \abr{Upper limits are usually in the highest energy bands. 61 sources have flag=2, 51 sources have flag=3 and only 5 sources have flag=4.}

Flag=5 indicates that AGILE Science Tools are not able to optimise the position and spectral shape parameters of the source at the same time \abr{during the step 2 of the refined analysis. 33 sources have this flag.} 


\begin{table*}[!ht]
\caption{List of  extended sources analysed for 2AGL Catalog.}
\label{tab:catextended}

\begin{tabular}{|l r r r r| l r r r r|}


\hline
{Region Name} &
{$l_\mathrm{e}$} &
{$b_\mathrm{e}$} &
{$r_1$} &
{$r_2$} &
{Region Name} &
{$l_\mathrm{e}$} &
{$b_\mathrm{e}$} &
{$r_1$} &
{$r_2$} \\
\hline

Boomerang & 106.57 & 2.91 & 1.00 & 2.00 &
HESS J1834-087 & 23.24 & -0.33 & 0.09 & 0.18 \\ 
CTA1 & 119.60 & 10.40 & 0.30 & 0.60 &
HESS J1837-069 (PWN) & 25.18 & -0.12 & 0.12 & 0.33 \\ 
CTB37A & 348.38 & 0.10 & 0.07 & 0.14 & 
HESS J1841-055 (PWN) & 25.87 & -0.36 & 0.40 & 0.62 \\ 
CTB37B & 348.63 & 0.38 & 0.06 & 0.12 &
HESS J1843-033 & 29.03 & 0.36 & 1.00 & 2.00 \\ 
CenA Lobes & 309.17 & 18.98 & 1.25 & 2.50 & 
HESS J1848-018 & 31.00 & -0.17 & 0.32 & 0.64 \\ 
Cygnus Cocoon & 80.95 & 1.80 & 1.80 & 3.00 & 
HESS J1857+026 & 36.00 & -0.07 & 0.20 & 0.40 \\ 
Cygnus Loop & 73.98 & -8.56 & 3.00 & 10.00 & 
HESS J1858+020 & 35.57 & -0.59 & 0.08 & 0.16 \\ 
G106.3+2.7 & 106.34 & 2.71 & 0.27 & 0.52 & 
HESS J1912+101 & 44.39 & -0.08 & 0.27 & 0.52 \\ 
G327.1-1.1 & 327.15 & -1.08 & 0.03 & 0.06 &
IC 443 & 189.07 & 2.92 & 0.16 & 0.27 \\ 	
Galactic Center Ridge & 359.94 & -0.05 & 2.00 & 4.00 &
IGR J18490-0000 & 32.63 & 0.52 & 1.00 & 2.00 \\ 
Geminga & 195.33 & 3.77 & 1.30 & 2.60  &
Kookaburra-Pulsar & 313.55 & 0.26 & 0.06 & 0.11 \\ 
HB 21 & 88.75 & 4.67 & 0.59 & 1.19 &
Kookaburra-Rabbit & 313.24 & 0.14 & 0.08 & 0.16 \\ 
HESS J1018-589B & 284.11 & -1.90 & 0.15 & 0.30 & 
LMC & 279.55 & -31.75 & 0.14 & 1.87 \\ 
HESS J1026-582 & 284.79 & -0.53 & 0.14 & 0.28 &
MGRO J1908+06 & 40.28 & -0.69 & 0.44 & 0.88 \\ 
HESS J1303-631 (PWN) & 304.21 & -0.33 & 0.19 & 0.24 & 
MGRO J2019+37 & 74.82 & 0.41 & 0.75 & 1.50 \\ 
HESS J1356-645 & 309.81 & -2.50 & 0.20 & 0.40 & 
MGRO J2031+41 & 79.53 & 0.63 & 1.80 & 3.60  \\
HESS J1427-608 & 314.40 & -0.15 & 0.04 & 0.08 &
MSH 15-52 & 320.33 & -1.19 & 0.11 & 0.25 \\ 
HESS J1457-593 & 318.36 & -0.44 & 0.31 & 0.62 & 
Puppis A & 260.32 & -3.28 & 0.16 & 0.37 \\ 
HESS J1458-608 & 317.74 & -1.71 & 0.17 & 0.34 & 
RCW 86 & 315.41 & -2.31 & 0.41 & 0.82 \\
HESS J1503-582 & 319.61 & 0.29 & 0.26 & 0.52 & 
RX J1713.7-3946 & 347.34 & -0.47 & 0.56 & 0.65 \\ 
HESS J1507-622 & 317.94 & -3.50 & 0.15 & 0.30 & 
S 147 & 180.24 & -1.50 & 0.75 & 1.50 \\
HESS J1614-518 & 331.52 & -0.58 & 0.23 & 0.42 &
SMC & 302.14 & -44.42 & 0.67 & 1.35 \\ 
HESS J1616-508 (PWN) & 332.39 & -0.14 & 0.18 & 0.32 & 
SN 1006-NE & 327.84 & 14.56 & 0.12 & 0.24 \\ 
HESS J1626-490 & 334.77 & 0.04 & 0.10 & 0.20 & 
SN 1006-SW & 327.86 & 15.34 & 0.13 & 0.26 \\ 
HESS J1632-478 (PWN) & 336.38 & 0.19 & 0.21 & 0.35 &
SNR G292.2-00.5 & 292.10 & -0.49 & 0.10 & 0.20 \\ 
HESS J1634-472 & 337.10 & 0.21 & 0.11 & 0.22 &
TeVJ2032+4130 & 80.24 & 1.17 & 0.16 & 0.32 \\ 
HESS J1640-465 & 338.31 & -0.03 & 0.01 & 0.04 &
Terzan 5 & 3.78 & 1.72 & 0.16 & 0.32 \\ 
HESS J1702-420 & 344.30 & -0.19 & 0.30 & 0.60 & 
VELA Jr & 266.28 & -1.24 & 1.00 & 1.12 \\ 
HESS J1708-410 & 345.68 & -0.48 & 0.08 & 0.16 &
VELAPS & 263.58 & -2.84 & 0.10 &   \\ 
HESS J1708-443 & 343.05 & -2.38 & 0.29 & 0.60 &
VELAX & 263.86 & -3.09 & 0.48 & 0.91 \\ 
HESS J1718-385 & 348.83 & -0.49 & 0.15 & 0.30 &
VER J2019+368 & 75.04 & 0.28 & 0.34 & 0.68 \\ 
HESS J1729-345 & 353.44 & -0.13 & 0.12 & 0.24 &
VER J2019+407 & 78.33 & 2.49 & 0.23 & 0.46 \\ 
HESS J1731-347 & 353.54 & -0.68 & 0.27 & 0.34 &
W28-HESS J1800-240ABC & 5.96 & -0.39 & 0.32 & 0.64 \\ 
HESS J1745-303 & 358.70 & -0.65 & 0.21 & 0.42 &
W44 & 34.65 & -0.39 & 0.15 & 0.30 \\ 
HESS J1804-216 & 8.35 & -0.01 & 0.27 & 0.35 &
W30 & 8.40 & -0.03 & 0.27 & 0.37 \\ 
HESS J1808-204 & 9.95 & -0.25 & 0.14 & 0.28 & 
W51C & 49.12 & -0.36 & 0.12 & 0.38 \\ 
HESS J1809-193 & 11.18 & -0.09 & 0.53 & 1.06 & 
Westerlund 1 & 339.54 & -0.36 & 1.10 & 2.20 \\ 
HESS J1825-137 & 17.71 & -0.70 & 0.13 & 0.16 & 
Westerlund 2 & 284.21 & -0.41 & 0.18 & 0.36 \\ 
HESS J1813-178 & 12.81 & -0.03 & 0.04 & 0.08 & 
gammaCygni & 78.33 & 2.49 & 0.23 & 0.63 \\ 
HESS J1831-098 & 21.85 & -0.11 & 0.15 & 0.30 \\ 

\hline
\end{tabular}
\caption*{The first column reports the region name, the second and third columns reports the Galactic coordinates of the center of the region, the fourth and the fifth columns reports the two radii $r_1$ and $r_2$ used for analysis indicating the dispersion for 2D Gaussian sources.}

\end{table*}



\section{Second AGILE-GRID \gray Sources List}
\label{sec:list}

The \abr{Second AGILE Catalog of \gray sources (2AGL)} includes 175 high-confidence sources detected using the AGILE-GRID data during the `pointing mode' period, with the methods and criteria described in Sect.~\ref{sec:construction}.
{\cpbis An interactive web page of the 2AGL Catalog 
and its FITS file version are publicly available at 
SSDC\footnote{https://www.ssdc.asi.it/agile2agl}.}

\abb{In this section we present a description of the 2AGL Catalog, the criteria used for association and identification of sources with known counterparts, the content of the main tables, and a comparison with previous AGILE Catalogs. Sect. \ref{sec:notes} reports notes on individual 2AGL sources, where also details used for associations and identifications are described.}

\subsection{Catalog description}
 \label{sect:catdes2}
\begin{figure*}
    \includegraphics[width=\linewidth]{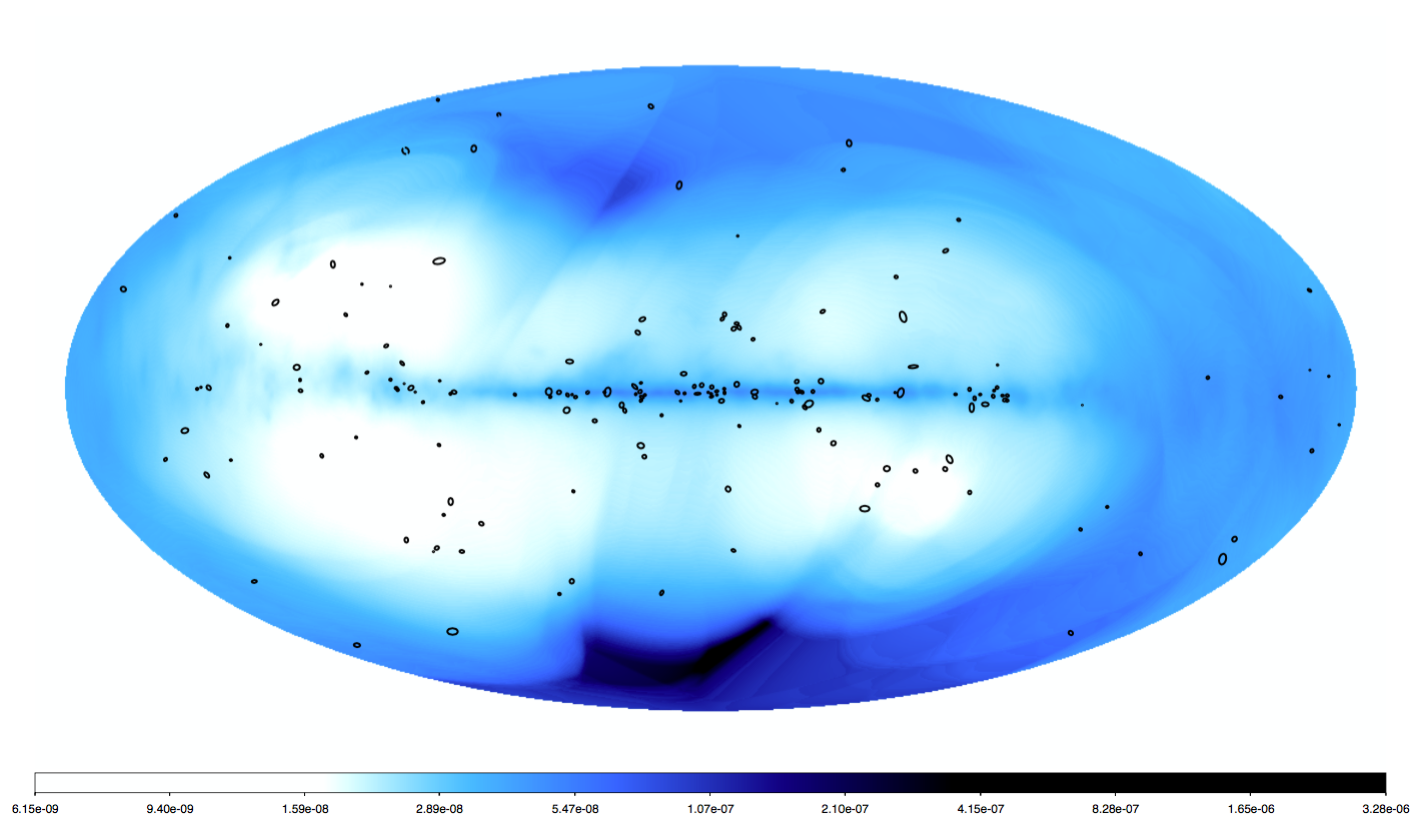}
    \caption{Sky map of 4$\sigma$ sensitivity in the 100 MeV -- 10 GeV energy range in Galactic coordinates and Hammer-Aitoff projection, with the hypothesis of a PL spectral model with $\alpha=2.1$, in $\mathrm{ph\ cm^{-2} s^{-1}}$. The $95\%$ elliptical confidence regions of the 2AGL sources are superimposed in black.}
    \label{fig:sensitivitycat2}
\end{figure*}

 \abb{
 The validated sources in the Catalog are \abr{listed in Table~\ref{tab:cat2}, including both confirmed and possible associations, and}  plotted in Fig.~\ref{fig:sensitivitycat2}
in Galactic sky coordinates. Table~\ref{tab:cat2columns} reports the description of the columns of Table~\ref{tab:cat2}.
}

The source designation is \textit{2AGL JHHMM+DDMM[c/e]} where the \textit{2} indicates that this is the second AGILE-GRID Catalog, \textit{AGL} represents the AGILE-GRID. The name of the sources potentially confused with the Galactic diffuse emission \abr{or with a large uncertainty on its location} is appended with \textit{c}, and caution should be used in interpreting these sources; an appended \textit{e} indicates sources associated with a spatially extended emission. 

It is important to note that each source is observed for a different number of days much smaller than 2.3 years of the `pointing mode'. The column `Exp' of Table \ref{tab:cat2} reports a rough estimate of days of observations, obtained dividing exposure by a mean $A_\mathrm{eff} = 300$ $\mathrm{cm}^2$ and 86400 seconds.
Association and identification of 2AGL sources are described in Sect.~\ref{sec:assoc}. 

\begin{figure*}
	\begin{multicols}{2}
    	\includegraphics[width=\columnwidth]{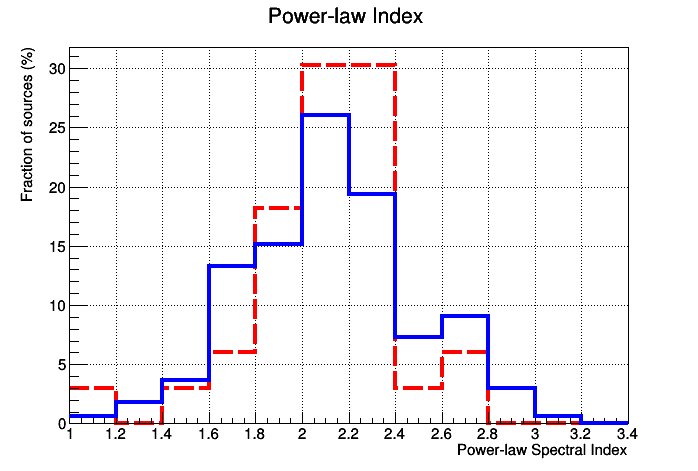}\par 
    	\includegraphics[width=\columnwidth]{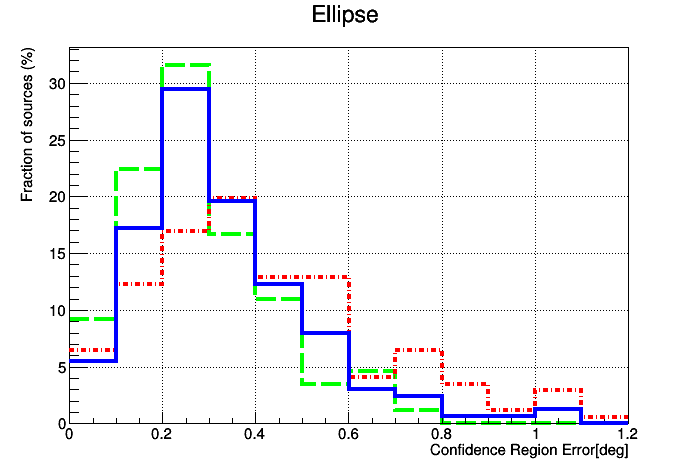}\par 
    \end{multicols}
	\begin{multicols}{2}
   	 	\includegraphics[width=\columnwidth]{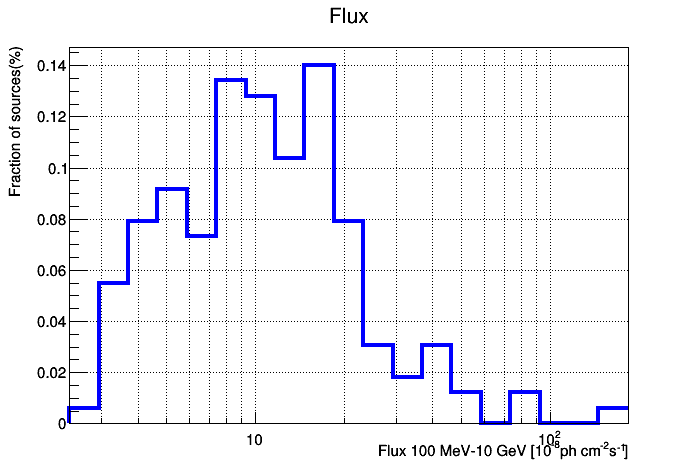}\par
  		\includegraphics[width=\columnwidth]{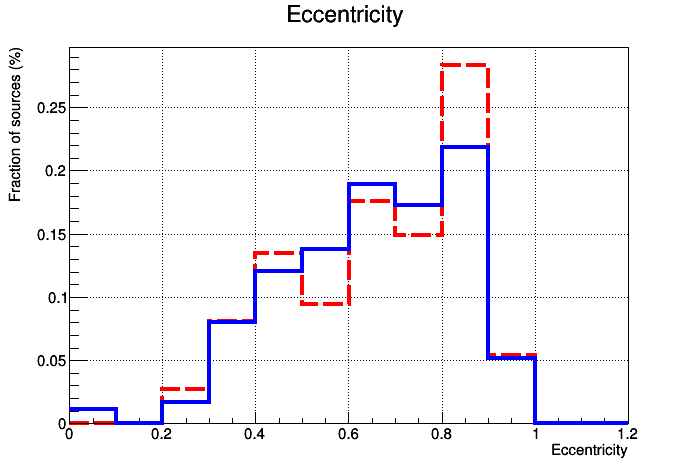}\par	
	\end{multicols}
    \begin{multicols}{2}
  		\includegraphics[width=\columnwidth]{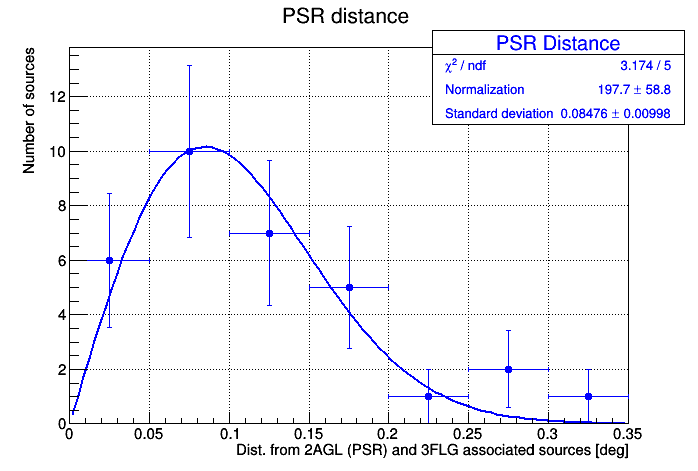}\par	
        \includegraphics[width=\columnwidth]{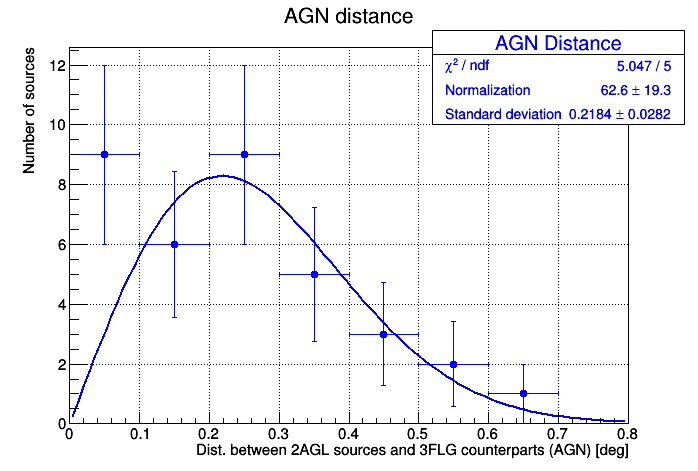}\par	
	\end{multicols}
    \begin{multicols}{2}
	\end{multicols}
\caption{Some distributions of 2AGL source parameters. Top left figure shows the spectral index distributions for sources with Power Law spectral model, all sources (blue) and only for sources with high latitude ($|b| > 10$) (red). Top right figure shows the 95\% confidence region error radius (blue), the semi-major axes of 95\% elliptical confidence region (red) and the semi-minor axes (green). Middle left figure shows the integral flux in the 100 MeV -- 10 GeV energy range, for sources with Power Law spectral model. Middle right figure shows the eccentricity of the 95\% elliptical confidence region for all sources (blue), and for high latitude sources ($|b| > 10$) (red). Bottom left figure shows the distance between the 2AGL sources classified as pulsars and their 3FGL counterparts for sources with $\sqrt{TS}\geq 5$. Bottom right figure shows the distance between 2AGL sources classified as AGN and their 3FGL counterparts, for  sources with $\sqrt{TS}\geq 5$. In the latter two figures the data are fit with a Rayleigh function.}\label{fig:stat}
\end{figure*}

Figure~\ref{fig:stat} reports some distributions of 2AGL source parameters (spectral indexes, fluxes, 95\% elliptical confidence regions \abr{parameters distributions, and} distances from known counterparts). 
\abr{AGNs and pulsars are the most important classes of the 2AGL Catalog. The median value of the spectral index for AGN classes is $\alpha=2.10 \pm 0.30$, and for pulsar class is $\alpha=1.98 \pm 0.30$; the two distributions are compatible. The distributions of distances of 2AGL from 3FGL counterparts for the two different classes are shown in the bottom figures together with fits to the Rayleigh function. This fit assumes implicitly that the errors on the source positions are constant. In reality we expect that the position error on each source depends on the statistical and systematic errors of AGILE-GRID and FERMI-LAT, which vary from source to source, and therefore that a single Rayleigh function is not fully adequate. Nevertheless the values of $\chi^2/ndf$ close to one demonstrate that, under the simplified assumption of constant errors, the two distributions are consistent.\\
}

\begin{table*}[htbp]
\begin{center}
\caption{AGILE-GRID Second Catalog columns description of Table \ref{tab:cat2}, Table \ref{tab:agileonly} and Table \ref{tab:gunid}}
\label{tab:cat2columns}
\begin{tabularx}{\textwidth}{| c | c | X |}
\hline
\multicolumn{1}{|c}{Column} & 
\multicolumn{1}{c}{Units} & 
\multicolumn{1}{c|}{Description} \\[3 pt]
\hline
 & & \\
Name &  & 2AGL JHHMM+DDMM[c/e] constructed according to IAU Specifications for Nomenclature; in the name, R.A. and Decl. are truncated at $0.1^\prime$ and $1^\prime$, respectively; `c' indicates that based on the region of the sky the source is potentially confused with Galactic diffuse emission; `e' indicates a source modelled as spatially extended  \\
R.A. & deg & Right Ascension, J2000\\
Dec. & deg & Declination, J2000\\
\textit{l} & deg &  Galactic Longitude \\
\textit{b} & deg & Galactic Latitude\\
$\theta_\mathrm{a}$ & deg & Semi-major axis of $95\%$ c.l. elliptical confidence region, deg. Statistical error only \\
$\theta_\mathrm{b}$ & deg & Semi-minor axis of $95\%$ c.l. elliptical confidence region, deg. Statistical error only \\
$\phi$ & deg & \par Position angle of $95\%$ confidence region, clockwise\\
SM & & Spectral model. PL indicates power-law fit to the energy spectrum; PC indicates power-law with exponential cut-off fit to the energy spectrum; PS indicates power-law with super exponential cut-off fit to the energy spectrum; LP indicates log-parabola fit to the energy spectrum\\
$\sqrt{TS}$ & & Significance derived from likelihood Test Statistic in the 100 MeV -- 10 GeV energy range\\
$\alpha$ & & Spectral index for PL, PC and PS spectral models, first index for LP spectral model, in the 100 MeV -- 10 GeV energy range, see Sect.~\ref{sect:spectralshape} for the definition of $\alpha$\\
$\Delta\alpha$ & & Statistical $1\sigma$ uncertainty of $\alpha$, in the 100 MeV -- 10 GeV energy range\\
$F_\gamma$ & $\mathrm{10^{-8}\ ph\ cm^{-2} s^{-1}}$ & Photon flux in the 100 MeV -- 10 GeV energy range, summed over 4 bands\\
$\delta F_\gamma$ & $\mathrm{10^{-8}\ ph \ cm^{-2} s^{-1}}$ & Statistical 1$\sigma$ uncertainty of $F$ \\
$F_\mathrm{e}$ & $\mathrm{10^{-9}\ erg\ cm^{-2} s^{-1}}$ & Integrated energy flux in the 100 MeV -- 10 GeV energy range \\
$\delta F_\mathrm{e}$ & $\mathrm{10^{-9}\ erg\ cm^{-2} s^{-1}}$ & Statistical 1$\sigma$ uncertainty of $S$ \\
VI & & Variability index equal to 1 indicates $< 1\%$ chance of being a steady source. \abr{No value indicates that there are not enough 4-days time periods to calculate the index. See Sect. \ref{variab} for more details.}\\
Exp & $\mathrm{cm^2 Ms}$ [days] &  Exposure in $\mathrm{cm^2 Ms}$ and days. Days are obtained dividing exposure by the mean $A_\mathrm{eff}=300$ $\mathrm{cm}^2$ and 86400 s\\
Flag & &  Flag (see Table \ref{tab:cat2flags})\\
AGL assoc. & & \abr{Positional associations with sources from  AGILE Catalogs listed in Table \ref{tab:catalogs}. For AGILE-TeVCat we have added the AGL prefix to the Catalog name} \\
\gray assoc. & & Positional associations with sources from not AGILE Catalogs listed in Table \ref{tab:catalogs}. \abb{For the same source in the FGL Catalogs, only the 3FGL Catalog is kept} \\
ID or assoc. & & Designator of identified or associated source \\
Class & & Class (see Table \ref{tab:cat2classes})\\
\hline
\end{tabularx}
\end{center}
\end{table*}

\subsection{Source association and identification}
\label{sec:assoc}
 
\begin{figure*}
    \includegraphics[trim=3cm 5.cm 3cm 5.cm, clip=true,width=\linewidth]{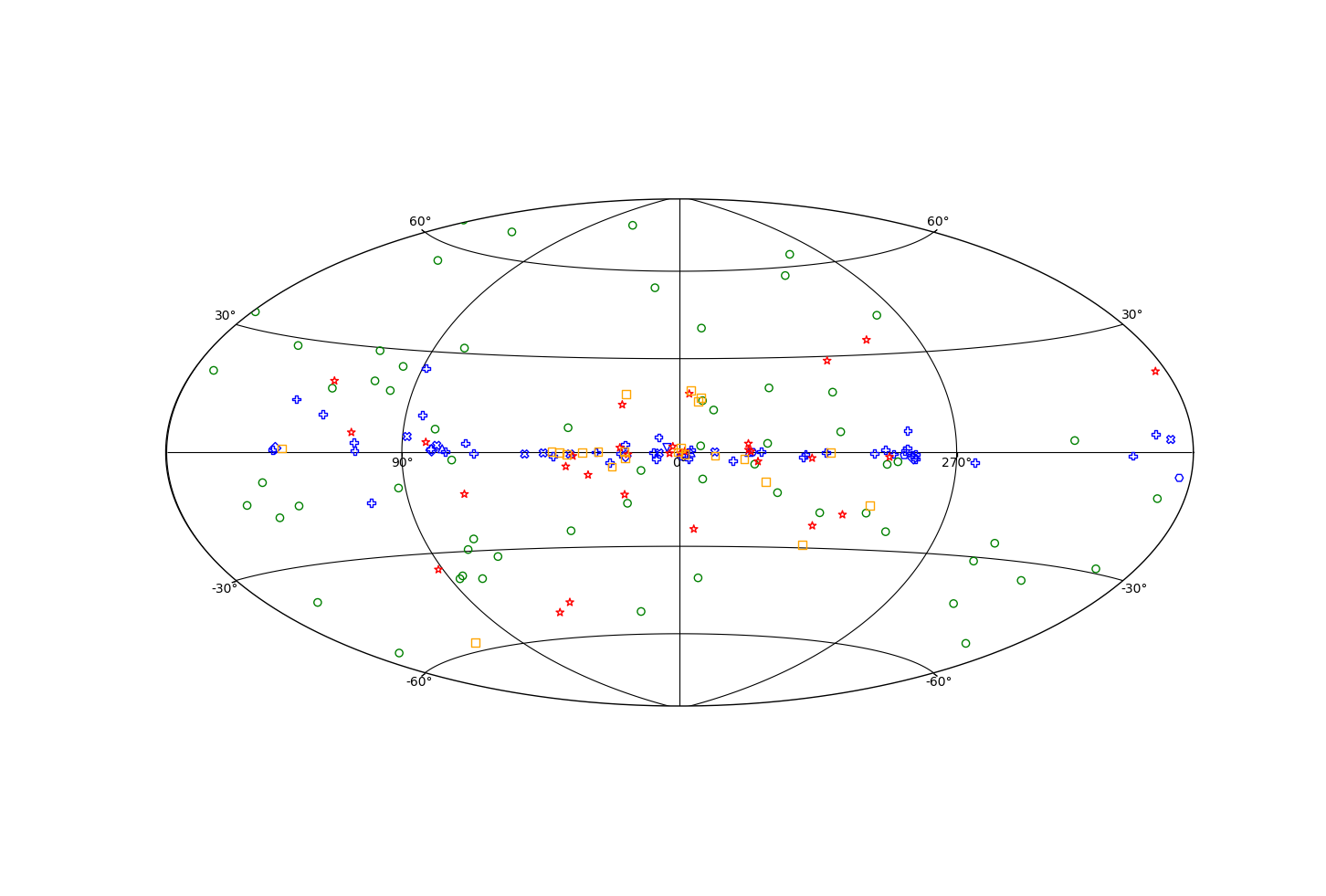}
    \includegraphics[trim=1cm 1.5cm 1cm 1.cm, clip=true,width=\linewidth]{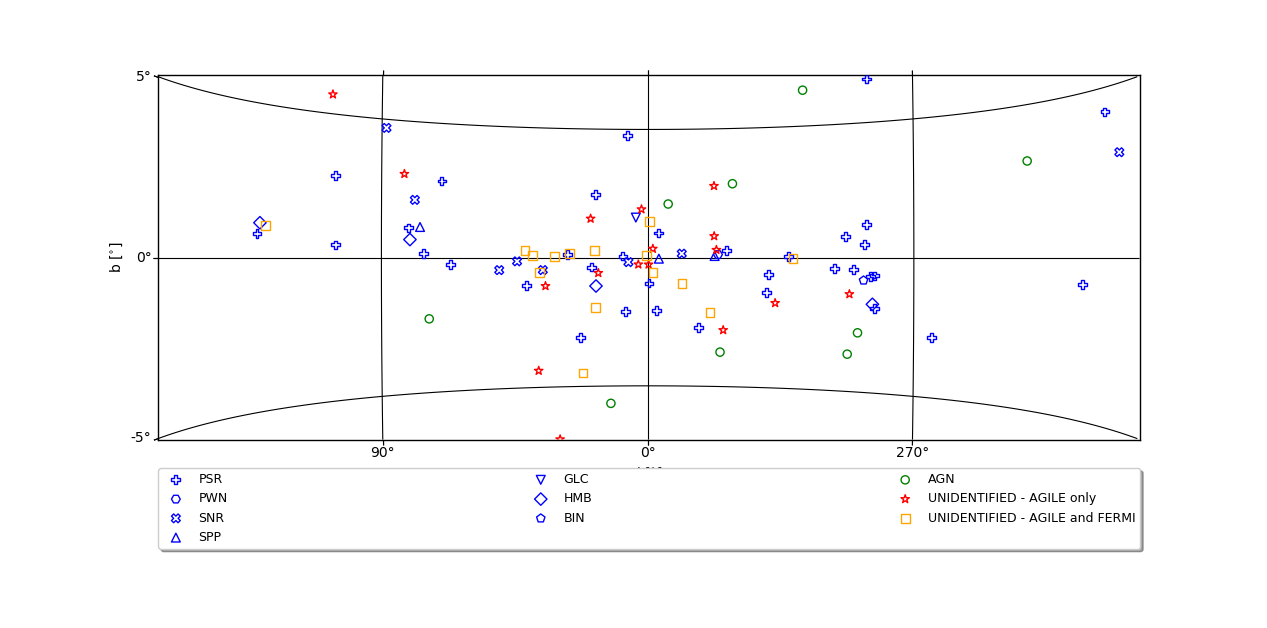}
     \caption{Positions, in Galactic coordinates, of the 2AGL sources labelled by different symbols according to their class for the full sky (top) and a blow-up of the inner Galactic region (bottom). All AGN classes are plotted with the same symbol.}\label{fig:aitof_marker}
\end{figure*}

In the 2AGL Catalog we define three different classes of identified, associated and unidentified sources. Table \ref{tab:cat2classes} reports a summary of the 2AGL classes. Designations shown in capital letters are firm identifications; lower case letters indicate associations. \abb{Fig.~\ref{fig:aitof_marker} reports a full sky map showing sources labelled by source class. A blow-up of the inner Galactic region is reported in the bottom of Fig.~\ref{fig:aitof_marker}. }

Associations are {\cp defined} using a spatial cross-correlations procedure based on spatial coincidence of the $95\%$ confidence region with various updated public catalogs of specific mission or of specific source classes {\cp known to be potential \gray emitters}. The list of catalogs used for association is reported in Table \ref{tab:catalogs}. Two sources are {\cp considered} associated if there is a partial or total overlapping between the $95\%$ elliptical confidence regions, taking into account the statistical error plus a systematic error of $0.1\degmark$ \citep{chen13}.

\begin{table*}[htbp]
\begin{center}
\caption{2AGL source classes.}

\begin{tabular}{|c c c c c|}
\hline
 & & & & \\
\multicolumn{1}{|c}{Description} & 
\multicolumn{2}{c}{Identified} &
\multicolumn{2}{c|}{Associated} \\[7 pt]
\multicolumn{1}{|c}{} &
\multicolumn{1}{c}{Designator} &
\multicolumn{1}{c}{Number} &
\multicolumn{1}{c}{Designator} &
\multicolumn{1}{c|}{Number} \\[3 pt]
\hline
 & & & &\\[3 pt]

Pulsars, identified by pulsations & PSR & 7 & ... & ...	\\[7 pt] 
Pulsars, no pulsations seen yet & ... & ... & psr & 33	\\[7 pt] 
Pulsars wind nebula & PWN & 1 & pwn & 1	\\[7 pt] 
Supernova remnants & SNR & 4 & snr & 4	\\[7 pt] 
Supernova remnants / Pulsars wind nebula & ... & ... & spp & 3	\\[7 pt] 
Globular clusters & GLC & 0 & glc & 1	\\[7 pt] 
High-mass X-ray binaries & HMXB & 1 & hmxb & 3	\\[7 pt] 
Binaries & BIN & 0 & bin & 1	\\[7 pt] 
BL Lac type of blazars & BLL & 6 & bll & 13	\\[7 pt] 
FSRQ type of blazars & FSRQ & 15 & fsrq & 18	\\[7 pt] 
Radio galaxies & RDG & 0 & rdg & 2	\\[7 pt] 
Blazars candidate of uncertain type & BCU & 2 & bcu & 6	\\[7 pt] 
Total & ... & 36 & ... & 85	\\[7 pt] 
\hline
 & & & & \\[3pt]
Unidentified & ... & ... & unid & 31	\\[7 pt] 
Gamma Unidentified & ... & ... & gunid & 23	\\[7 pt] 
\hline
 & & & & \\[3pt]
Total Sources & ... & ... & ... & 175	\\[7 pt] 
Total AGILE Only sources & ... & ... & ... & 29	\\[7 pt]

\hline
\end{tabular}

\caption*{ Designations shown in capital letters are firm identifications; lower case letters indicate associations. The designation spp indicates potential association with SNR or PWN.}

\label{tab:cat2classes} 
\end{center}

\end{table*}

\abb{We have used different criteria to establish a firm identification of a 2AGL source with a known counterpart. }

{\cpbis 
AGN of the blazar type dominate the extragalactic \gray sky and are known to be highly variable in $\gamma$-rays, not always showing simultaneous correlated variability at other wavelengths. So positional consistency plus \gray variability were used for the identification of 2AGL sources with AGN counterparts.}
{\cp
Identification with} AGNs is {\cp established} if at least one flaring episode with a significance $>4\sigma$ \abb{and with a peak flux at least 3 times the average flux} is found  on a  4-days time scale in the 100 MeV -- 10 GeV energy range, or at OB 
level (the full list of OBs is reported in Table~\ref{tab:pointings}). \abb{ToO observations on the source could strongly influence the average flux; in that cases, only detections with a significance $>4\sigma$ are considered.} The light curves are determined with the flux and position parameters allowed to 
{\cp 
vary, with the position of the 2AGL source used as the starting position.} Section~\ref{sect:agn} reports a discussion about the association or identification of some AGNs. 

\begin{table*}[htbp]
\begin{center}
\caption{Association with \gray catalogs}
\begin{tabular}{|c c c|}
\hline
 & & \\
\multicolumn{1}{|c}{Catalog name} & 
\multicolumn{1}{c}{Source number} &
\multicolumn{1}{c|}{Paper references} \\[7 pt]
\hline
 & & \\[3 pt]

0FGL & 205 &  \cite*{abdo09_eta} \\[7 pt] 
1AGL & 47 &  \cite*{pittori09}	\\[7 pt] 
1AGLR & 54 &  \cite*{verrecchia13} \\[7 pt] 
1FGL & 1450 &  \cite*{abdo10b} \\[7 pt] 
2FGL & 1872 &  \cite*{nolan2012}  \\[7 pt] 
3EG & 267 &  \cite*{hartmann1999} \\[7 pt] 
3FGL & 3033 &  \cite*{acero15}  \\[7 pt] 
3FHL & 1556 &  \cite*{ajello17}  \\[7 pt] 
AGILE-TEVCAT & 52 & \cite*{rappoldi16} \\[7 pt] 
TeGeV & 168 & \cite*{carosi15} 	\\[7 pt] 

\hline
\end{tabular}

\label{tab:catalogs} 

\end{center}
\end{table*}

We introduce the special {\cp subclass} of `AGILE-only' sources, i.e. unidentified source that are only present in this \gray Catalog but not \abr{positionally consistent with sources} in 1FGL, 2FGL or 3FGL Catalogs: details are reported in Sect.~\ref{sect:agileonly}.

The `\gray unidentified' sources are a {\cp subclass} of sources detected in \gray by the AGILE-GRID {\cp and} the Fermi-LAT but that are unidentified {\cp in other wavelengths}: details are reported in Sect.~\ref{sect:gammaonly}.


{\fv Regarding candidate AGN sources, `AGILE-only' sources, and `\gray unidentified' sources, are also identified with the VOU-BLAZAR tool (hereafter VOUblaz)  specifically designed to identify blazar sources based on a multi-frequency study in large error regions, and through time resolved spectral energy distribution (SED) creation {\cp
and analysis}. This tool is developed within the Italian Space Agency (ASI) web portal for the "Open Universe" initiative\footnote{http://www.openuniverse.asi.it/} \citep{giommi18, padovani18}, an initiative under the auspices of the United Nations Committee On the Peaceful Uses of Outer Space (COPUOS). \abr{The multi-wavelength information used by the VOUblaz tool, both in the phase of the identification of blazar candidates and for the construction of the SED of a given candidate, is generally non simultaneous as it is obtained through VO queries to a large number (> 50) of catalogs and archival spectral data. A blazar found within the error region is considered as a viable counterpart of an AGILE source when the ratio between the low and high-energy humps (Compton dominance) is well within the range observed in previous gamma-ray catalogs.} Details of the analysis with this tool is reported in the discussion of each source.
}

Pulsars are firmly identified if pulsation is found in the AGILE-GRID data: {\cp details are} reported in Sect. \ref{sect:pulsar}.

\abb{
\abr{As a general rule for all classes}, for {\cp ``firmly} identified" counterparts we \abr{have included} \gray sources for which there are peer reviewed publications demonstrating high-confidence associations with refined analysis methods. \abr{In the SNR class we have IC 443, W28, W44 and Gamma Cygni (Sect.~\ref{sect:snrs} for more details), Crab Nebula as PWN (Sect.~\ref{sect:pwn} for more details), and Cygnus X-3 as an HMXB (Sect.~\ref{sect:bin} for more details). Also some AGNs has been identified in this way.}
}

\subsection{Extended sources}

{\cpbis We report in Table \ref{tab:extended} some details for 2AGL sources also classified  as extended \gray emission; they are marked with an `e' appended to the 2AGL name. 
Note that there could be a difference in positioning of the same source if evaluated as extended (as reported in Table \ref{tab:extended}) or point-like (as reported in Table~\ref{tab:cat2}).

}

\begin{table*}[htbp]
\begin{center}
\caption{Extended sources}
\label{tab:extended} 
\begin{tabular}{|l r r r r c|}
\hline
 &  & & & & \\
\multicolumn{1}{|c}{AGILE name} & 
\multicolumn{1}{c}{$\sqrt{TS}$} &
\multicolumn{1}{c}{$l_\mathrm{e}$} &
\multicolumn{1}{c}{$b_\mathrm{e}$} &
\multicolumn{1}{c}{r} &
\multicolumn{1}{c|}{Counterp.} \\ [3 pt]
\hline
 & & &  &  & \\[3 pt]
2AGL J0617+2239e   & 9.7 & 189.07 & 2.92 & 0.27 & IC 443  \\ 
2AGL J1856+0119e   & 10.2      & 34.65 & -0.39 & 0.30 & W44  \\ 
2AGL J1634-4734e   & 10.2 & 336.38 & 0.19 & 0.35 & HESS J1632-478  \\ 
2AGL J2044+5012e  & 4.5 & 88.75 & 4.67 & 0.59 & HB 21 \\[3 pt] 
\hline
\end{tabular}

\caption*{Extended sources modelled as 2D Gaussian in the 2AGL analysis. $l_\mathrm{e}$ and $b_\mathrm{e}$ are the centre of the extended region, in Galactic coordinates, that could be slightly different from the point-like source position. The $r$ column indicates the dispersion for Gaussian sources. The $\sqrt{TS}$ in the table is related to the result of the fitting with the extended shape with a Power Law spectral model with $\alpha=2.1$. Fluxes and spectra are reported in Table~\ref{tab:cat2} and Table~\ref{tab:cat2spectra}.}
\end{center}
\end{table*}

\subsection{Spectral models}
Each source is analysed fitting the data with the four spectral models described in Sect.~\ref{sect:spectralshape}. The final selection is based on the $TS_\mathrm{curved}$ criteria described in the same section. The 2AGL sources with curved spectra are listed in Table \ref{tab:curvedspectra}.

 Table~\ref{tab:cat2spectracolumns} reports a description of the columns of Table~\ref{tab:cat2spectra}, that presents the fluxes in individual bands as defined in Sect.~\ref{sect:refanal}.



\begin{figure*}
      \begin{subfigure}{0.5\textwidth}
        \centering\includegraphics[width=0.9\linewidth]{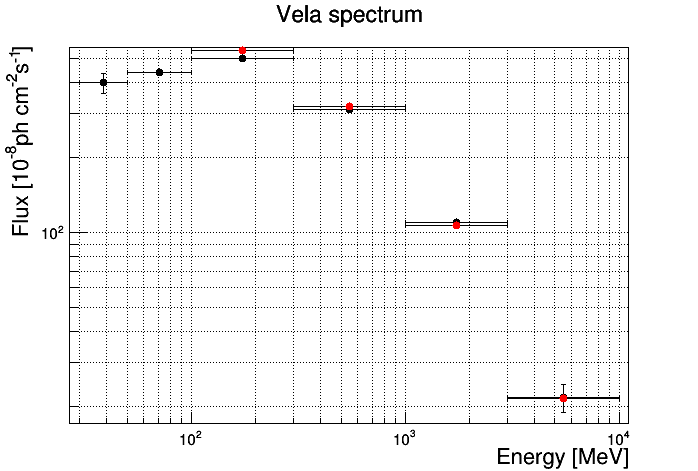}
          \caption{Vela pulsar}
     \end{subfigure}
     \begin{subfigure}{0.5\textwidth}
        \centering\includegraphics[width=0.9\linewidth]{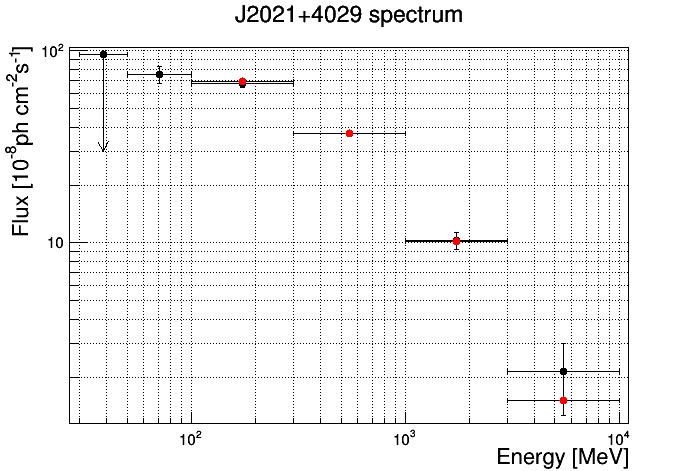}
          \caption{2AGL J2021+4029 (Gamma Cygni and LAT PSR J2021+4026)}
     \end{subfigure}
     \begin{subfigure}{0.5\textwidth}
    	\centering\includegraphics[width=0.9\linewidth]{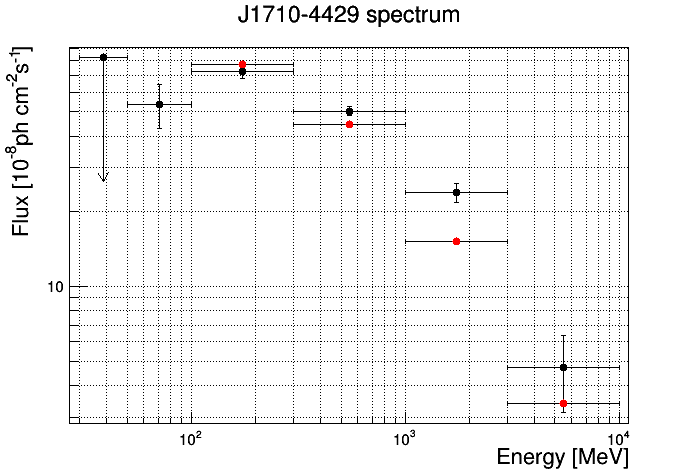}
        \caption{2AGL J1710-4429}
    \end{subfigure}
    \begin{subfigure}{0.5\textwidth}
    	\centering\includegraphics[width=0.9\linewidth]{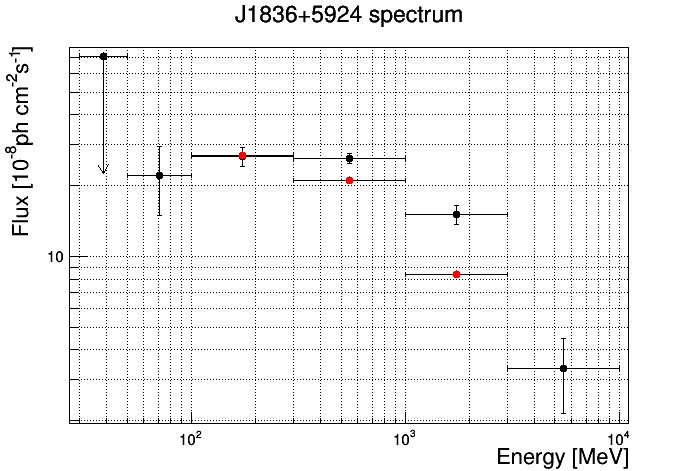}
        \caption{2AGL J1836+5924 (PSR J1836+5925)}
    \end{subfigure}
\caption{\abr{Comparison between the 2AGL (black spectra, 30 MeV -- 10 GeV) and 3FGL (red spectra, 100 MeV -- 10 GeV) spectra for some of the most highly exposed AGILE-GRID sources. }
Error bars are $1\sigma$ statistical error; upper limits are $2\sigma$. Different values in flux at the highest energy bands depend on the evaluation of the cut-off energies due to the statistics.}
\label{fig:spectra1}
\end{figure*}

\begin{table*}[ ht]
\begin{center}
\caption{2AGL sources with curved spectra.}
\label{tab:curvedspectra} 
\begin{tabular}{|l r r c c c r r r c c r l|}
\hline
 & & & & & & & & & & & & \\
\multicolumn{1}{|c}{2AGL Name} & 
\multicolumn{1}{c}{$F_\gamma$} &
\multicolumn{1}{c}{$\Delta F_\gamma$} &
\multicolumn{1}{c}{$\sqrt{TS}$} &
\multicolumn{1}{c}{SM} &
\multicolumn{1}{c}{$\alpha$} &
\multicolumn{1}{c}{$\Delta\alpha$} &
\multicolumn{1}{c}{$E_\mathrm{c}$} &
\multicolumn{1}{c}{$\Delta E_\mathrm{c}$} &
\multicolumn{1}{c}{$\beta$} &
\multicolumn{1}{c}{$\Delta \beta$} &
\multicolumn{1}{c}{$TS_\mathrm{curved}$} &
\multicolumn{1}{c|}{Counterpart} \\ [3 pt]
\hline
 & & & & & &  & & & & & &\\[3 pt]
J0835-4514 & $969.5$ & 13.8 & 168.9 & PS & 1.71 & 0.12 & 3913.1 & 533.6 & 1.35 & 0.3 & 127.8 & PSR J0835-4510 \\
J0634+1749 & $426.1$ & 9.8 & 76.3 & PC & 1.71 & 0.09 & 1232.0 & 69.9 &   &   & 82.2 & PSR J0633+1746 \\
J1710-4429 & $154.0$ & 8.4 & 43.5 & PC & 1.51 & 0.09 & 3025.2 & 957.4 &   &   & 16.0 & PSR J1709-4429 \\
J1836+5924 & $71.8$ & 5.2 & 41.3 & PC & 1.21 & 0.16 & 1988.4 & 662.3 &   &   & 30.4 & LAT PSR J1836+5925 \\
J2021+4029 & $119.3$ & 5.9 & 43.8 & PC & 1.76 & 0.10 & 3307.8 & 1335.6 &   &   & 16.0 & LAT PSR J2021+4026 \\
J2021+3654 & $70.9$ & 7.9 & 21.3 & PC & 1.38 & 0.30 & 950.6 & 361.1 &   &   & 17.4 & PSR J2021+3651 \\
J0007+7308 & $41.6$ & 4.2 & 26.2 & PC & 1.29 & 0.22 & 2003.1 & 1010.3 &   &   & 16.0 & PSR J0007+7303 \\
J1856+0119e & $58.8$ & 8.1 & 13.2 & LP & 2.54 & 0.30 & 1018.6 & 59.0 & 1.00 & 0.1  & 15.9 & SNR W44 \\
J1801-2334 & $35.8$ & 10.2 & 7.5 & LP & 3.37 & 0.46 & 2935.1 & 543.3 & 0.68  & 0.2  & 21.6 & SNR W28 \\[3 pt]
\hline
\end{tabular}
\caption*{2AGL sources with curved spectra. $F_\gamma$ and $\Delta F_\gamma$ is the photon flux expressed in $\mathrm{10^{-8}\ ph\ cm^{-2} s^{-1}}$ and related statistical error in the 100 MeV -- 10 GeV energy range, summed over 4 bands, and $\sqrt{TS}$ is the significance derived from likelihood Test Statistic in the same energy band. \textit{SM} is the spectral model: PC indicates Power Law with exponential cut-off fit to the energy spectrum, and PS indicates Power Law with super exponential cut-off fit to the energy spectrum. $\alpha$, $\beta$ and $E_\mathrm{c}$ are the indexes and the cut-off energies expressed in MeV described in Sect.~\ref{sect:spectralshape} with related $1\sigma$ errors. $TS_\mathrm{curved}$ is used for selection of curved spectral shape (see Sect.~\ref{sect:spectralshape}). Last column reports the association with known counterparts.}
\end{center}
\end{table*}

\begin{table*}[htbp]
\begin{center}
\caption{Spectral information for the 2AGL Catalog columns description of Table \ref{tab:cat2spectra} and Table \ref{tab:spactra_det_50_100}.} 
\label{tab:cat2spectracolumns} 
\begin{tabularx}{\textwidth}{| c | c | X |}
\hline
\multicolumn{1}{|c}{Column} & 
\multicolumn{1}{c}{Units} & 
\multicolumn{1}{c|}{Description} \\[3 pt]
\hline
 & & \\
Name &  & \abr{See Table \ref{tab:cat2columns} for a detailed description. The full list of 2AGL sources is reported in Table \ref{tab:cat2} } \\
$\sqrt{TS}$ & & Significance derived from likelihood Test Statistic for in the  100 MeV -- 10 GeV energy range\\
SM & & Spectral model. PL indicates power-law fit to the energy spectrum; PC indicates power-law with exponential cutoff fit to the energy spectrum; PS indicates power-law with super exponential cutoff fit to the energy spectrum; LP indicates log-parabola fit to the energy spectrum \\
$\alpha$ & & Spectral index for PL, PC and PS spectral models, first index for LP spectral model, in the 100 MeV -- 10 GeV energy range, see Sect.~\ref{sect:spectralshape} for the definition of $\alpha$\\
$\Delta\alpha$ & & Statistical $1\sigma$ uncertainty of $\alpha$, in the  100 MeV -- 10 GeV energy range\\

$F_a$ & $\mathrm{10^{-8}\ ph\ cm^{-2} s^{-1}}$ & Photon flux in the 30 MeV - 50 MeV energy band, with asymmetric errors\\
$\sigma_a$ & & Significance derived from likelihood Test Statistic in the 30 MeV - 50 MeV energy band\\

$F_b$ & $\mathrm{10^{-8}\ ph\ cm^{-2} s^{-1}}$ & Photon flux in the 50 MeV - 100 MeV energy band, with asymmetric errors\\
$\sigma_b$ & & Significance derived from likelihood Test Statistic in the 50 MeV - 100 MeV energy band\\

$F_1$ & $\mathrm{10^{-8}\ ph\ cm^{-2} s^{-1}}$ & Photon flux in the 100 MeV - 300 MeV energy band, with asymmetric errors\\
$\sigma_1$ & & Significance derived from likelihood Test Statistic in the 100 MeV - 300 MeV energy band\\
$F_2$ & $\mathrm{10^{-8}\ ph\ cm^{-2} s^{-1}}$ & Photon flux in the 300 MeV -- 1000 MeV energy band, with asymmetric errors\\
$\sigma_2$ & & Significance derived from likelihood Test Statistic in the 300 MeV -- 1000 MeV energy band \\
$F_3$ & $\mathrm{10^{-8}\ ph\ cm^{-2} s^{-1}}$ & Photon flux in the 1000 MeV -- 3000 MeV energy band, with asymmetric errors\\
$\sigma_3$ & &  Significance derived from likelihood Test Statistic in the 1000 MeV -- 3000 MeV energy band \\
$F_4$ & $\mathrm{10^{-8}\ ph\ cm^{-2} s^{-1}}$ & Photon flux in the 3000 MeV -- 10000 MeV energy band, with asymmetric errors \\
$\sigma_4$ & &  Significance derived from likelihood Test Statistic in the 3000 MeV -- 10000 MeV energy band\\
ID or ass. & & Designator of identified or associated source \\
Class & & Class (see Table \ref{tab:cat2classes})\\
\hline
\end{tabularx}
\end{center}
\end{table*}

\subsection{Comparison with the AGILE Astronomer's Telegrams}

\abb{The list of Astronomer's Telegrams in `pointing mode', obtained during the \gray flare monitoring program (see Sect. \ref{sect:obs}) and associated with 2AGL source are reported in Table \ref{tab:agileatelspointing}.}

 \subsection{Comparison with 1AGL and 1AGLR Catalogs}
Three  of the 47 1AGL Catalog sources are not present in the 2AGL Catalog: 1AGL J1222+2851, 1AGL J1238+0406 and 1AGL J1815-1732. We note that 1AGL J1222+2851 is associated with WComae (ON+231) (with an associated Astronomer's Telegram, see Table~\ref{tab:agileatelspointing}) but due to the longer integration time of the 2AGL Catalog this source goes below the significance threshold.

Four of the 54 1AGLR Catalog sources are not present in the 2AGL Catalog: 1AGL J1238+0406, 1AGLR J1807-2103, 1AGLR J2016+3644, and 1AGLR J2030-0617.

These differences are mainly due to a different integration time or to splitting a 1AGL/1AGLR sources in different 2AGL sources.

\subsection{Comparison with AGILE TeVCat}

Thirteen of the 52 sources of the AGILE TeVCat \cite{rappoldi16} are not present in the 2AGL catalog. These sources are not confirmed due to a different analysis procedure, different Science Tools, different background cuts ($albrad=85$, $fovradmax=60$) and improved IRFs, or to splitting the source in different 2AGL sources. The AGILE TeVCat sources not present in the 2AGL Catalog are: (1) TeVJ0232+202, associated with 1ES 0229+200, below  the 2AGL Catalog threshold, (2) TeVJ0521+211 (VER J0521+211), (3)  TeVJ0835-455 (Vela X), 
(4) TeVJ0852-463 (RX J0852.0-4622), (5) TeVJ1729-345 (HESS J1729-345), (6) TeVJ1732-347 (HESS J1731-347), (7) TeVJ1745-303 (HESS J1745-303), (8) TeVJ1813-178 (HESS J1813-178), (9) TeVJ1825-137 (HESS J1825-137), (10)  TeVJ1841-055 (HESS J1841-055), (11) TeVJ1912+101 (HESS J1912+101), (12) TeVJ2323+588 (Cassiopeia A), (13)  TeVJ2359-306 (HESS J2359-306).
 
\section{Notes on individual sources}
\label{sec:notes}

\abb{In this section we 
comment on some specific 2AGL sources, divided by classes or sky regions, including AGILE-GRID identifications or associations for individual sources based on criteria described in Section \ref{sec:assoc}.}

\abb{Sect.~\ref{sect:agn} reports a description of associated or identified AGNs, Sect.~\ref{sect:agileonly} on AGILE-only sources, Sect.~\ref{sect:gammaonly} on \gray only sources, Sect.~\ref{sect:pulsar} on pulsars,
Sect.~\ref{sect:cyg} reports a discussion on the Cygnus region, Sect. \ref{sect:carinareg} on  the Carina region, Sect.~\ref{sect:pwn} on PWNs, Sect.~\ref{sect:snrs} on SNRs, Sect.~\ref{sect:bin} on binaries, and Sect.~\ref{sect:confused} on confused sources.}

\subsection{Notes on AGN sources}
\label{sect:agn}

In the following we report identifications or associations for AGNs. In particular, we report the flaring episodes {\cpr detected with our variability analysis} used to establish an 
identification between the 2AGL source and a known counterpart, {\cpr and/or the reference to MultiWavelength (MW) information obtained by the use of the VOUblaz tool demonstrating high-confidence identifications.} 

\textbf{2AGL J0135+4754} Integrating in the time interval MJD 55037.5-55041.5, a MLE analysis yields a detection of $4.6\sigma$ and a flux $F=(55 \pm 17) \times 10^{-8} \mathrm{ph \: cm^{-2} s^{-1}}$, establishing a firm identification of 2AGL J0135+4754 source with OC 457. 

\textbf{2AGL J0252+5038} Integrating in the time interval MJD 54813.5-54817.5 a MLE analysis yields a detection of $4.0\sigma$ and a flux $F=(70 \pm 21) \times 10^{-8} \mathrm{ph \: cm^{-2} s^{-1}}$, positionally consistent with the FSRQ NVSS J025357+510256, establishing a firm identification with 2AGL J0252+5038.

\textbf{2AGL J0221+4250} The $95\%$ elliptical confidence region contains two source: 3C 66A and PSR J0218+4232, but the variability analysis excludes the PSR. Integrating in the OB 5820 (MJD 54632.5-54647.5), a MLE analysis yields a detection of $4.3\sigma$ and a flux $F=(32 \pm 8) \times 10^{-8} \mathrm{ph \: cm^{-2} s^{-1}}$. The best detection in the OB 5820 is in the time interval MJD 54641.5-54645.5, with a detection of $4.0\sigma$ and a flux $F(E>100 \: \MeV)=(50 \pm 15) \times 10^{-8} \mathrm{ph \: cm^{-2} s^{-1}}$, with a statistical $95\%$ c.l. elliptical confidence region that include only 3C 66A, establishing a firm identification of 2AGL J0221+4250 source with this BLL. 

\textbf{2AGL J0429-3755}.  Integrating in the time interval MJD 54497.5-54501.5  MJD a MLE analysis yields a detection of $4.9\sigma$ and a flux $F=(50 \pm 15) \times 10^{-8} \mathrm{ph \: cm^{-2} s^{-1}}$, positionally consistent with the BLL PKS 0426-380, establishing a firm identification with  2AGL J0429-3755 source.

{\cpr  
\textbf{2AGL J0531+1334}. A low level significance 2AGL source with a positional association with the FSRQ PKS 0528+134,
which could be promoted to firm identification by the use of the VOUblaz tool thanks to MW information. 
}


\textbf{2AGL J0538-4401}. Integrating in the OB 6210 (MJD 54749.5-54756.5), a MLE analysis yields a detection of $5.4\sigma$ and a mean flux $F=(35 \pm 9) \times \mathrm{10^{-8} ph \: cm^{-2} s^{-1}}$, positionally consistent with PKS 0537-441, establishing a firm identification of this AGN with 2AGL J0538-4401 source.
{\cpbis
The source PKS 0537-441 has an high average flux value in the 
2AGL Catalog because it has been mainly observed in a ToO pointing during a flaring state.}

\textbf{2AGL J0723+7122}. The 2AGL J0723+7122 source is firmly identified with BLL S5 0716+714. The highest detection of 2AGL J0723+7122 with a time resolution of 4-days is during the time interval MJD 54349.5-54353.5: an MLE analysis yields a detection of $6\sigma$ and a flux $F(E>100\: \mathrm{MeV})=(70 \pm 17) \times \mathrm{10^{-8} ph \: cm^{-2} s^{-1}}$. Additional observations are reported in \citep{giommi08, chen08, vittorini09}. In particular \citep{chen08} reports  a peak level of $F(E>100\: \mathrm{MeV})=(193 \pm 42) \times \mathrm{10^{-8} ph \: cm^{-2} s^{-1}}$ in MJD 54353.5-54354.5 (1-day timescale), and shows an increase in flux by a factor of four in three days. An ATEL is reported for this source (see Table \ref{tab:agileatelspointing}) during the `pointing mode'.

\textbf{2AGL J1052-6234}. We detect a flare during the time interval  MJD 54649.5-54653.5: a MLE analysis yields a detection of $4\sigma$ and a flux $F=(40 \pm 12) \times \mathrm{10^{-8} ph \: cm^{-2} s^{-1}}$, identifying the PMN J1047-6217 (that is inside the $95\%$ c.l. elliptical confidence region) with the 2AGL J1052-6234 source.

\textbf{2AGL J1228+4910}. The 2AGL J1228+4910 is  identified with FSRQ TXS 1226+492 {\cpr  
(also known as BZQ J1228+4858), }
because it shows a \gray flare during the period 54461.5-54465.5: an MLE analysis yields a detection of $4.0\sigma$ and a flux $F=(103 \pm 37) \times \mathrm{10^{-8} ph \: cm^{-2} s^{-1}}$. In addition, during the OB 6710 (MJD 54850.75-54890.5)  an MLE analysis yields a detection of $4.6\sigma$ and a mean flux $F=(89 \pm 23) \times \mathrm{10^{-8} ph \: cm^{-2} s^{-1}}$.

\textbf{2AGL J1228+0154}. The 2AGL J1228+0154 source is identified with FSRQ 3C 273. \gray activity is reported in \citep{pacciani2009}.

\textbf{2AGL J1255-0543}. The 2AGL J1255-0543 source is identified with  FSRQ 3C 279 \citep{giuliani2009c, pittori18}. In the automated analysis we detect  2AGL J1255-0543 during the time interval OB 5010 (MJD 54450.5-54473.5): an MLE analysis yields a detection of $4.9\sigma$ and a flux $F=(26 \pm 6) \times \mathrm{10^{-8} ph \: cm^{-2} s^{-1}}$. The \gray emission exhibited the largest amplitude variability on both long (months) and short (days) time scales.
{\cpbis
The source 3C 279 has an high average flux value in the 
2AGL Catalog because it has been mainly observed in ToO pointings during flaring states.}

\textbf{2AGL J1507+1019}. This source is identified with PKS 1502+106. We detect  2AGL J1507+1019 during the OB 6800 (MJD 54890.5-54921.5): an MLE analysis yields a detection of $6.1\sigma$ and a flux $F=(60 \pm 12) \times \mathrm{10^{-8} ph \: cm^{-2} s^{-1}}$. 

\begin{figure}
    \includegraphics[width=\linewidth]{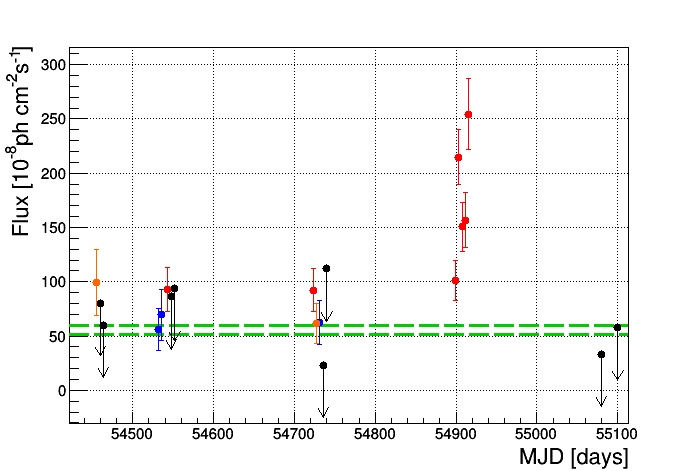}
    \caption{Light curve of PKS 1510-089 in the 100 MeV -- 10 GeV energy range, with a 4-days resolution. The black downward arrows represent $2\sigma$ upper limits. The blue, orange and red circles refer to a $\sqrt{TS} \geq 3$, $\geq4$ and $\geq 5$ respectively. The green dashed lines are the average flux plus/minus the error from Table~\ref{tab:cat2}.}
    \label{fig:lc_PKS1510M089}
\end{figure}

\textbf{2AGL J1513-0905}. The 2AGL J1513-0905 is identified with  PKS 1510-089. This FSRQ shows a strong variability during the AGILE pointing mode \citep{pucella2008, dammando2009, dammando2011, dammando2011b}. The 2AGL Catalog light curve is reported in Fig.~\ref{fig:lc_PKS1510M089}. Many ATELs are reported for this source (see Table~\ref{tab:agileatelspointing}).

\textbf{2AGL J1626-2943}. The 2AGL J1626-2943 source is identified with  PKS 1622-29. We detect the 2AGL J1626-2943 during the time interval MJD 54329.5-54333.5: an MLE analysis yields a detection of $5\sigma$ and a flux $F=(95 \pm 26) \times \mathrm{10^{-8} ph \: cm^{-2} s^{-1}}$. 
Another detection of $3.8\sigma$ with a flux of $F=(117 \pm 35) \times \mathrm{10^{-8} ph \: cm^{-2} s^{-1}}$ is detected in MJD 54517.5-54521.5.

{\cpr  
\textbf{2AGL J1741+5126}. The $95\%$ c.l. elliptical confidence region is marginally compatible with FSRQ 4C +51.37. 
The association can be promoted to identification by the use of the VOUblaz tool thanks to MW information.
}

\textbf{2AGL J1803-3935}. The 2AGL J1803-3935 source is identified with FSRQ PMN J1802-3940. We detect 2AGL J1803-3935 during the time interval  MJD 55077.5-55081.5; a MLE analysis yields a detection of $4.5\sigma$ and a flux $F=(46 \pm 14) \times \mathrm{10^{-8} ph \: cm^{-2} s^{-1}}$.

\textbf{2AGL J1833-2104}. The 2AGL J1833-2104 source is identified with FSRQ PKS 1830-211 \citep{donnarumma11}. We detect 2AGL J1833-2104 during many time intervals in the 4-days timescale light curve. The two main flares are: (i) MJD 54529.5-54533.5, where a MLE analysis yields a detection of $4.0\sigma$ and a flux $F=(48 \pm 15) \mathrm{\times 10^{-8} ph \: cm^{-2} s^{-1}}$; (ii) MJD 55113.5-55117.5, where a MLE analysis yields a detection of $5.5\sigma$ and a flux $F=(50 \pm 12) \times \mathrm{10^{-8} ph \: cm^{-2} s^{-1}}$. An ATEL is reported for this source (see Table \ref{tab:agileatelspointing}) during the `pointing mode'.

\textbf{2AGL J1849+6706}. The 2AGL J1849+6706 source is identified with FSRQ S4 1849+67. We detect 2AGL J1849+6706 during many time intervals in the 4-days timescale light curve. The two main flares are: (i) MJD 54637.5-54641.5, where a MLE analysis yields a detection of $4.2\sigma$ and a flux $F=(42 \pm 14) \times \mathrm{10^{-8} ph \: cm^{-2} s^{-1}}$; (ii) MJD 54853.5-54857.5 , where a MLE analysis yields a detection of $4.5\sigma$ and a flux $F=(44 \pm 14) \times \mathrm{10^{-8} ph \: cm^{-2} s^{-1}}$.

\textbf{2AGL J1913-1928}. We found two interesting sources within the 2AGL J1913-1928 $95\%$ c.l. elliptical confidence region: PMN J1911-1908 (an already known \gray emitter present in the 3FGL Catalog - 3FGL J1911.4-1908) and PKS B1908-201 (3FGL J1911.2-2006). We detect  2AGL J1913-1928  during the  OB 4800 (MJD 54406.5-54435.5); a MLE analysis yields a detection of $5.7\sigma$ and a flux $F=(80 \pm 16) \times \mathrm{10^{-8} ph \: cm^{-2} s^{-1}}$, but this detection still excludes the two sources. Another detection is reported during the time interval MJD 54725.5-54729.5; a MLE analysis yields a detection of $4.0\sigma$ and a flux $F=(40 \pm 12) \times \mathrm{10^{-8} ph \: cm^{-2} s^{-1}}$ that is associable only with the PMN J1911-1908.  
{\cpr    
By the use of the VOUblaz tool, thanks to spectral MW information, the most probable association would be with the PKS B1908-201.} 
In any case, due to the uncertainties this source is classified as unidentified.

\textbf{2AGL J2025+3352}. The 2AGLJ2025+3352 source is identified with BCU B2 2023+33. We detect 2AGL J2025+3352 during the time interval MJD 54425.5-54429.5; a MLE analysis yields a detection of $5.6\sigma$ and a flux $F=(109 \pm 25) \times \mathrm{10^{-8} ph \: cm^{-2} s^{-1}}$.

\begin{figure}
    \includegraphics[width=\linewidth]{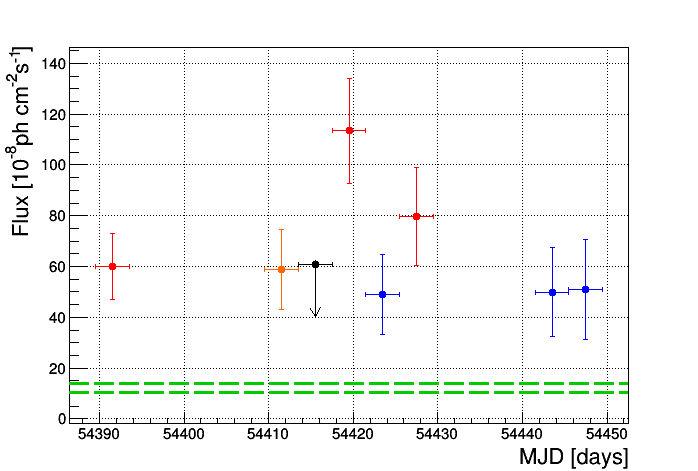}
    \caption{Light curve of PKS 2023-07 in 100 MeV -- 10 GeV energy range, with a 4-days resolution. The black downward arrows represent $2\sigma$ upper limits. The blue, orange and red circles refer to a $\sqrt{TS} \geq 3$, $\geq 4$ and $\geq 5$ respectively. The green dashed lines are the average flux plus/minus the error from Table~\ref{tab:cat2}}\label{fig:lc_PKS2023-07}
\end{figure}

\textbf{2AGL J2027-0740}. The 2AGL J2027-0740 source is identified with PKS 2023-07. We detect 2AGL J2027-0740 during many  time intervals (see Fig.~\ref{fig:lc_PKS2023-07}): (i) MJD 54389.5-54393.5 with a detection of $6\sigma$ and a flux $F=(60 \pm 13) \times \mathrm{10^{-8} ph \: cm^{-2} s^{-1}}$; (ii)  MJD 54417.5-54421.5  with a detection of $7.6\sigma$ and a flux $F=(113 \pm 21) \times \mathrm{10^{-8} ph \: cm^{-2} s^{-1}}$; (iii) MJD 54425.5-54429.5 with a detection of $5.4\sigma$ and a flux $F=(80 \pm 19) \times \mathrm{10^{-8} ph \: cm^{-2} s^{-1}}$.

{\cpr
\textbf{2AGL J2202+4214}. This source is firmly identified with BL Lacertae. Firm identification of this source has been established by the use of the VOUblaz tool thanks to MW information.
}

\begin{figure}
    \includegraphics[width=\linewidth]{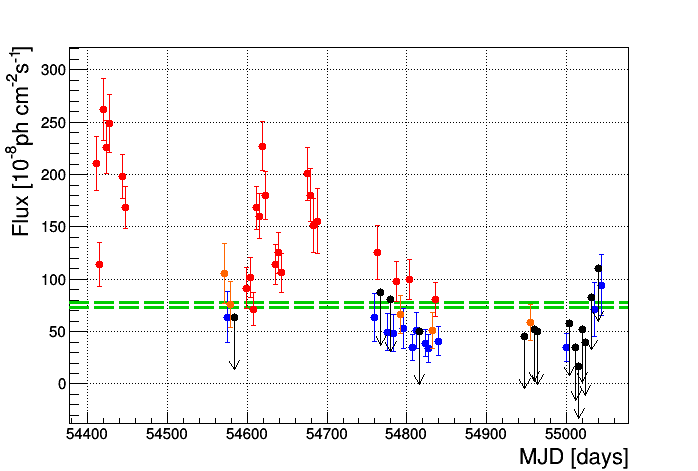}
    \caption{Light curve of 3C454.3 in the 100 MeV -- 10 GeV energy range, with a 4-days resolution. The black downward arrows represent $2\sigma$ upper limits. The blue, orange and red circles refer to a $\sqrt{TS}$ value $\geq 3$, $\geq 4$ and $\geq 5$ respectively. The green dashed lines are the average flux plus/minus the error from Table~\ref{tab:cat2}}\label{fig:lc_3C4543}
\end{figure}

\textbf{2AGL J2254+1609}. This source is firmly identified with 3C454.3. Fig.~\ref{fig:lc_3C4543} reports the light curve in `pointing period'. Firm identification of this source has been established thanks to different MW campaigns \citep{vercellone2008a, vercellone2008b, donnarumma09, vercellone10, pacciani10, striani10, vercellone11}. Many ATELs are reported for this source (see Table \ref{tab:agileatelspointing}).

\subsection{AGILE-only sources}
\label{sect:agileonly}

In Table~\ref{tab:agileonly} we report the full list of AGILE-only sources, i.e. 2AGL sources that do not have any counterpart on 1FGL, 2FGL or 3FGL catalogs. 
For AGILE-only sources, the light curves over 1-, 4- and 7-day time-scales have been calculated, in addition to the light curves estimated over OB timescales. All the \gray fluxes reported below for each source are estimated through the AGILE-GRID MLE analysis in the 100 MeV -- 10 GeV energy range. \abb{Some detections with a significance of less than $4\sigma$ are also reported 
for a possible identification}. 

\textbf{2AGL J0714+3318}. This source is already part of 1AGL and 1AGLR catalogs (based on single OB data analysis); we have a detection in the OB 4200 (MJD 54377.5-54385.5), as in 1AGLR: a MLE analysis over this period yields a detection of $4.0\sigma$ and a flux $F=(30 \pm 9) \times \mathrm{10^{-8} ph \: cm^{-2} s^{-1}}$. Just outside the $95\%$ c.l. elliptical confidence region we note the 5BZQ J0719+3307/B2 0716+33 classified as a FSRQ and associated to the Fermi-LAT source 3FGL J0719.3+3307.
{\fvr 
Using the VOUblaz tool, we investigated some candidate {\cpr associations in other wavelengths}:
just outside the 95\% error ellipse, there is the 3HSP J071223.5+331333, with a SED compatible with the AGILE-GRID sensitivity.
}

{\cpr  
}
\textbf{2AGL J1120-6222}. The 2AGL J1120-6222 shows a high confidence detection over the period MJD 55055.5-55062.5 (7 days); a MLE analysis yields a detection of $4.6\sigma$ and a flux $F=(67 \pm 18) \times \mathrm{10^{-8} ph \: cm^{-2} s^{-1}}$. 
{\cpr Interestingly,
we note that just inside the $95\%$ c.l. elliptical confidence region it is located the HMXB 1A 1118-615 (WRAY 15-793), a Be X-ray Binary system included in the INTEGRAL General Reference Catalog (INTREFCAT, version 41 of the 22 June 2018; \citep{ebisawa03}). 
}

\textbf{2AGL J1138-1724}. The  2AGL J1138-1724 source is identified with the BL Lac 5BZB J1137-1710, associated also with 2FHL J1137.9-1710. We detect a \gray flare during the time interval MJD 54968.5-54969.5 (1-day timescale): an MLE analysis yields a detection of $4\sigma$ and a flux $F=(130 \pm 50) \times \mathrm{10^{-8} ph \: cm^{-2} s^{-1}}$.

\textbf{2AGL J1203-2701c}. This source is marked with `c' because the position could be influenced by another \gray excess just outside the $95\%$ c.l. elliptical confidence region. No candidate source within the ellipse is present from flat spectrum radio source of other catalogs of interest. {\cpr The VOUblaz tool allows to find some candidate sources to be associated to this 2AGL, of which 5BZQ J1205-2634, an LBL just outside the 95\% error ellipse, is the most probable counterpart.
}

\textbf{2AGL J1312-3403}. 
{\cpbis {We note that outside the AGILE 95\% error ellipse of this AGILE-only source, at about $0.46\degmark$, there is the 3FGL J1311.8-3430, which is a LAT PSR J1311-3430 (associated to the 0FGL J1311.9-3419 and to the 3EG J1314-3431).
}
In MJD 54699.5-54701.5 time interval (2-days timescale) we detect a spatially coincident  event at $4.0\sigma$ with a flux of $F=(97 \pm 38) \times \mathrm{10^{-8} ph \: cm^{-2} s^{-1}}$. 
}
{\fvr
The VOUblaz tool shows 
{\cpr  also an IBL candidate just outside the 95\% error ellipse, with a single radio point in the SED, weak XMM X-ray data, and no \grays; extrapolation from X-ray to \gray in the AGILE-GRID range is difficult.}}

{\cpr  
\textbf{2AGL J1402-8142}. This source has no known
candidate counterparts within the $95\%$ elliptical confidence region, either radio or from specific source classes. In  MJD 54751.5-54753.5 time interval (2-days timescale) we detect an event at $3.9\sigma$ with a flux of $F=(56 \pm 20) \times \mathrm{10^{-8} ph \: cm^{-2} s^{-1}}$. Just outside the ellipse at $\sim 1\degmark$, there is the 4U 1450-80 X-ray source of the INTREFCAT. 
}

\textbf{2AGL J1628-4448}. Low latitude source without any known \gray source within its elliptical confidence region. A detection appears in the OB 6200 (MJD 54719.5-54749.5) at $3.6\sigma$ and a flux $F=(121 \pm 35) \times \mathrm{10^{-8} ph \: cm^{-2} s^{-1}}$. \abr{Fermi-LAT (which was running during the OB 6200) did not report the detection of the source because it was just for a $14\%$ of the time below $50\degmark$ off-axis; more details in Appendix A. }

{\cpr  

\textbf{2AGL J1631-2039}. This is an high latitude source whose elongated ellipse does not include any known \gray source. We detect this source as active over the OB 6800 period  (MJD 54890.5-54921.5), at $3.6\sigma$ and a flux $F=(16 \pm 5) \times \mathrm{10^{-8} ph \: cm^{-2} s^{-1}}$. 
} {\fvr Using the VOUblaz tool an interesting 408 mJy (1.4 GHz) NVSS radio source, PMN J1629-2039, is found at 13$\arcmin$ (R.A.,Dec.: 247.26267,-20.44881), but the possible extrapolated \gray emission into the AGILE-GRID range for a possible LBL \citep{padovani95} blazar is too weak for the AGILE-GRID sensitivity.}

{\fvr
}

\textbf{2AGL J1636-4610} Low latitude source 
with a detection at $4.1\sigma$ and a flux $F=(16 \pm 4) \times \mathrm{10^{-8} ph \: cm^{-2} s^{-1}}$ appears over the  OB 6800  period (MJD 54890.5-54921.5). 

\textbf{2AGL J1640-5050c}. Low latitude source showing a very soft spectral index of $2.84 \pm 0.17$. Two detections appear from the OB time-scale light curve: one at 3.9$\sigma$ and a flux $F=(19.2 \pm 5.2) \times \mathrm{10^{-8} ph \: cm^{-2} s^{-1}}$ on MJD 54647.5-54672.75 (OB 5900); the other one at $3.4\sigma$ and a flux $F=(12.4 \pm 3.8) \times \mathrm{10^{-8} ph \: cm^{-2} s^{-1}}$ on MJD 55055.5-55074.5 (OB 7800). Its large elliptical confidence location region partially overlaps with the unassociated 3FGL J1643.6-5002 error ellipse. Another Fermi-LAT unassociated source, 3FGL J1639.4-5146, appears at around $1\degmark$ from the source centroid. This source is marked with `c' due to the large source confidence location region.

\textbf{2AGL J1731-0527}. High latitude source without any known \gray source within its elliptical confidence region. It shows a detection at $3.7\sigma$ and flux $F=(27.6 \pm 8.7) \times \mathrm{10^{-8} ph \: cm^{-2} s^{-1}}$ over the OB period MJD 55104.5-55119.5 (OB 8300), plus other detections over 2-, 4- and 7-day time-scales. Within the confidence location region, at 15$\arcmin$ from the source centroid, lays the flat-spectrum radio source CRATES J173032-051505.

\textbf{2AGL J1736-7819}. Low latitude source with no known \gray source within its elliptical confidence region. Partially overlapping with the 95\% confidence contour of the EGRET source  3EG J1720-7820. The source is also detected on a 4-day integration starting at MJD=54653.5 at $3.1\sigma$, with a flux $F=(21 \pm 8) \times \mathrm{10^{-8} ph \: cm^{-2} s^{-1}}$. At 19$\arcmin$ from the 2AGL J1736-7819 centroid lays the radio source PKS 1723-78.

\textbf{2AGL J1740-3013}. This is a Galactic Centre region source, positionally consistent with TeVJ1741-302 (HESS J1741-302). Despite several attempts to constrain its nature, no plausible counterpart has been found \citep{abdalla18}. We note an increasing flux from this source during the 7-days time interval MJD 54740.5-54747.5; a MLE analysis yields a detection of $4.3\sigma$ and a flux $F=(86 \pm 23) \times \mathrm{10^{-8} ph \: cm^{-2} s^{-1}}$.

\textbf{2AGL J1743-2613c}. The 2AGL J1743-2613c is an unidentified source of the Galactic Centre region. Within $0.8\degmark$ we found 3FGL J1740.5-2642, and 3FGL J1741.9-2539 associable with the BCU NVSS J174154-253743. This source is marked with `c' due to the complexity of the region.

\textbf{2AGL J1746-2921}. The 2AGL J1746-2921 is a flaring source of the Galactic Centre. The source had shown a increasing \gray emission in the OB 6200 (MJD 54719.5-54749.5). More refined light curves of 1-, 4- and 7- days time scale showed flaring activities in these periods: 54719.5-54726.5 (7-days), 54730.5-54731.5 (1-day), 54737.5-54741.5 (4-days).

\textbf{2AGL J1754-2626}. The 2AGL J1754-2626 is an unidentified source near the Galactic Centre. At about 11$\arcmin$ appears the high-mass X-ray binary IGR J17544-2619 \citep{liu09}, seen by INTEGRAL  \citep{ebisawa03} and also by Swift-BAT \citep{swiftbat105}. The source shows two flaring episodes on the OB 5600 (MJD 54566.5/54586.5) and OB 6200 (MJD 54719.5-54749.5), both with a significance above $5\sigma$ and a flux $F=(58 \pm 12) \times \mathrm{10^{-8} ph \: cm^{-2} s^{-1}}$ and $F=(23 \pm 5) \times \mathrm{10^{-8} ph \: cm^{-2} s^{-1}}$, respectively.

\textbf{2AGL J1820-1150c}. Low latitude source with no known \gray source within its elliptical confidence region. It shows a strong flaring episode at $6.1\sigma$ and a flux $F=(85 \pm 16) \times \mathrm{10^{-8} ph \: cm^{-2} s^{-1}}$ over the OB 4800 period (MJD 54406.5-54435.5). No clear possible association has been identified. This source is marked with `c' due to the large uncertainty on its location.

\textbf{2AGL J1823-1504}. The 2AGL J1823-1504 is an unidentified source of the Galactic plane. The 3EG J1824-1514 was partially inside the $95\%$ elliptical confidence region. We detect a \gray flare from 2AGL J1823-1504 region in the time interval MJD 54552.5-54553.5; a MLE analysis yields a detection of $4.1\sigma$ and a flux $F=(281 \pm 86) \times \mathrm{10^{-8} ph \: cm^{-2} s^{-1}}$. 

\textbf{2AGL J1857+0015}. Low latitude source without any known \gray source within its elliptical confidence  region, close to the W44 region. It shows only a detection at $3.4\sigma$ and a flux $F=(105 \pm 39) \times \mathrm{10^{-8} ph \: cm^{-2} s^{-1}}$ over the 4-day time-scale MJD 54773.5-54777.5. At 46$\arcmin$ from the source centroid lays the pulsar PSR B1853+00.

\textbf{2AGL J1910-0637}. This source is just outside the Galactic plane and we have two detection on week time scale, the most significant at $3.6\sigma$ in MJD 54572.5-54579.5 time interval. No \gray source  neither flat spectrum radio source are within the $95\%$ elliptical confidence  region.

\textbf{2AGL J1913+0050}. This AGILE-GRID-only source appears just below the Galactic plane. It shows two detections over OB 6400 (MJD 54770.5-54800.5): one at $3.5\sigma$ and flux $F=(8.0 \pm 2.5) \times \mathrm{10^{-8} ph \: cm^{-2} s^{-1}}$, and the other one at $3.2\sigma$ and flux $F=(14.0 \pm 4.6) \times \mathrm{10^{-8} ph \: cm^{-2} s^{-1}}$ on OB 6500 (MJD 54800.5-54820.5). Interestingly, within the $95\%$ elliptical confidence location region, at 27$\arcmin$ from the source centroid, appears the low-mass X-ray binary Aql X-1 (also known 4U 1908+005), seen also by SWIFT-BAT (SWIFT J1911.2+0036 \citep{swiftbat105}).

\textbf{2AGL J1927-4318}. High latitude source without any known \gray source within its elliptical confidence  region. It shows a detection at $3.3\sigma$ and a flux $F=(17.2 \pm 6.0) \times \mathrm{10^{-8} ph \: cm^{-2} s^{-1}}$ over the OB 6200 period (MJD 54719.5-54749.5) plus a detection at $3.1\sigma$ over a short time-scale of 4-days on MJD 54533.5-54537.5. Within the AGILE-GRID elliptical source contour, there is the radio source PKS B1922-430 (at 25$\arcmin$ from the source centroid).

\textbf{2AGL J2029+4403}. 
 Low latitude source in the complex Cygnus region without any known \gray source within its elliptical confidence location region. We note that at about 28$\arcmin$ there is the LAT pulsar 3FGL J2030.8+4416.

\textbf{2AGL J2055+2521}. The PSR 3FGL J2055.8+2539 is at 22$\arcmin$ from the source centroid, just outside the $95\%$ elliptical confidence region. It shows a detections in MJD 54426.5-54427.5 (1-day timescale) at $4.1\sigma$ and flux $F=(191 \pm 71) \times \mathrm{10^{-8} ph \: cm^{-2} s^{-1}}$. 

\textbf{2AGL J2206-1044}. We have a \gray detection on OB 8300 (MJD 55104.5-55111.5) at $4.7\sigma$  and a flux $F=(63 \pm 16) \times \mathrm{10^{-8} ph \: cm^{-2} s^{-1}}$.
A 390 mJy NVSS (1.4GHz) radio source is just outside the $95\%$ c.l. elliptical confidence region. \abr{The source has a $VI=1$ at 4 days timescale.}

\textbf{2AGL J2227+6418}. We detect a \gray flare from 2AGL J2227+6418 region over the time interval MJD 54425.5-54429.5 (4-days). An MLE analysis yields a detection of $4.9\sigma$ and a flux $F=(32 \pm 8) \times \mathrm{10^{-8} ph \: cm^{-2} s^{-1}}$.

\textbf{2AGL J2228-0818}. The  2AGL J2228-0818 source is an high latitude AGILE-only source with a variable behaviour. We detect  2AGL J2228-0818 during the time interval MJD 54425.5-54429.5 (4-days); a MLE analysis yields a detection of $4.2\sigma$ and a flux $F=(61 \pm 18) \times \mathrm{10^{-8} ph \: cm^{-2} s^{-1}}$. Another interesting episode is in the time interval MJD 54573.5-54577.5: a MLE analysis yields a detection of $4.1\sigma$ and a flux $F=(80 \pm 22) \times \mathrm{10^{-8} ph \: cm^{-2} s^{-1}}$. \abr{The source has a $VI=1$ at 4 days timescale.}

\textbf{2AGL J2303+2120}. This high latitude weak source, with a large error ellipse, has been detected in a single 2-days integration on MJD 54571.5-54573.5 time interval at $3.7\sigma$ and a flux $F=(150 \pm 50) \times \mathrm{10^{-8} ph \: cm^{-2} s^{-1}}$  and at $3\sigma$ on a single 1 week exposure. 

\textbf{2AGL J2332+8215c}. 
We find a marginal overlap of 2AGL J2332+8215c $95\%$ c.l. elliptical confidence region with FSRQ S5 2353+81. We detect a \gray flare from 2AGL J2332+8215c region on the time interval MJD 54425.5-54429.5 (4-days): an MLE analysis yields a detection of $5\sigma$ and a flux $F=(37 \pm 8) \times \mathrm{10^{-8} ph \: cm^{-2} s^{-1}}$. 
{\cpr 
No firm association can be established} and the source is marked as `c' since the region appears noisy.

\subsection{\gray only sources}
\label{sect:gammaonly}

In this section we report notes for some \gray only sources. Table \ref{tab:gunid} reports the full list.

\textbf{2AGL 0032+0512c}. This source is marginally {\cpr spatially coincident} with 3FGL J0030.4+0451, associated with the PSR J0030+0451. This source is marked with `c' because the evaluation of its position is strongly influenced by the spectral index $\alpha$. For this reason we keep this source as \gray only without association with the PSR.
{ \fvr
}
{ \fvr
The VOUblaz tool allows to find some possible blazar candidates among which: 
(i) 5BZQ J0029+0554 (CRATES J002945+055443), a possible LBL. The SED created with the tool shows a hint of synchrotron bump but the extrapolated \gray emission is too weak for the AGILE-GRID sensitivity; (ii) a candidate NVSS radio source (J003047+044038). Again considering the source SED the possible synchrotron peak is too low for the extrapolated \gray emission in the AGILE-GRID band.
}

\textbf{2AGL J0221+6208c}. The region of 2AGL J0221+6208c includes 3FGL J0217.3+6209 (TXS 0213+619), 3FGL J0220.1+6202c, 3FGL J0223.6+6204 (unass.) and, partially, 3FGL J0224.0+6235. Integrating in the time interval MJD 54641.5-54645.5 (4-days timescale), a MLE analysis yields a detection of $4.1\sigma$ and a flux $F=(45 \pm 13) \times 10^{-8} \mathrm{ph \: cm^{-2} s^{-1}}$ in the  100 MeV -- 10 GeV energy range, but the $95\%$ confidence region already includes the mentioned 3FGL sources. \abr{For this reason we classify this source as \gray only but without association with a specific 3FGL source.}

{\fvr
\textbf{2AGL J1103-7747}. This source is spatially associated to the unclassified \gray source 3FGL J1104.3-7736c. Using the VOUblaz tool at least three candidate radio source are found, among which CRATES J105731-772424 at $4.9\arcmin$ from the Fermi-LAT source: the SED seems compatible with a weak \gray emission in the AGILE-GRID range.
}

\textbf{2AGL J1718-0432}. \abr{The 2AGL J1718-0432 source is spatially associated to the unclassified \gray source 3FGL J1720.3-0428,} and shows a detection at $4.0\sigma$ and a flux $F=(35 \pm 10) \times \mathrm{10^{-8} ph \: cm^{-2} s^{-1}}$ in MJD 54897.5-54899.5.
{ \fvr
The VOUblaz tool shows a known blazar with no radio/X-ray could be associated, 5BZQ J1716-0452, at a distance of 27.316\,$\arcmin$,  a probable FSRQ; the SED shows a clear synchrotron bump, but without X-ray or \gray counterparts. 
}

{\fvr
}

\textbf{2AGL J1737-3206}. The 2AGL J1737-3206 shows a full overlapping with 3FGL J1737.3-3214c.

\textbf{2AGL J1847-0157} The 2AGL J1847-0157 shows a full overlapping with 3FGL J1848.4-0141. We note the source TeV J1848-017 (HESS J1848-018) inside the $95\%$ c.l. elliptical confidence region, already present in the AGILE TeVCat \citep{rappoldi16}. 

\subsection{Pulsars}
\label{sect:pulsar}

\abr{Forty pulsars are associated with 2AGL sources. Thirteen  pulsars have been identified with pulsed timing analysis on AGILE data, but only seven of them are included in the 2AGL Catalog because for these we also obtained spatial identification using the MLE analysis.}

PSR J2021+3651 (2AGL J2021+3654) has been the first pulsar identified by pulsation in the AGILE-GRID \gray data \citep{halpern2008}. 
In \citep{pellizzoni09} the pulsation of Vela (2AGL J0835-4514), Geminga (2AGL J0634+1749, Fig.~\ref{fig:spectra1} shows the spectra), Crab (2AGL J0534+2205) and PSR B1706-44 (2AGL J1710-4429) has been reported. 

In \citep{pellizzoni09b} the detection of pulsation for 7 additional pulsar has been reported; of them, (i) four pulsars are identified only by pulsation because no spatial identification using the MLE analysis is possible: PSR J1357-6429, PSR J1524-5625, PSR J1824-2452, and PSR J2043+2740; (ii)  PSR J1016-5857 (2AGL J1015-5852), PSR J1513-5908 (2AGL J1517-5909), and PSR J2229+6114 (2AGL J2230+6113)  have been identified with both pulsed emission and MLE analysis. An additional comment about PSR J1824-2452 is necessary, because \citep{pellizzoni09b} reports also an MLE identification of this source: PSR J1824-2452, inside the globular cluster M28, was observed by the AGILE-GRID with good significance ($>4\sigma$) but only within a time interval of 5-days during a targeted pointing (for details see \citep{pellizzoni09b}). No increase in significance of the detection was obtained with the addition of more data. The source was also detected in the imaging data relative to the same time interval. The apparent single burst of emission observed by the AGILE-GRID might be related to the fact that this millisecond pulsar is known to experience variability episodes \citep{cognard04}.

The pulsation of PSR J2021+4026, that is part of the extended source pulsar + nebula 2AGL J2021+4029, has been identified. 

\abr{To summarise, the seven pulsars with identified pulsed emission that are firmly identified as 2AGL sources are}: PSR J2021+3651, Vela, Geminga, PSR B1706-44,  PSR J1016-5857, PSR J1513-5908, and PSR J2229+6114. The PSR J2021+4026 is classified as SNR and Crab region is classified as PWN (see Sect.~\ref{sect:pwn}) because the analysis considers the integrated flux of the regions.

In the following we report additional notes.



\textbf{2AGL J1029-5834}. The source is associated with  PSR J1028-5819 but there is only a partial overlap of the $95\%$ elliptical confidence region. 

\textbf{2AGL J1458-6055}. The $95\%$ elliptical confidence region of 2AGL J1458-6055 source partially overlaps the $95\%$ elliptical confidence region of LAT PSR J1459-6053 and 3FGL J1456.7-6046. It is not possible to establish a direct association with the pulsar.

\textbf{2AGL J1517-5909}. PSR J1513-5908 \citep{pellizzoni09b, pilia10} is included in the 2AGL J1517-5909 $95\%$ elliptical confidence region. In the same region  the associated extended source PWN MSH 15-12 is not detected by the AGILE-GRID.

\textbf{2AGL J1710-4429}. PSR J1709-4429 (PSR B1706-44) \citep{pellizzoni09} is firmly identified with 2AGL J1710-4429. 
Fig.~\ref{fig:spectra1} reports the spectra of this 2AGL source.


\textbf{2AGL J1836+5924}. The 2AGL J1836+5924 \citep{bulgarelli08} is associated with LAT PSR J1836+5925 and shows an exponential cut-off Power Law spectral shape. 



\subsection{The Cygnus region}
\label{sect:cyg}
\begin{figure*}
    \includegraphics[width=\linewidth]{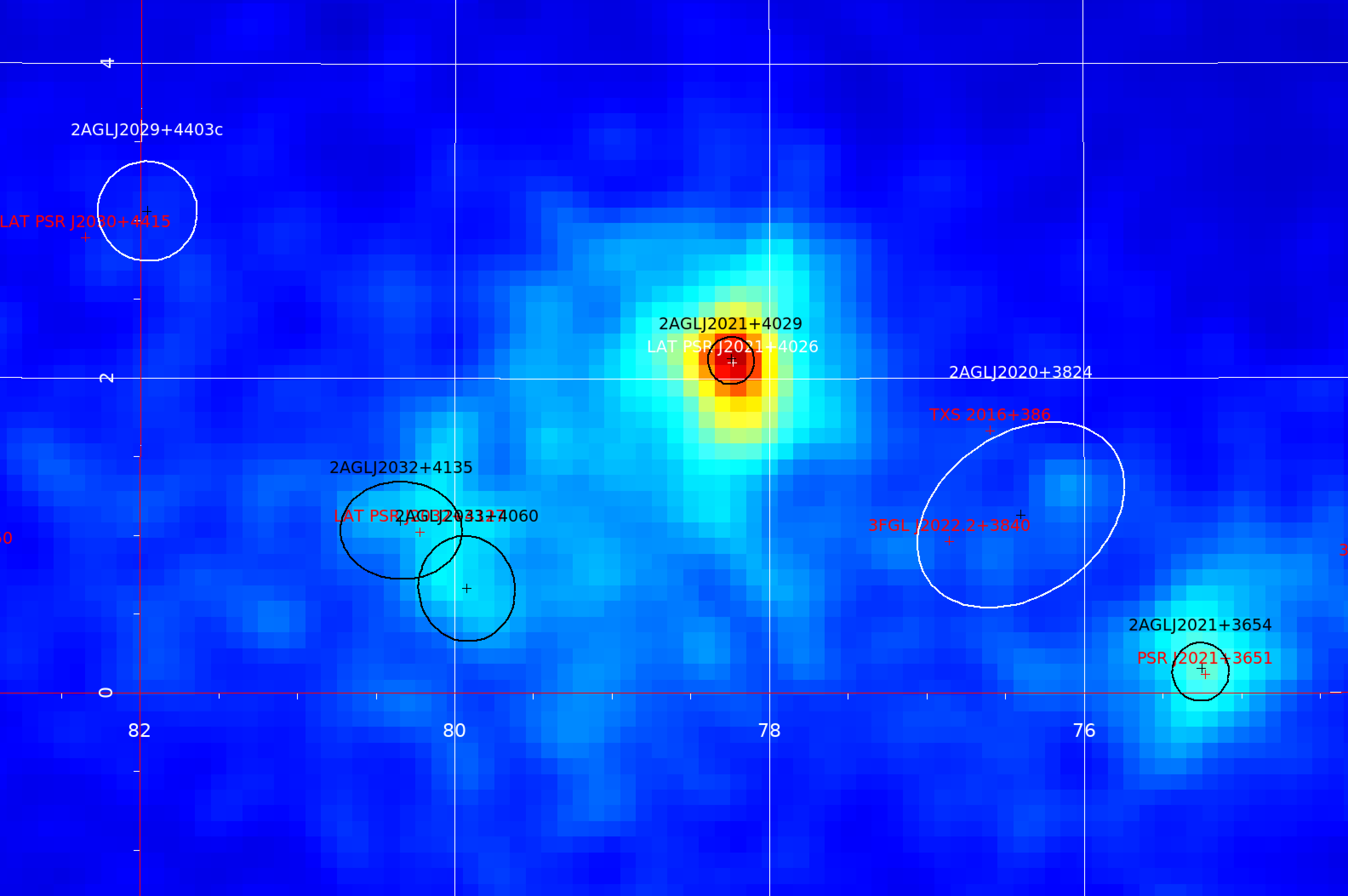}
    \caption{The count map of the Cygnus region in Galactic coordinates with superimposed the 2AGL Catalog $95\%$ c.l. elliptical confidence regions. We report also some associations and 3FGL  Catalog $95\%$ c.l. elliptical confidence regions inside the 2AGL regions.}\label{fig:cygnus}
\end{figure*}

The Cygnus region (Fig.~\ref{fig:cygnus}) is a site of bright diffuse emission, transient and persistent point-like and extended sources in \gray. The most prominent persistent \gray point-like sources are the three pulsars: PSR J2021+3651 (2AGL J2021+3654, see Sect.~\ref{sect:pulsar}), PSR J2021+4026 (2AGL J2021+4029, see Sect.~\ref{sect:snrs}), PSR J2032+4127 (2AGL J2032+4135). 
Furthermore, three microquasars were detected in this region with variable \gray emission: Cygnus X-1 \citep{sabatini_2010,sabatini_2013}, Cygnus X-3 (2AGL J2033+4060, \citealp{tavani_2009b,fermi_2009,bulgarelli12b,piano_2012}), V404 Cygni \citep{piano_2017}. \abr{Cygnus X-1 and V404 Cygni are not part of this Catalog.}
Many Astronomer's Telegrams are associated with sources of the Cygnus region, in particular with PSR J2021+4026 and Cygnus X-3 (see Table \ref{tab:agileatelspointing}).

Two main extended \gray sources are present in this region but not present in the 2AGL Catalog because not detected with extended model: the Cygnus cocoon \citep{ackermann_2011} and the SNR Gamma Cygni (G78.2+2.1, \citealp{fraija_2016}).

\subsection{Carina region}
\label{sect:carinareg}

\begin{figure*}
    \includegraphics[width=\linewidth]{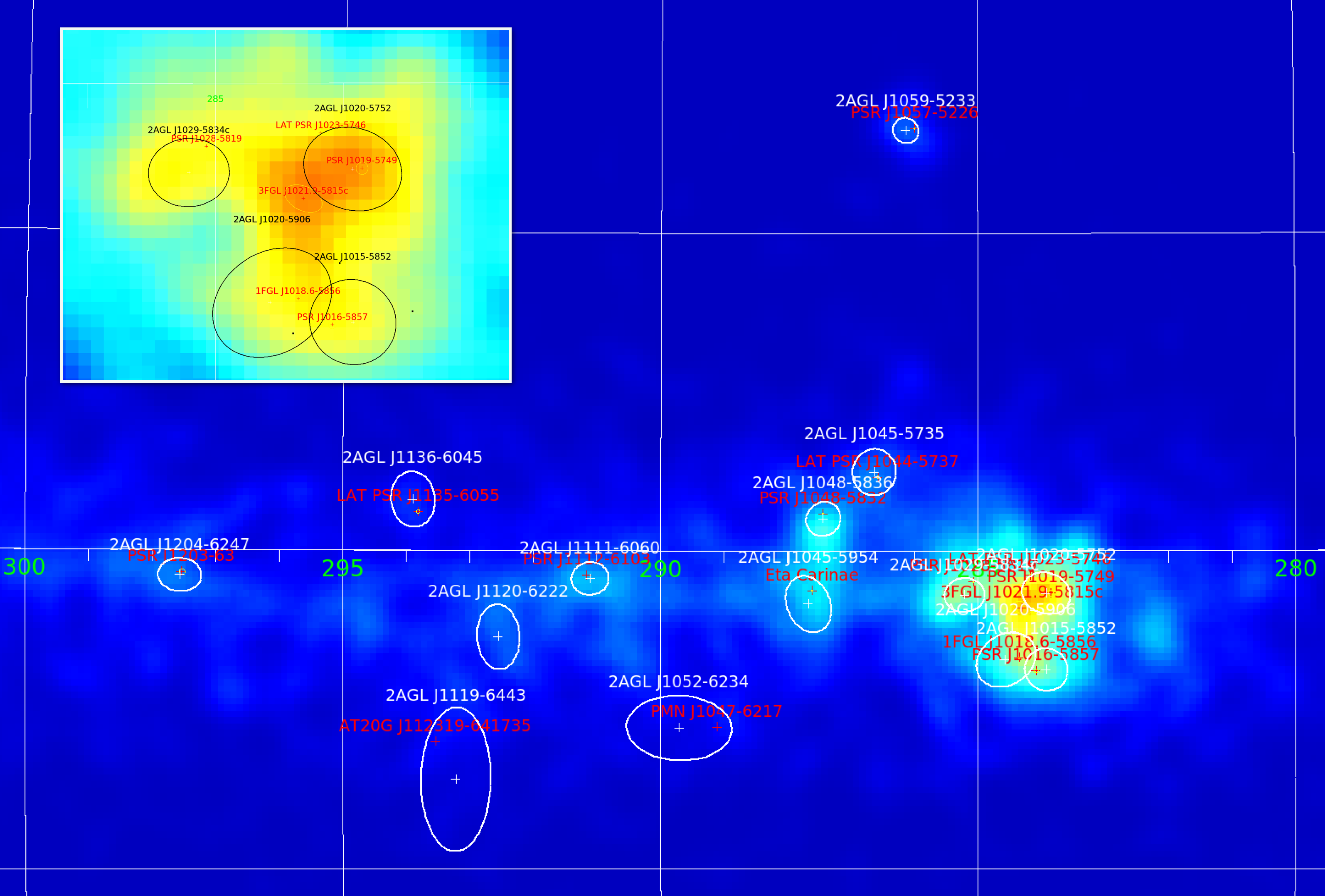}
    \caption{The count map of the Carina region in Galactic coordinates with superimposed the 2AGL Catalog $95\%$ c.l. elliptical confidence regions. We report also some associations and 3FGL Catalog $95\%$ c.l. elliptical confidence regions inside the 2AGL regions. On top and left there is a zoom of the region around $(l,b)=(285\degmark,0\degmark)$. }\label{fig:carina}
\end{figure*}

The Carina region is shown in Fig.~\ref{fig:carina}. We have detected some remarkable sources. We detect $\eta$-Carinae (2AGL J1045-5954) binary system, the HMXB 1FLG J1018.6-5856 \citep{ackermann12} (2AGL J1020-5906), 8 pulsars, i.e.  PSR J1016-5857 (2AGL J1015-5852, see Sect.~\ref{sect:pulsar}), PSR J1019-5749 (2AGL J1020-5752), PSR J1028-5819 (2AGL J1029-5834), LAT PSR J1044-5737 (2AGL J1045-5735), PSR J1048-5832 (2AGL J1048-5836), PSR J1112-6103 (2AGL J1111-6060), LAT PSR J1135-6055 (2AGL J1136-6045), PSR J1203-63 (2AGL J1204-6247), a marginal association with the PSR J1028-5819 (2AGL J1029-5834, see Sect.~\ref{sect:pulsar}),  and two AGNs ATG20G J112319-641735 (associate with 2AGL J1119-6443) and PMN J1047-6217 (identified with 2AGL J1052-6234, see Sect.~\ref{sect:agn}).

$\eta$-Carinae (2AGL J1045-5954) is a colliding wind binary in the same Galactic region of the Carina Nebula. The system was observed from radio to X-ray and, in 2009, the AGILE-GRID detected for the first time \gray emission correspondent to $\eta$-Carinae \citep{TavaniEtacarinae, sabatini11}. The source was then detected also by Fermi-LAT \citep{abdo09_eta, abdo10} with a flux consistent with the AGILE-GRID estimation. The absence of \gray emission variation correlated with the X-ray variability cannot provide an unambiguous identification of the detected source with $\eta$-Carinae. However, the AGILE-GRID revealed a flare activity that could be associated with particle acceleration and, in the last months, a Nustar observation has detected non-thermal variable X-ray emission with the same spectral index 1.65 as in the \gray band \citep{hamaguchi18}. This could be the first evidence that $\eta$-Carinae is emitting high energies photons (in both hard X-ray and \gray bands) likely due to particle acceleration.

\subsection{Pulsar wind nebulae (PWN)}
\label{sect:pwn}

\begin{figure}
    \includegraphics[width=\linewidth]{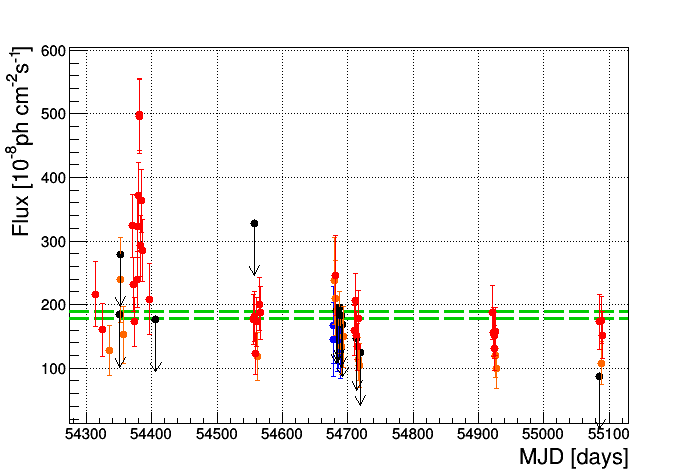}
    \caption{Light curve of the Crab pulsar+nebula in the 100 MeV -- 10 GeV energy range, with a 1-day resolution. The black downward arrows represent $2\sigma$ upper limits. The blue, orange and red circles refer to a $\sqrt{TS} \geq3$, $\geq 4$ and $\geq 5$ respectively. The green dashed lines are the average flux plus/minus the error from Table~\ref{tab:cat2}}\label{fig:lc_crab}
\end{figure}

\textbf{2AGL J0534+2205}. The Crab pulsar has been identified by pulsation in \citep{pellizzoni09}. Crab is the most prominent PWN of the 2AGL Catalog. The Anti-Centre region (including Crab and Geminga) was observed for $\sim45$ days, mostly in 2007 September and in 2008 April (during AO1) with the addition of other sparse short Crab pointings for SuperAGILE calibration purpose during the SVP. Fig.~\ref{fig:lc_crab} shows the light curve with 1-day resolution, where the first recorded \gray flare from the Crab nebula is evident \citep{tavani11, vittorini11, striani11, striani13}. The flux reported in the 2AGL Catalog includes the contribution from both pulsar and nebula, excluding the first \gray flare from Crab nebula, and no separate analysis is made for this Catalog because we rely on the already cited papers. 

\textbf{2AGL J0835-4514}. In the 2AGL Catalog the Vela region is reported as pulsar (see Sect.~\ref{sect:pulsar}). The detection of the Vela Pulsar Wind Nebula with the AGILE-GRID is reported in \citep{pellizzoni10}.

\textbf{2AGL J1634-4734e} HESS J1632-478 \citep{balbo2010}, associated with 2AGL J1634-4734e, is an energetic pulsar wind nebula with an age of the order of $10^4$ years in the Norma region. The nebula has a size of 1 pc and is modelled as an extended source in the 2AGL Catalog.

\subsection{SuperNova Remnants}
\label{sect:snrs}
The AGILE-GRID has detected many confirmed \gray SNRs, all middle-aged and very bright in the radio band.


\textbf{2AGL J0617+2239e}. This source is associated with the SNR IC 443. In angular size, it is the largest SNR detected in the Galaxy in the radio band. Its radio morphology is composed of two interacting synchrotron emitting shells. The AGILE-GRID detects \gray emission from IC 443 in correspondence of the more external radio shell \citep{tavani10}. In 2013, Fermi-LAT published a \gray spectrum down to energies below 200 MeV, confirming the presence of high energy CR at the SNR shock \citep{ackermann13}. TeV \gray emission was detected at the centre of the radio shell, in correspondence of a likely associated Molecular Cloud (MC) \citep{acciari09}. The spectral index of the remnant is steeper than 2.0 and the cut-off is at energies below 100 TeV.

\textbf{2AGL J1715-3815}. This extended source is associated with the TeV emitting SNR CTB 37A \citep[HESS J1714-385,]{aharonian08_CT} located at a distance of $\sim$ 13 kpc. The TeV emission is correlated with detected MCs \citep{maxted13} and could be associated with the non-thermal X-ray source CXOU J171419.8-383023. The detection of this SNR with Fermi-LAT in 2013 \citep{brandt13} in the GeV band is correlated with TeV and X-ray emissions, supporting the likely hypothesis of the presence of energised particles in the SNR shock with surrounding MCs.

\textbf{2AGL J1801-2334}. This source is associated with the SNR W28, another mixed morphology SNR with a shell radio structure described by a set of large, coherent arcs running along the edge of the remnant. The AGILE-GRID detected \gray emission from two different regions of this remnant, correlated with two MCs at different distances and with an anti-correlation in brightness with the TeV emission \citep{giuliani10, aharonian08}. In 2014, Fermi-LAT confirmed the AGILE-GRID detection \citep{hanabata14}. The \gray emission is interpreted as accelerated CR interacting with two distant MCs. The parameters in this Catalog are found assuming a log-parabola shape for the spectrum (see Table~\ref{tab:curvedspectra}), with a very steep spectral index. \textcolor{black} {We are unable to detect W28 with an extended model, only the point-like model provides a detection.}

\textbf{2AGL J1856+0119e}. This source is associated with the very radio-bright middle-aged SNR W44. The AGILE-GRID detects a \gray spectrum down to energies below 200 MeV, confirming for the first time the presence of energised CRs in correspondence of a SNR/MC interaction shock \citep{giuliani11, cardillo14}. This spectrum was confirmed by Fermi-LAT in 2013 \citep{ackermann13}. No TeV emission has been detected from this source. The parameters in this Catalog are found assuming a log-parabola shape for the spectrum with a steep spectral index (see Sect.~\ref{tab:curvedspectra}).

\textbf{2AGL J1911+0907}. This source is in correspondence of the TeV J1911+0905/ SNR W49B. The VHE \gray emission is found to be point-like, as well as in the AGILE-GRID data. The SNR originated from a core-collapse supernova that occurred between one and four thousand years ago, and has evolved into a mixed morphology remnant that is interacting with MCs \citep{abdalla18}.

\textbf{2AGL J1924+1416}. This source is associated with the SNR W51c, a component of the MC complex W51, together with the two star forming regions W51A and W51B. W51C shows a thick incomplete shell-like structure in the radio band, explained by electron synchrotron emission. Fermi-LAT detected \gray emission from this remnant, for the first time, in 2009 \citep{abdo09} and then, thanks to the improvement of the analysis software, the satellite data were re-analysed showing a spectrum down to energies below 200 MeV also for this source \citep{jogler16}. The MAGIC telescope detected also TeV emission from W51c \citep{aleksic12} but, as like as in all the other \gray detected SNRs, the spectrum has a spectral index steeper than the theoretical value of 2, with a cut-off energy below 100 TeV.

\textbf{2AGL J2021+4029}. The radio source G78.2+2.1 (Gamma Cygni) is a typical shell-type SNR located within the extended emission of the Cygnus X region \textcolor{black} {and it has the pulsar PSR J2021+4026 at its centre. The 2AGL J2021+4029 includes both the SNR Gamma Cygni and the PSR J2021+4026. Fig. \ref{fig:spectra1} reports the spectra in the 100 MeV -- 10 GeV energy range}. PSR J2021+4026 is found to be variable in \gray \citep{chen_2011,allafort_2013}; many ATELs are associated with this source (see Table \ref{tab:agileatelspointing}). Work is ongoing on the analysis of non-pulsed \gray emission, in order to separate the pulsar contribution from that due to the SNR \citep{piano18}. 

\textbf{2AGL J2044+5012e}. This source is associated with the mixed morphology SNR HB21, located in a dense environment with a radio shape suggesting interaction with a MC. Its \gray emission was detected by Fermi-LAT in 2013 \citep{pivato13} but there are no other detection in the \gray band. However, the behavior of the remnant seems to be the same of all the other middle-aged SNRs detected at high energies. \abr{We detect  HB21 as an extended source (see Table \ref{tab:extended})}.

\subsection{Binaries}
\label{sect:bin}

The 2AGL Catalog contains the detection of four High Mass X-ray Binaries (HMXB): LSI +61 303 (2AGL J0239+6120, \abr{that show variability}), Cygnus X-3 (2AGL J2033+4060, see Sect.~\ref{sect:cyg}), LS 5039 (2AGL J1826-1438), and 1FGL J1018.6-5856 (2AGL J1020-5906, see Sect.~\ref{sect:carinareg}). Two other TeV emitters,  HESS J0632+057 and  PSR B1259-63, do not have {\cpr persistent} counterparts in 2AGL Catalog. Cygnus X-1, that has shown a flaring activity during the `pointing mode' (see Sect.~\ref{sect:cyg}), have no 2AGL counterpart.
\\
Eta Carinae, identified as a binary system, is included as 2AGL J1045-5954 (see Sect.~\ref{sect:carinareg}).

\subsection{Confused sources}
\label{sect:confused}

In this section we comment on some confused sources (see Sect.~\ref{sect:catdes2}). \abr{The 2AGL J1203-2701c, 2AGL J1640-5050c, 2AGL J1743-2613c,  2AGL J1820-1150c, and 2AGL J2332+8215c are discussed in Sect.~\ref{sect:agileonly}. The 2AGL 0032+0512c and 2AGL J0221+6208c are discussed in Sect.~\ref{sect:gammaonly}.}

\textbf{2AGL J1407-6136c}. The $95\%$ elliptical confidence region of this 2AGL source partially overlaps the $95\%$ elliptical confidence region of 3FGL J1405.4-6119 and 3FGL J1409.7-6132 (PSR J1410-6132). In the 1AGL Catalog a source was centred at a distance of $1.21\degmark$ from the current position of this 2AGL source. An ATEL is reported for this source (see Table \ref{tab:agileatelspointing}) during the `pointing mode'.

\textbf{2AGL J1838-0623c}. This is a Galactic plane source with a large $95\%$ c.l. elliptical confidence region, which leads to multiple spatial coincidences. A refined analysis is not able to reduce the c.l. confidence region.

\section{2AGL \gray Sources List detected below 100 MeV}
\label{sec:list100}

Table \ref{tab:spactra_det_50_100} reports an extension of the 2AGL Catalog that lists the sources detected also in the 50 -- 100 MeV energy range. The spectral model is the same as in the 2AGL Catalog and the flux is calculated fixing all the spectral model parameters. Galactic and diffuse emission model parameters are determined independently for each energy band. In addition, the low confidence extension  in the 30 -- 50 MeV energy range is also reported in the same table with the same assumptions. 

\section{Conclusions}
\label{sec:concl}

{\mt 
This paper presents an exhaustive report of the results obtained by the AGILE-GRID during the `pointing mode' period of operations starting in mid-2007 and ending in October 2009 (2.3 years). The AGILE-GRID turns out to be very effective in studying \gray sources above 100 MeV. The AGILE mission has been the first one to operate in 2007 after the end of operations in 1996 of the EGRET \gray instrument on board of the Compton Gamma-Ray Observatory. The AGILE-GRID was the first \gray detector capable of obtaining an unprecedentedly large FoV of 2.5 sr with much improved angular resolution compared with the previous generation of instruments.
The pointing capability and observation strategy during the pointing phase of operations provided an excellent exposure of \gray sources for each individual pointing period.

The detector optimisation in the 100 MeV -- a few hundreds of MeV energy range  (with significant capability in the 50-100 MeV energy range) is among the most relevant assets of the AGILE-GRID. Despite its relatively small size, the AGILE-GRID turns out to be quite effective in detecting both short-lived and persistent \gray sources. 
%
{\cpbis 
This paper presents the 2AGL Catalog of \emph{persistent} \gray sources
detected by the AGILE-GRID during the pointing phase which
includes 175 high-confidence sources, both of Galactic and extragalactic origin.
Many source classes are represented:
pulsars, PWNe, SNRs, compact objects, and blazars. In
addition to confirming previously known relatively strong \gray 
sources, the 2AGL Catalog includes a substantial number of new
sources compared to EGRET, as well as 29 sources that are detected
only by the AGILE-GRID.

A comparison with the Fermi-LAT instrument (operating
since 2008 and overlapping in time during the second half of the
pointing period considered here) has been performed. 
Different observing strategy, effective area, sky region exposures and 
viewing angle during gamma-ray transient episodes may account for 
different results (see Appendix A for a direct comparison
between the AGILE-GRID and Fermi-LAT visibility of
some of the 2AGL sources detected only by AGILE-GRID).
}

Starting in 2010, AGILE have continued operations in orbit in "spinning mode" therefore changing several important details concerning the exposure of \gray observations. An account of the \gray sources detected by the AGILE-GRID in spinning mode will be presented elsewhere.
}

\begin{acknowledgements}
The AGILE Mission is funded by the Italian Space Agency (ASI) with
scientific and programmatic participation by the Italian Institute
of Astrophysics (INAF) and the Italian Institute of Nuclear
Physics (INFN). {\mt Investigation supported by the ASI grant I/089/06/2.
We thank the ASI management for unfailing support during AGILE operations.
We acknowledge the effort of ASI and industry personnel in operating 
the  ASI ground station in Malindi (Kenya), and the data processing done 
at the ASI/SSDC in Rome: the success of AGILE scientific operations depend on the effectiveness of the data flow from Kenya to SSDC as well as on the data analysis and software management.}
{\cpbis We also acknowledge the use of innovative tools 
that are under development within the United Nations
"Open Universe" initiative.
}
\end{acknowledgements}

\bibliographystyle{aa}  

\clearpage

\onecolumn

\begin{landscape}
\fontsize{7}{7}\selectfont
{\setlength{\tabcolsep}{2pt}

\begin{longtable}{lcccccccccccccccccccccc}
  
	\caption{2AGL Catalog sources. See Table \ref{tab:cat2columns} for a description of the columns}\\
  \midrule[0.6pt]
  \toprule[1pt]
 
 & & & &  & & & & & & & & \\
\multicolumn{1}{l}{Name 2AGL} & 
\multicolumn{1}{c}{R.A.} &
\multicolumn{1}{c}{Dec.} &
\multicolumn{1}{c}{\textit{l}} &
\multicolumn{1}{c}{\textit{b}} &
\multicolumn{1}{c}{$\theta_a$} &
\multicolumn{1}{c}{$\theta_b$} &
\multicolumn{1}{c}{$\phi$} &
\multicolumn{1}{c}{SM} &
\multicolumn{1}{c}{$\sqrt{TS}$} & 
\multicolumn{1}{c}{$\alpha$} &
\multicolumn{1}{c}{$\Delta \alpha$} & 
\multicolumn{1}{c}{$F_\gamma$} &
\multicolumn{1}{c}{$\delta F_\gamma$} & 
\multicolumn{1}{c}{$F_e$} &  
\multicolumn{1}{c}{$\delta F_e$} &
\multicolumn{1}{c}{$VI$}& 
\multicolumn{1}{c}{$Exp$}& 
\multicolumn{1}{c}{$Flag$}& 
\multicolumn{1}{c}{AGL assoc.} &
\multicolumn{1}{c}{\gray assoc.} &
\multicolumn{1}{c}{ID or assoc.}  &
\multicolumn{1}{c}{$Class$}      \\[3 pt]
%
%
\toprule[0.6pt]

 & & & &  & & & & & & & & & & & & & & & & & & \\[3 pt]
 \endhead
\toprule[1pt]

\label{tab:cat2} 
\endfoot

J0007+7308 & 1.77 & 73.13 & 119.68 & 10.54 & 0.07 & 0.06 & 2.98 & PC & 26.2 & 1.29 & 0.22 & 41.6 & 4.2 & 67.2 & 6.9 &  0 & 6954 [268] &  & \makecell*{1AGL J0006+7311\\1AGLR J0007+7307\\AGL TeV J0006+727} &   \makecell*{TeV J0006+7259\\3EG J0010+7309\\3FGL J0007.0+7302\\3FHL J0007.0+7303 }   & \makecell*{LAT PSR J0007+7303}  & psr  	\\[7 pt] 
J0032+0512c & 8.09 & 5.21 & 114.11 & -57.34 & 1.20 & 0.72 & 0.77 & PL & 4.0 & 2.18 & 0.27 & 6.6 & 1.9 & 10.7 & 3.0 &  0 & 1561 [60] & 3,5 & \makecell*{} &   \makecell*{3FGL J0030.4+0451}   & \makecell*{}  & gunid  	\\[7 pt] 
J0135+4754 & 23.72 & 47.90 & 130.43 & -14.34 & 0.17 & 0.14 & 14.28 & PL & 10.7 & 1.98 & 0.11 & 12.1 & 2.1 & 19.5 & 3.4 &  0 & 4278 [165] &  & \makecell*{1AGLR J0135+4759} &   \makecell*{0FGL J0137.1+4751 \\3FHL J0136.9+4751 }   & \makecell*{OC 457}  & FSRQ  	\\[7 pt] 
J0217+0127 & 34.29 & 1.46 & 162.19 & -54.74 & 0.68 & 0.39 & 6.79 & PL & 4.7 & 1.85 & 0.20 & 17.0 & 6.7 & 27.4 & 10.8 &   & 680 [26] & 3 & \makecell*{} &   \makecell*{3FGL J0217.8+0143\\3FHL J0217.8+0143 }   & \makecell*{PKS 0215+015}  & fsrq  	\\[7 pt] 
J0221+4250 & 35.17 & 42.84 & 139.86 & -17.08 & 0.31 & 0.72 & -39.70 & PL & 9.8 & 2.20 & 0.13 & 10.5 & 1.9 & 17.0 & 3.1 &  0 & 3499 [135] &  & \makecell*{1AGLR J0222+4305\\AGL TeV J0222+430} &   \makecell*{TeV J0222+4302\\TeV J0223+4259\\3EG J0222+4253\\3FGL J0218.1+4233\\3FGL J0222.6+4301\\3FHL J0218.3+4230 \\3FHL J0222.6+4302 }   & \makecell*{3C 66A}  & BLL  	\\[7 pt] 
J0221+6208c & 35.20 & 62.13 & 133.17 & 1.05 & 0.36 & 0.57 & -36.16 & PL & 7.6 & 2.33 & 0.14 & 9.4 & 1.8 & 15.2 & 3.0 &  0 & 4343 [168] &  & \makecell*{} &   \makecell*{3FGL J0217.3+6209\\3FGL J0220.1+6202c\\3FGL J0221.8+6138c\\3FGL J0223.6+6204\\3FGL J0224.0+6235\\3FGL J0225.8+6159}   & \makecell*{}  & gunid  	\\
\end{longtable}

}
\end{landscape}
\normalsize

\twocolumn

\onecolumn

\fontsize{7}{7}\selectfont
{\setlength{\tabcolsep}{2pt}

\begin{longtable}{lcccccccccccccccc}
  
	\caption{Spectral information of 2AGL Catalog sources. See Table \ref{tab:cat2spectracolumns} for a description of the columns.}\\
  \midrule[0.6pt]
  \toprule[1pt]
 
 & & & &  & & & & & & & & & & & \\
\multicolumn{1}{l}{} & 
\multicolumn{1}{c}{} & 
\multicolumn{1}{c}{} &
\multicolumn{1}{c}{} & 
\multicolumn{1}{c}{} & 
\multicolumn{2}{c}{0.1 -- 0.3 GeV} & 
\multicolumn{2}{c}{0.3 -- 1 GeV} & 
\multicolumn{2}{c}{1 -- 3 GeV} & 
\multicolumn{2}{c}{3 -- 10 GeV} & \\[3 pt]
 \cmidrule{6-16}
& & & &  & & & & & & & & & & & \\
\multicolumn{1}{l}{Name 2AGL} & 
\multicolumn{1}{c}{$\sqrt{TS}$} & 
\multicolumn{1}{c}{SM} &
\multicolumn{1}{c}{$\alpha$} &
\multicolumn{1}{c}{$\Delta \alpha$} & 
\multicolumn{1}{c}{$F_1$} & 
\multicolumn{1}{c}{$\sqrt{TS}_1$} & 
\multicolumn{1}{c}{$F_2$} & 
\multicolumn{1}{c}{$\sqrt{TS}_2$} & 
\multicolumn{1}{c}{$F_3$} & 
\multicolumn{1}{c}{$\sqrt{TS}_3$} & 
\multicolumn{1}{c}{$F_4$} & 
\multicolumn{1}{c}{$\sqrt{TS}_4$} &
\multicolumn{1}{c}{ID or ass.} &
\multicolumn{1}{c}{Flag} &
\multicolumn{1}{c}{Class.}\\[3 pt]
%
%
\toprule[0.6pt]

& & & &  & & & & & & & & & & \\
 \endhead
\toprule[1pt]

\label{tab:cat2spectra} 
\endfoot

J0007+7308 & 26.2 & PC & 1.29 & 0.22 & $16.0_{-2.2}^{+2.2}$ & 8.3 & $15.5_{-1.0}^{+1.0}$ & 22.0 &  $6.4_{-0.9}^{+1.0}$  & 11.1 &  $2.4_{-0.8}^{+1.0}$  & 4.1 & \makecell*{LAT PSR J0007+7303} & & psr \\[7 pt] 
J0032+0512c & 4.0 & PL & 2.18 & 0.27 & $<8.1_{}^{}$ & 1.7 & $1.7_{-0.5}^{+0.6}$ & 3.6 &  $<0.9_{}^{}$  & 1.3 &  $<0.8_{}^{}$  & 0.0 & \makecell*{} & 3,5 & gunid \\[7 pt] 
J0135+4754 & 10.7 & PL & 1.98 & 0.11 & $6.1_{-1.5}^{+1.6}$ & 4.3 & $3.4_{-0.5}^{+0.5}$ & 9.0 &  $0.7_{-0.3}^{+0.3}$  & 3.2 &  $<2.0_{}^{}$  & 2.2 & \makecell*{OC 457} &  & FSRQ \\[7 pt] 
J0217+0127 & 4.7 & PL & 1.85 & 0.20 & $<17.2_{}^{}$ & 1.6 & $<8.9_{}^{}$ & 2.8 &  $2.5_{-1.0}^{+1.3}$  & 3.2 &  $<3.1_{}^{}$  & 1.5 & \makecell*{PKS 0215+015} & 3 & fsrq \\[7 pt] 
J0221+4250 & 9.8 & PL & 2.20 & 0.13 & $7.6_{-1.4}^{+1.5}$ & 5.8 & $2.0_{-0.4}^{+0.4}$ & 6.8 &  $0.7_{-0.2}^{+0.2}$ & 4.6 &  $<0.3_{}^{}$  & 0.0 & \makecell*{3C 66A} &  & BLL \\[7 pt] 
J0221+6208c & 7.6 & PL & 2.30 & 0.14 & $6.7_{-1.5}^{+1.5}$ & 4.8 & $1.6_{-0.4}^{+0.4}$ & 5.1 & $0.5_{-0.2}^{+0.2}$  & 3.8 & $<0.2_{}^{}$ & 0.0 & \makecell*{} & & gunid \\[7 pt] 
J0237+1653 & 6.9 & PL & 2.26 & 0.16 & $10.7_{-2.3}^{+2.4}$ & 5.1 & $2.0_{-0.6}^{+0.6}$ & 3.8 & $<0.7_{}^{}$  & 0.2 &  $<1.2_{}^{}$ & 1.0 & \makecell*{AO 0235+164} & 2 & bll \\[7 pt] 
J0239+6120 & 22.9 & PL & 2.02 & 0.05 & $24.4_{-2.3}^{+2.4}$ & 11.8 & $9.9_{-0.8}^{+0.8}$ & 16.8 & $3.6_{-0.5}^{+0.5}$  & 10.5 & $<2.4_{}^{}$  & 1.2 & \makecell*{LS I +61 303} &  & hmxb \\
\end{longtable}
}

\normalsize

\twocolumn

\onecolumn

\fontsize{7}{7}\selectfont
{\setlength{\tabcolsep}{2pt}

\begin{longtable}{lccccccccc}
 	
	\caption{AGILE Astronomer's Telegrams in pointing mode and associated 2AGL Catalog sources. \abb{The first column reports possible associations with other counterparts, or only the AGL name if no association was possible, the second column reports a classification of the source, the third column the number of Astronomer’s Telegram on this source, the fourth and fifth column the period of integration, the sixth column shows the ID of the Telegram, followed by the $95\%$ circular confidence region parameters in  Galactic coordinates $(l,b)$,  and radius $r$ in degree. The last column reports an association with a 2AGL source or published paper if the source is not present in the 2AGLCatalog. If the circular confidence region parameters are not present, a direct association has done with the source in the text of the Telegram. }}\\
  	\midrule[0.6pt]
  	\toprule[1pt]
 & & & & & & & & &   \\

\multicolumn{1}{l}{Positionally Consistent} & 
\multicolumn{1}{c}{Object} & 
\multicolumn{1}{c}{ATels} & 
\multicolumn{1}{c}{Start} & 
\multicolumn{1}{c}{Stop} & 
\multicolumn{1}{c}{ATel ID} & 
\multicolumn{1}{c}{l} &
\multicolumn{1}{c}{b} &
\multicolumn{1}{c}{r} &
\multicolumn{1}{c}{2AGL Assoc. or} \\
\multicolumn{1}{c}{(association)} &
\multicolumn{1}{c}{Class} & 
\multicolumn{1}{c}{} & 
\multicolumn{1}{c}{(UTC)} & 
\multicolumn{1}{c}{(UTC)} &  
\multicolumn{1}{c}{} & 
\multicolumn{1}{c}{} &
\multicolumn{1}{c}{} &
\multicolumn{1}{c}{} &
\multicolumn{1}{c}{published papers} \\[3pt]
\toprule[0.6pt]

 & & & & & & & & &  \\

\endhead
\toprule[1pt]
\endfoot

\multirow{8}{*}{3C 454.3} &  \multirow{8}{*}{Blazar}& \multirow{8}{*}{8}  & 2007 Jul 24 14:30& 2007 Jul 27 05:27& 1160 & &  & &  \\[7 pt] 
 &  & &  2007 Jul 24 14:30& 2007 Jul 30 11:40& 1167 & & &  &\\[7 pt] 
 &  & &  2007 Nov 02 13:50& 2007 Nov 12 17:01&  1278& & & &\\[7 pt] 
 &  & &  2007 Nov 02 12:00& 2007 Nov 22 18:33 &  1300& & &  &\\[7 pt] 
 &  & &  2008 May 24 03:10 & 2008 May 27 04:17 & 1545& & &  &  J2254+1609\\[7 pt] 
 &  & & 2008 Jun 15 10:46& 2008 Jun 16 07:11 & 1581& & & & \\[7 pt] 
 &  & &  2008 Jun 25 03:00 & 2008 Jun 26 03:00 &  1592& & & & \\[7 pt] 
 &  & &  2008 Jul 25 18:00 & 2008 Jul 28 03:00 & 1634& & & &  \\
\end{longtable}
}

\normalsize

\twocolumn

\onecolumn

\fontsize{7}{7}\selectfont
{\setlength{\tabcolsep}{2pt}

\begin{longtable}{lcccccccccccc}
 
 \caption{AGILE only sources. A * in the 'ID or Assoc.' column indicates a possible marginal positional association. See Table~\ref{tab:cat2columns} for a description of the columns.}\\
  \midrule[0.6pt]
  \toprule[1pt]
 
 & & & & & & & & &  &  \\
\multicolumn{1}{l}{Name 2AGL} & 
\multicolumn{1}{c}{R.A.} &
\multicolumn{1}{c}{Dec.} &
\multicolumn{1}{c}{\textit{l}} &
\multicolumn{1}{c}{\textit{b}} &
\multicolumn{1}{c}{$\theta_a$} &
\multicolumn{1}{c}{$\theta_b$} &
\multicolumn{1}{c}{$\phi$} &
\multicolumn{1}{c}{$\sqrt{TS}$} & 
\multicolumn{1}{c}{$Flag$} & 
\multicolumn{1}{c}{ID or Assoc.} &
\multicolumn{1}{c}{Class} \\[3 pt]
\toprule[0.6pt]

  & & & & & & &  & & & \\

 \endhead
\toprule[1pt]

\label{tab:agileonly}
\endfoot

J0714+3318 & 108.42 & 33.30 & 184.44 &  18.81 & 0.45 & 0.17 & -54.48 & 7.3 &  &  \makecell*{}	& unid \\[7 pt] 
J1120-6222 & 169.98 &-62.37 & 292.54 &  -1.36 & 0.41 & 0.23 & -3.77 & 4.9 & 2 &  \makecell*{}	& unid \\[7 pt] 
J1138-1724 & 174.48 &-17.39 & 279.03 &  42.04 & 0.32 & 0.23 & 24.24 & 5.7 & 2 &  \makecell*{2FHL J1137.9-1710\\5BZB J1137-1710}	& BLL \\[7 pt] 
J1203-2701c& 180.85 &-27.02 & 289.91 &  34.63 & 0.59 & 0.31 & -25.51 & 4.6 & 2,5 &  \makecell*{}	& unid \\[7 pt] 
J1312-3403 & 198.09 &-34.05 & 307.87 &  28.62 & 0.31 & 0.27 & 15.07 & 6.6 & 2 &  \makecell*{}	& unid \\[7 pt] 
J1402-8142 & 210.49 &-81.70 & 305.58 & -19.20 & 0.41 & 0.36 & 25.06 & 4.4 & 4 &  \makecell*{}	& unid \\[7 pt] 
J1628-4448 & 247.01 &-44.80 & 338.09 &   2.78 & 0.43 & 0.35 & -43.83 & 7.7 & 2 &  \makecell*{}	& unid \\[7 pt] 
J1631-2039 & 247.64 &-20.64 & 356.71 &  18.64 & 0.45 & 0.24 & 41.00 & 5.9 &  &  \makecell*{}	& unid \\
\end{longtable}
}

\normalsize

\twocolumn
\onecolumn

\fontsize{7}{7}\selectfont
{\setlength{\tabcolsep}{2pt}

\begin{longtable}{lcccccccccccc}
  
\caption{Unidentified \gray sources. See Table \ref{tab:cat2columns} for a description of the columns.}\\
  \midrule[0.6pt]
  \toprule[1pt]
 
 & & & & & & & & & &    \\
\multicolumn{1}{l}{Name 2AGL} & 
\multicolumn{1}{c}{R.A.} &
\multicolumn{1}{c}{Dec.} &
\multicolumn{1}{c}{\textit{l}} &
\multicolumn{1}{c}{\textit{b}} &
\multicolumn{1}{c}{$\theta_a$} &
\multicolumn{1}{c}{$\theta_b$} &
\multicolumn{1}{c}{$\phi$} &
\multicolumn{1}{c}{$\sqrt{TS}$} & 
\multicolumn{1}{c}{$Flag$} & 
\multicolumn{1}{c}{ID or Assoc.} &
\multicolumn{1}{c}{Class} \\[3 pt]
\toprule[0.6pt]

 & & & & & & & & & &  \\
 \endhead
\toprule[1pt]

\label{tab:gunid}
\endfoot

J0032+0512c &   8.09 &  5.21 & 114.11 &-57.34 & 1.20 & 0.72 & 0.77 & 4.0 & 3,5 & \makecell*{3FGL J0030.4+0451}	& gunid \\[7 pt] 
J0221+6208c &  35.20 & 62.13 & 133.17 &  1.05 & 0.57 & 0.36 & -36.16 & 7.6 &  & \makecell*{}	& gunid \\[7 pt] 
J1103-7747  & 165.71 &-77.78 & 297.16 &-16.14 & 0.98 & 0.54 & -27.99 & 4.4 & 3 & \makecell*{3FGL J1104.3-7736c}	& gunid \\[7 pt] 
J1407-6136c & 211.69 &-61.59 & 311.77 & -0.02 & 1.10 & 0.62 & 18.72 & 4.8 & 2 & \makecell*{3FGL J1405.4-6119\\PSR J1410-6132\\LAT PSR J1413-6205}	& gunid \\[7 pt] 
J1624-2406  & 246.12 &-24.10 & 353.00 & 17.45 & 0.43 & 0.24 & 3.50 & 6.0 & 3 & \makecell*{0FGL J1625.9-2423\\1FGL J1623.5-2345c}	& gunid \\[7 pt] 
J1625-2021  & 246.32 &-20.35 & 356.11 & 19.77 & 0.43 & 0.34 & -4.23 & 5.2 & 4 & \makecell*{1FGL J1626.2-2038}	& gunid \\[7 pt] 
J1631-2425  & 247.74 &-24.41 & 353.77 & 16.13 & 0.51 & 0.38 & -27.45 & 5.9 & 3 & \makecell*{3FGL J1628.2-2431c}	& gunid \\[7 pt] 
J1654-4704 & 253.54 &-47.06 & 339.40 & -2.12 & 0.33 & 0.23 & -20.50 & 5.3 & 2 & \makecell*{3FGL J1652.2-4649\\3FGL J1655.1-4644\\3FGL J1655.7-4712}	& gunid \\
\end{longtable}
}

\normalsize

\twocolumn
\onecolumn

\fontsize{7}{7}\selectfont
{\setlength{\tabcolsep}{2pt}

\begin{longtable}{lccccccccccc}
  
  \caption{2AGL Sources with detection in the range 30-100 MeV. See Table \ref{tab:cat2spectracolumns} for a description of the columns.}\\

  \midrule[0.6pt]
  \toprule[1pt]
 
 & & & &  & & & & & & & \\
\multicolumn{1}{l}{} & 
\multicolumn{1}{c}{} & 
\multicolumn{1}{c}{} &
\multicolumn{1}{c}{} & 
\multicolumn{1}{c}{} & 
\multicolumn{2}{c}{0.03 - 0.05 GeV} & 
\multicolumn{2}{c}{0.05 - 0.1 GeV} \\[3 pt]
 \cmidrule{6-11}
  & & & &  & & & & & & & \\
\multicolumn{1}{l}{Name 2AGL} & 
\multicolumn{1}{c}{$\sqrt{TS}$} & 
\multicolumn{1}{c}{SM} &
\multicolumn{1}{c}{$\alpha$} &
\multicolumn{1}{c}{$\Delta \alpha$} & 
\multicolumn{1}{c}{$F_a$} & 
\multicolumn{1}{c}{$\sqrt{TS}_a$} & 
\multicolumn{1}{c}{$F_b$} & 
\multicolumn{1}{c}{$\sqrt{TS}_b$} & 
\multicolumn{1}{c}{Flag} &
\multicolumn{1}{c}{ID or assoc.} &\\[3 pt]
%
%
\toprule[0.6pt]

 & & & &  & & & & & &  \\[3 pt]
 \endhead
\toprule[1pt]
\label{tab:spactra_det_50_100}
\endfoot

J0007+7308 & 26.2 & PC & 1.29 & 0.22 & $58.06_{-19.85}^{+20.18}$ & 3.0 & $23.20_{-6.40}^{+6.54}$ & 3.7 &  & \makecell*{LAT PSR J0007+7303} \\[7 pt] 
J0534+2205 & 56.9 & PL & 2.32 & 0.03 & $383.61_{-27.76}^{+28.16}$ & 15.4 & $218.65_{-10.29}^{+10.44}$ & 25.9 &  & \makecell*{Crab} \\[7 pt] 
J0634+1749 & 76.4 & PS & 1.71 & 0.09 & $<177.42_{}^{}$ & 2.2 & $78.09_{-13.79}^{+14.16}$ & 6.1 &  & \makecell*{Geminga} \\[7 pt] 
J0835-4514 & 168.9 & PS & 1.71 & 0.04 & $399.01_{-35.95}^{+36.49}$ & 12.1 & $438.50_{-15.51}^{+15.69}$ & 35.7 &  & \makecell*{Vela  PSR} \\[7 pt] 
J1324-4348 & 5.4 & PL & 2.69 & 0.25 & $<44.25_{}^{}$ & 0.8 & $11.83_{-3.89}^{+3.95}$ & 3.1 & 2 & \makecell*{Cen A} \\[7 pt] 
J1458-6055 & 7.5 & PL & 2.31 & 0.18 & $<60.08_{}^{}$ & 1.3 & $20.97_{-5.81}^{+5.88}$ & 3.7 & 2 & \makecell*{LAT PSR J1459-6053\\3FGL J1456.7-6046} \\[7 pt] 
J1513-0905 & 25.0 & PL & 2.19 & 0.05 & $114.86_{-37.91}^{+38.73}$ & 3.1 & $61.72_{-10.76}^{+11.01}$ & 6.1 &  & \makecell*{PKS 1510-089} \\[7 pt] 
J1517-5909 & 5.4 & PL & 2.39 & 0.22 & $50.79_{-17.05}^{+17.22}$ & 3.0 & $17.98_{-5.45}^{+5.51}$ & 3.3 & 2,5 & \makecell*{PSR J1513-5908} \\
\end{longtable}
}

\normalsize

\twocolumn

\clearpage

\begin{appendix}
\label{appendix1}
\section{AGILE-only sources and the non-detection by Fermi-LAT}\label{AgilevsFermi}

\abr{In this Appendix, we show that the Fermi-LAT non-detection of some of 
the short flaring episodes observed by the AGILE-GRID in correspondence of the 
so-called {\it AGILE-only sources} (see Sect.~\ref{sect:agileonly}) might be due to poor exposure and non-optimal viewing angle of the source within the FoV of the instrument.}

\begin{figure*}[ht]
  \centering
  \includegraphics[width=9cm]{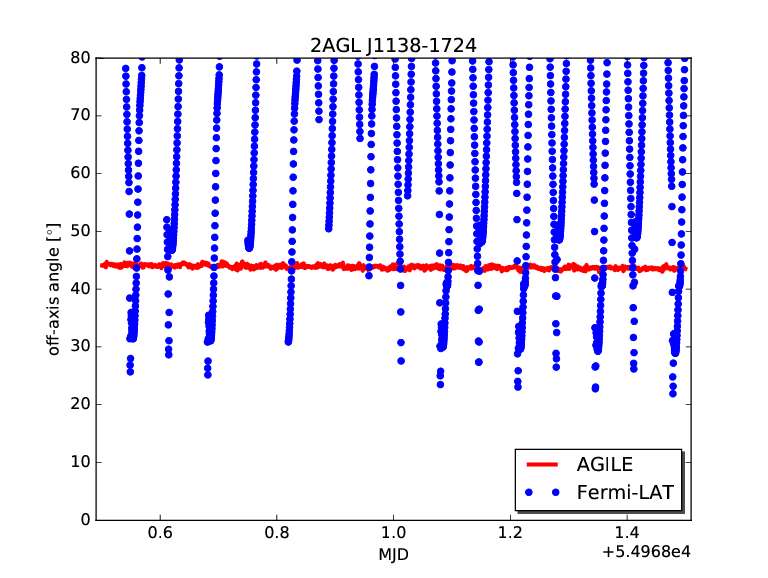}
  \includegraphics[width=9cm]{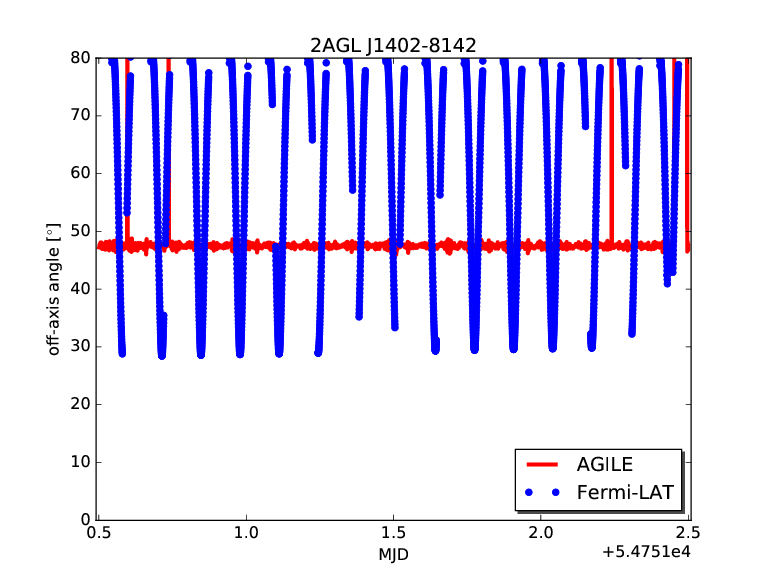}
  \caption{{\it Left panel:} Time-evolution of the off-axis viewing angles 
  for the AGILE-only source 2AGL J1138-1724, as observed by AGILE-GRID (red points) 
  and Fermi-LAT (blue points) during the 1-day time interval MJD 54968.5--54969.5. 
  {\it Right panel}: The same plot of the AGILE-GRID/Fermi-LAT off-axis viewing angles for the 
  2AGL J1402-8142 AGILE-only source, over the time interval MJD 54751.5--54753.5.}
\label{figure_agile-fermi_vis}
\end{figure*}

\abr{To verify that, we have compared the Fermi-LAT attitude data with the AGILE-GRID ones during the time intervals of two transient episodes observed only by AGILE-GRID for the 2AGL J1138-1724 and 2AGL J1402-8142 sources.}

\abr{For the former case, 2AGL J1138-1724, we compared the AGILE-GRID/FERMI-LAT visibility for the $\gamma$-ray flare observed by AGILE-GRID from his position during the 1-day time interval MJD 54968.5-54969.5 (see Sect.~\ref{sect:agileonly}). For this period we found
that the Fermi-LAT observed the 2AGL J1138-1724 sky region at an off-axis angle
greater than 50$^\circ$ for more than 80\% of its total exposure time, while for AGILE-GRID
the off-axis viewing angle was always below 50$^\circ$, due to the constant pointing
attitude (see Fig.~\ref{figure_agile-fermi_vis}, {\it left panel})\footnote{At high values of the off-axis angle ($>50^\circ$), the Fermi-LAT sensitivity is up to 50\% lower than the nominal on-axis value.} }

\abr{Analogously, we compared the AGILE-GRID/Fermi-LAT visibility of the 2AGL J1402-8142 sky region during a transient episode detected by AGILE-GRID on the 2-day timescale ranging from MJD 54751.5 to MJD 54753.5. Again, for the Fermi-LAT, the source viewing angle is mostly above 50$^\circ$ off-axis, while for the AGILE-GRID the source is almost all the time in good visibility (see Figure~\ref{figure_agile-fermi_vis}, {\it right panel}).}

\begin{figure*}[ht]
  \centering
  \includegraphics[width=9cm]{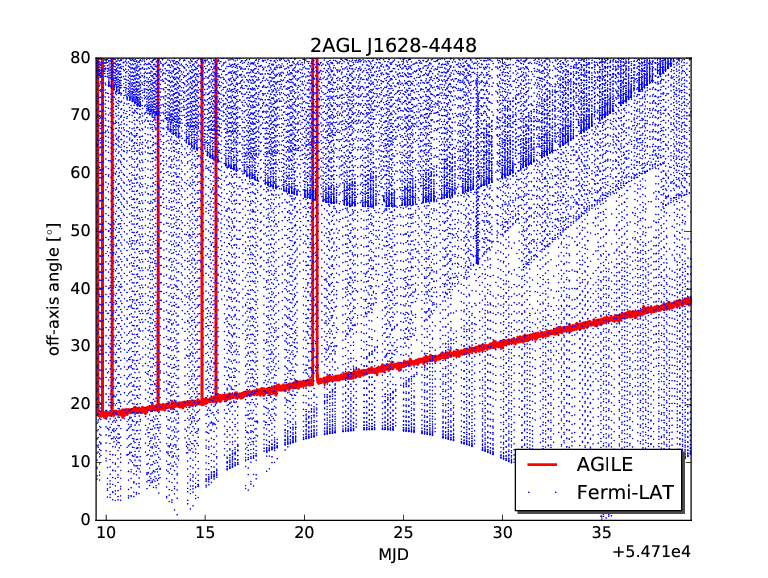}
  \caption{Time-evolution of the off-axis viewing angles for the AGILE-only source 2AGL J1628-4448, as observed by the AGILE-GRID (red points) and the Fermi-LAT (blue points) during the 30-day AGILE OB 6200 time-interval (MJD 54719.5-54749.5).}
\label{figure_agile-fermi_vis_2AGLJ1628}
\end{figure*}

\abr{Even for AGILE-only detections occurring on longer OB time-scales, the AGILE-GRID pointing attitude can always guarantee an optimum source viewing angle with respect to the Fermi-LAT. This is the case, for instance, of the 2AGL J1628-4448 detection during the OB 6200 (30-day long) (see Sect.~\ref{sect:agileonly}): 
also on this longer interval, the source visibility for the AGILE-GRID is, for most of the time, well below an off-axis angle of 40$^\circ$, while for the Fermi-LAT it is just below 50$^\circ$ for 14\% of the time (see Fig.~\ref{figure_agile-fermi_vis_2AGLJ1628}).}

\abr{Besides the source visibility/exposure due to the different observing modes
(AGILE-GRID in pointing, Fermi-LAT in all-sky survey observing mode), particularly relevant
over short time intervals, other reasons can be invoked to explain the Fermi-LAT non-detection of the AGILE-only sources. Among them, we can consider:
source variability; different spectral response of the instruments; event classification algorithms; background model (especially important for sources near the Galactic plane).}

\section{Evaluation of the Maximum Likelihood method for detection of 2AGL sources }

\label{tscat2}

\abr{
The detection process reported in Sect.~\ref{sec:construction} can be divided in the following steps: (i) the determination of the seeds, (ii) iterative analysis of seeds, (iii) refined analysis.} 

\abr{For the evaluation of the list of initial seeds we have considered that the most seeds found with the significance $TS$ maps method has been found also with the Wavelet method where only one trial for each seed has been done. A second trial could be considered during the iterative analysis of seeds.}

\begin{figure*}[ht]
  \centering
  \includegraphics[width=16cm]{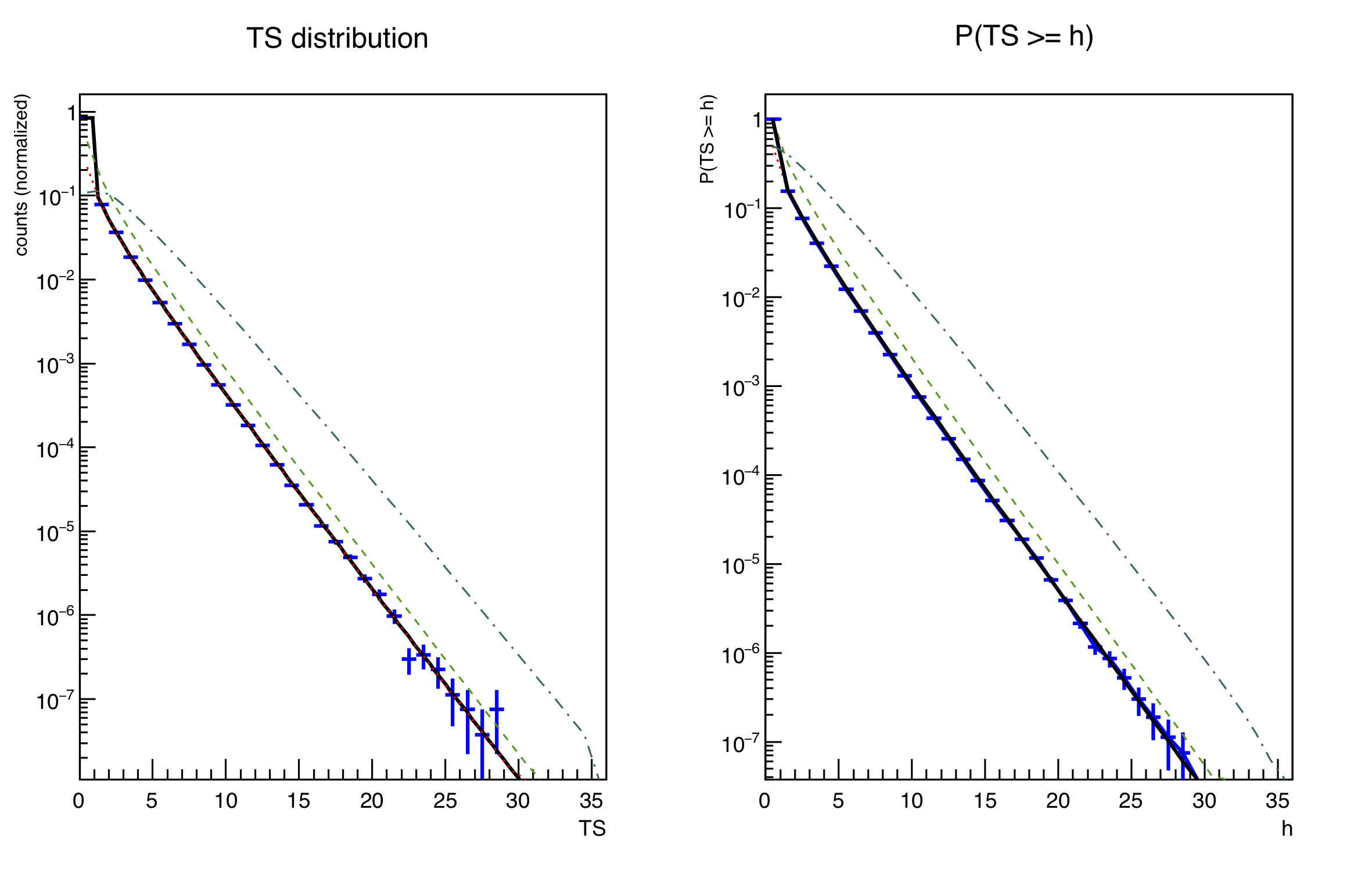}
  \caption{$TS$ distribution (left side) and p-distribution (right side) of a simulated empty field, with $g_\mathrm{gal} = 0$ and $g_\mathrm{iso} = 5$, flux free, and position and spectral parameters fixed. The blue crosses are the calculated distribution, the black line is the best fit, the red dotted line is the $0.5 \chi^2_1$ theoretical 
  distribution, the green dashed line is the $\chi_1$ theoretical distribution, and the cyan dotted-dashed line is the $0.5 \chi^2_3$ distribution.}
\label{figure_ts}
\end{figure*}

\abr{To evaluate the final number of spurious sources involved in the 2AGL procedure, that is related with the refined analysis, we have performed a set of Monte Carlo simulations of AGILE-GRID observations, to compare the data distribution of $TS$ produced by the analysis procedure with that predicted by Wilks’s theorem. The probability that the result of a trial in an empty field has $TS \ge h$ is the complement of the cumulative distribution of $TS$. Simulated data are generated using a background model and the AGILE-GRID instrument response functions described in Sect.~\ref{sec:prep}. The energy range used is 100 MeV--10 GeV. We have considered a typical mean value of the exposure, considering the ring centred at $(l=39.375, b=-22.024)$, simulating observation of an empty field (considering only  isotropic background model  with $g_\mathrm{iso}=5$ and putting $g_\mathrm{gal}=0$), adding Poisson-distributed noise to each pixel, and analysing each resulting sky map exactly as flight data. We performed a maximum likelihood analysis at the centre of the ring, keeping the position and spectral index of a candidate source fixed. Figure \ref{figure_ts} shows the resulting $TS$ distribution (left panel) and the related p-value distribution (right panel); if we fit the $TS$ distribution with the function $\eta \chi^2_1 (TS) $ if $TS > 1$ (see \citep{bulgarelli12a}) we get $\eta=0.5$, i.e. $TS=16$ corresponds to $4 \sigma$ significance.}

\end{appendix}



\end{document}